\documentclass[B5paper,12pt]{report} 
\intextsep
\floatsep                        
\abovecaptionskip
\belowcaptionskip
\usepackage[T1]{fontenc}
\usepackage{graphicx}
\usepackage{epsfig}
\usepackage{indentfirst}
\usepackage{textcomp}
\usepackage{amssymb}
\usepackage{amsmath}
\usepackage{url}
\usepackage{epsfig,graphicx,color,makeidx}
\usepackage{natbib}
\usepackage{threeparttable}
\usepackage{booktabs}
\usepackage{longtable}
\usepackage[skip=5pt]{caption}
\usepackage{tablefootnote}
\usepackage{nomencl}
\usepackage{aas_macros}
\usepackage[b5paper,left=25mm,right=20mm,top=20mm,bottom=20mm]{geometry}
\usepackage{pdfpages}

\graphicspath{{./figs/}}
\begin{document}
 \tracingmacros=0
\pagenumbering{}
\pagestyle{plain} 
\includepdf[pages={1-2},width=\textwidth]{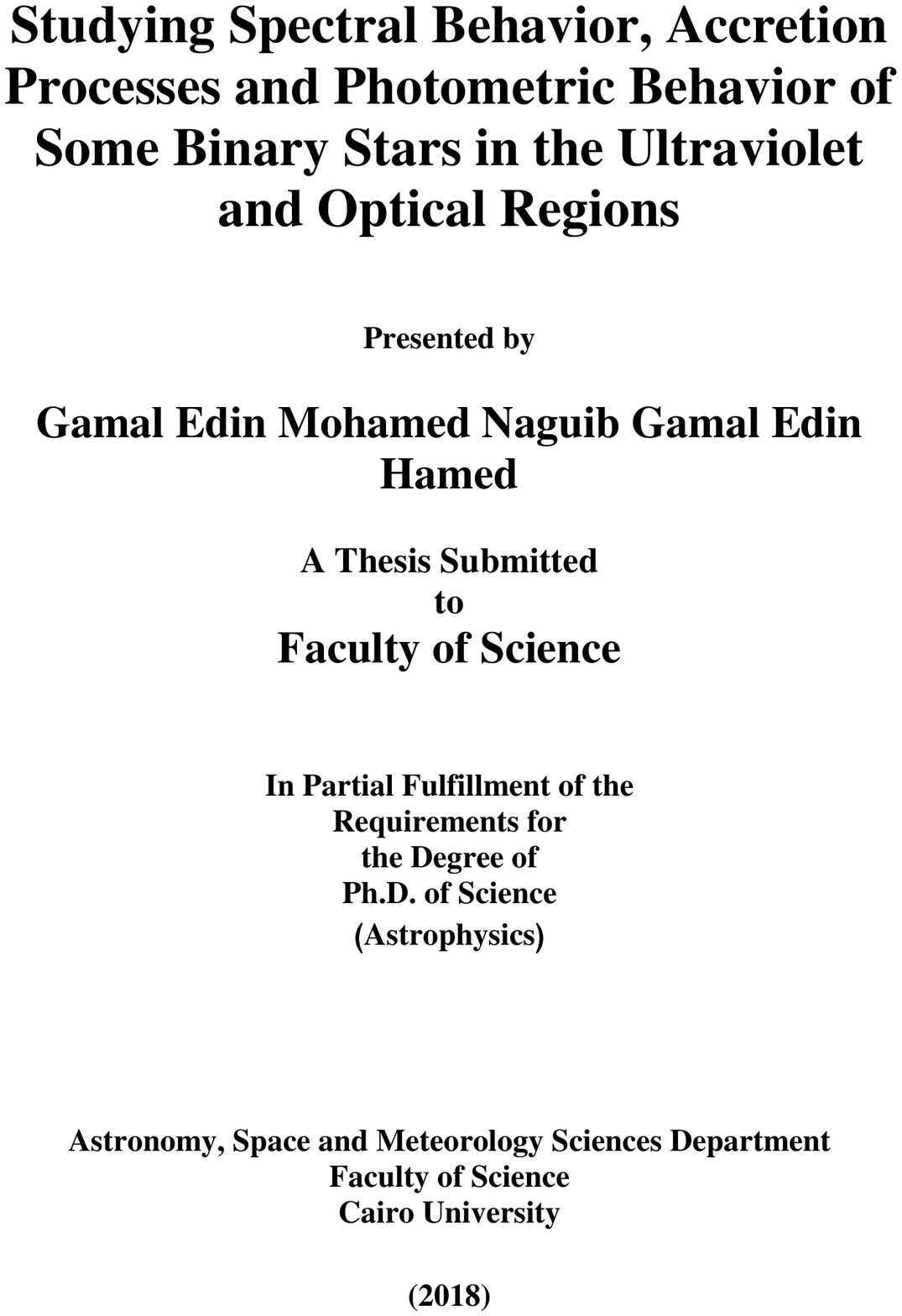}
\includepdf[pages={1},width=\textwidth]{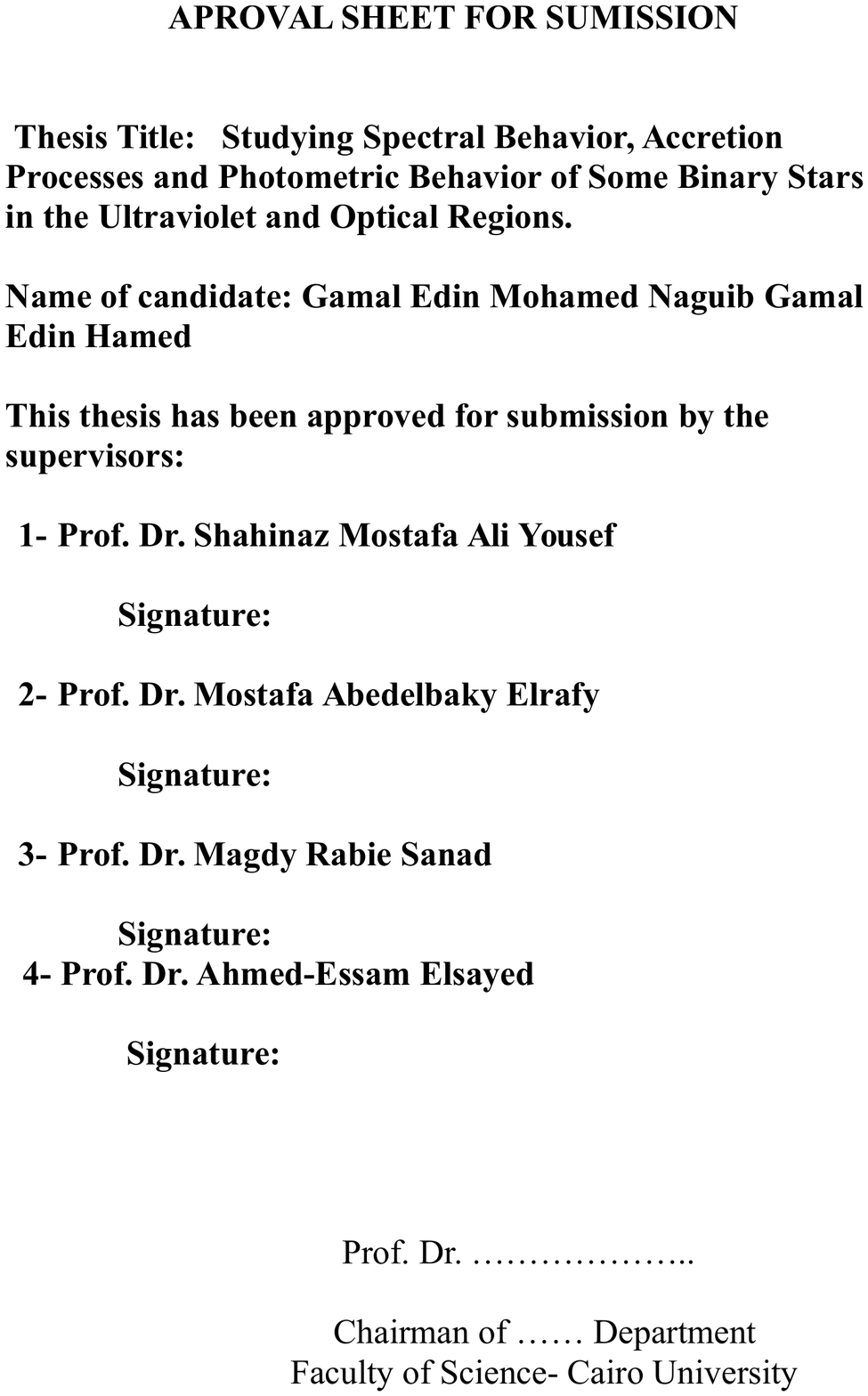}
\includepdf[pages=1,width=\textwidth]{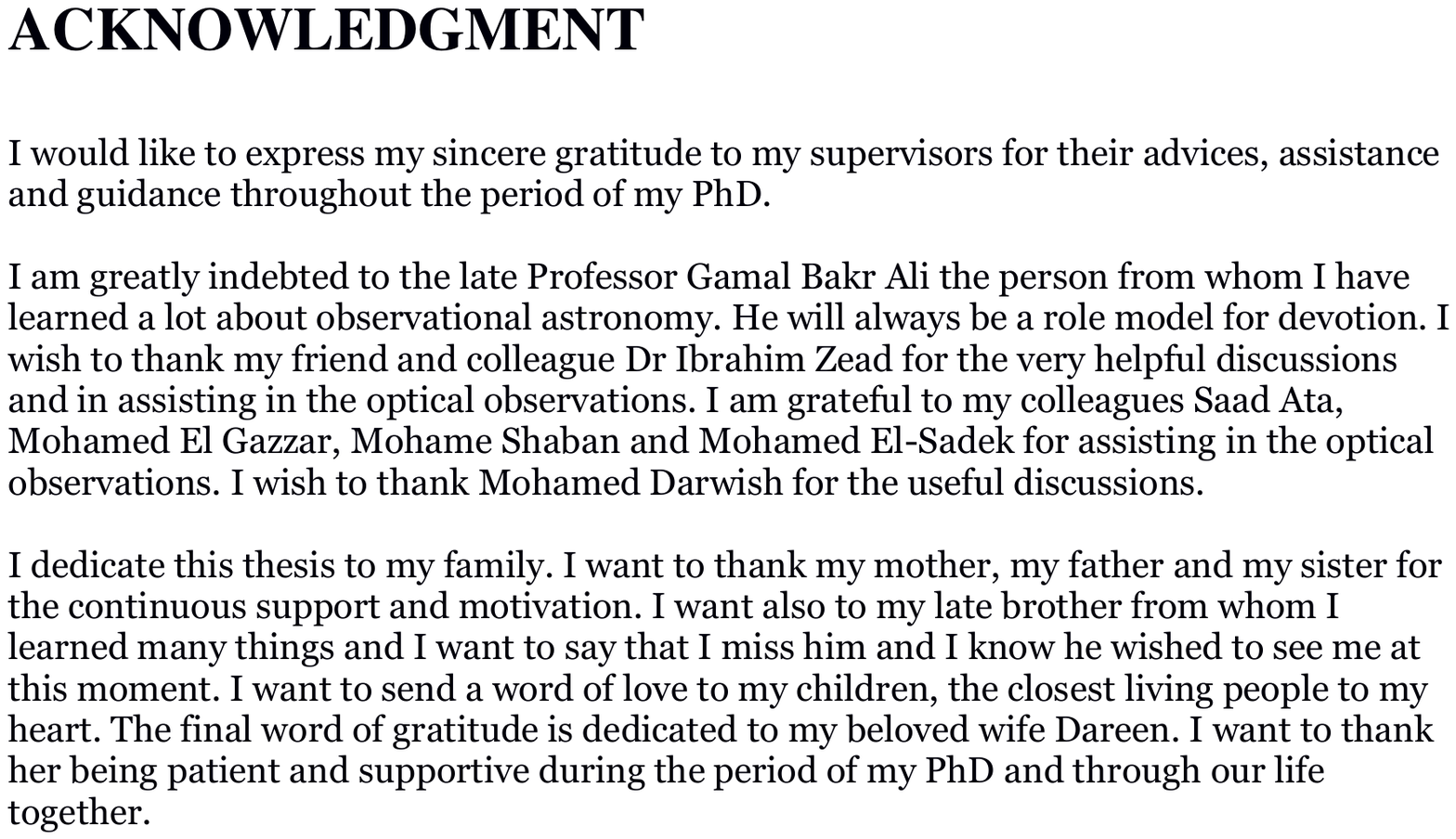}
\clearpage
\pagenumbering{arabic}
\tableofcontents \markboth{}{}
\listoftables  \markboth{}{}
\listoffigures \markboth{}{}
\makenomenclature \markboth{}{}
\renewcommand{\nomname}{List of Symbols and Abbreviations}
\printnomenclature \markboth{}{}
\chapter{Introduction}
\section{Binary Stars}
\begin{figure}
\centering
\includegraphics[height=15cm,width=17cm]{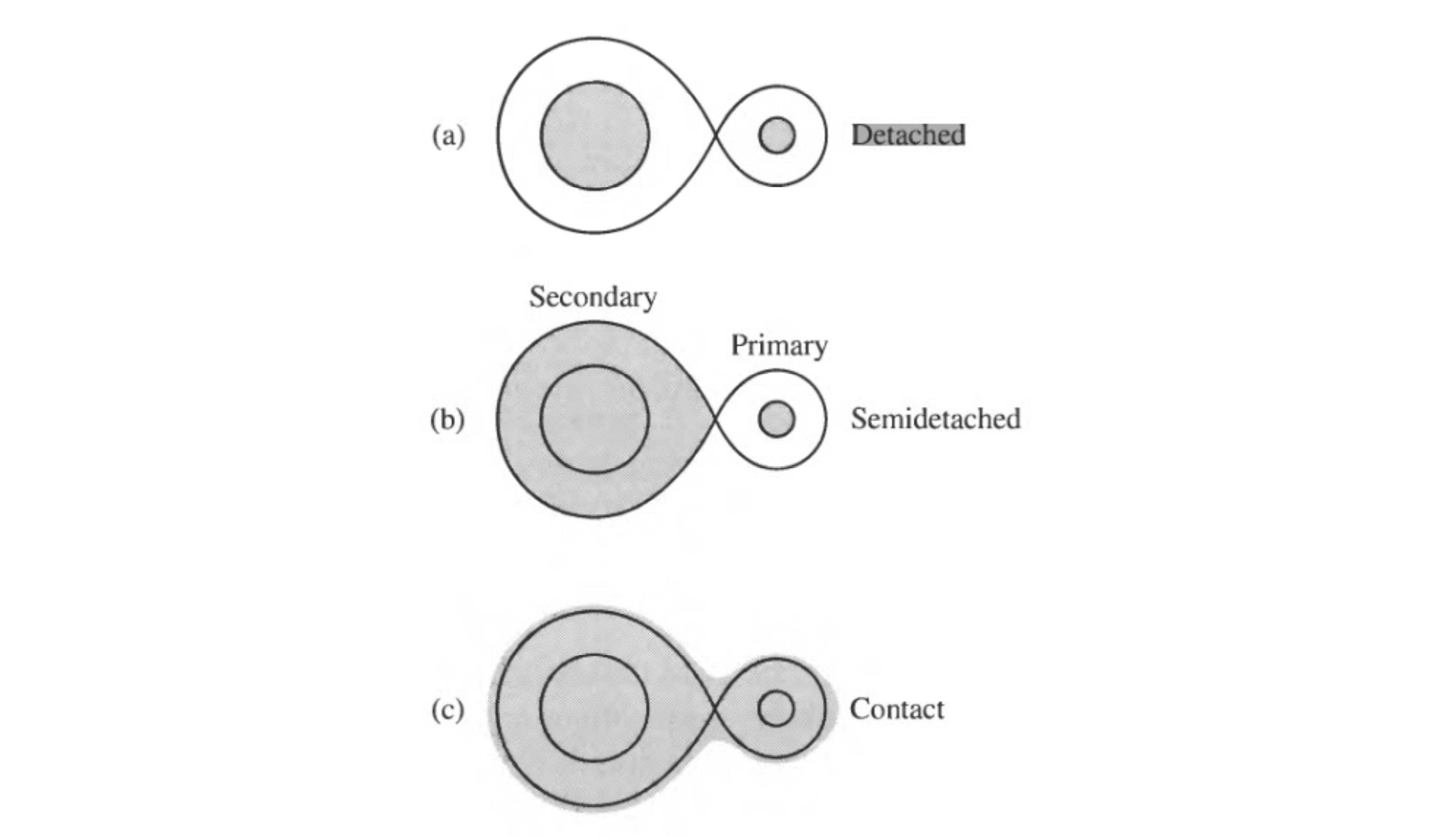} 
\caption[Classes of binary stars]{Classes of binary stars \citep{Carroll2007}}
\label{fig:binaries}
\end{figure}

Binary star system consists of two stars orbiting their mutual center of mass. The study of binary stars is highly important in Astrophysics since it provides us with multiple stellar characteristics. The study of gravitational interaction of a star with other objects is the only method capable of determining stellar masses directly. Depending on the level of interaction between the members, binary stars can be classified to:
Detached binaries, when the radius of each of the two stars is smaller than its Roche lobe radius (the region determining the equipotential surface of the system) and no mass transfer happens between the two stars. If one star fills its Roche lobe and matter starts to transfer through the inner Lagrangian point \nomenclature{$L_1$}{Inner Lagrangian Point} then the system is a semi-detached binary. When both stars fill or exceed their Roche lobe, they have one atmosphere and they are called contact binaries (see fig.~\ref{fig:binaries}).

\section{Cataclysmic Variables}
Cataclysmic variable stars (CVs)\nomenclature{CVs}{Cataclysmic Variables} are a class of semidetached binary stars composed of  a white dwarf ($M_1$\nomenclature{$M_1$}{Mass of The Primary Star}) accreting matter from a late type main sequence or giant star ($M_2$\nomenclature{$M_2$}{Mass of The Secondary Star}). Their variability is powered by the flow of material lost by the Roche lobe filling secondary star on the surface of the primary white dwarf. They are classified into two main categories  depending on the strength of the  magnetic field of the white dwarf. Systems with relatively strong magnetic field $\> 10^5 G$ are called magnetic cataclysmic variables where the magnetic field affects the flow of the accreted material from the secondary on the primary. Magnetic cataclysmic variables are subdivided into polars and intermediate polars. Non-magnetic cataclysmic variables are divided based on their light curves and the frequency of eruptions into Classical Novae, Dwarf Novae, Recurrent Novae and Nova-like \citep{2003cvs..book.....W}.

\subsection{Classical Novae}
In Classical novae, only one eruption has been observed for the system where the magnitude difference ranges from 6 to 19 magnitudes brighter than the prenova stage. There is a strong correlation between the amplitude of the nova and its fading rate where novae with the largest amplitudes fade the fastest while the ones with smallest amplitudes can remain bright for years. Hence we can classify novae according to their fading rate as shown in table ~\ref{tab:novaetypes}

\begin{table}[ht!]
\caption[Speed Classes of Classical Novae]{Speed Classes of Classical Novae \citep{1964gano.book.....P}\label{tab:novaetypes}}
\begin{center}
\begin{tabular}{|c|c|c|}\hline\hline
Speed Class&$t_2$ (days)&Rate of Decline (mag/day)\\
\hline
Very Fast&10 or less&$>$0.20\\
Fast&11 to 25&0.18 to 0.08\\
Moderately Fast&26 to 80&0.07 to 0.025\\
Slow &80 to 150&0.024 to 0.013\\
Very Slow&151 t0 250&0.012 to 0.008\\
\hline
\end{tabular}
\end{center}
\end{table}
\subsubsection{\textnormal{\textbf{Cause of the outburst}}}
\begin{figure}
\centering
\includegraphics[height=15cm,width=13cm]{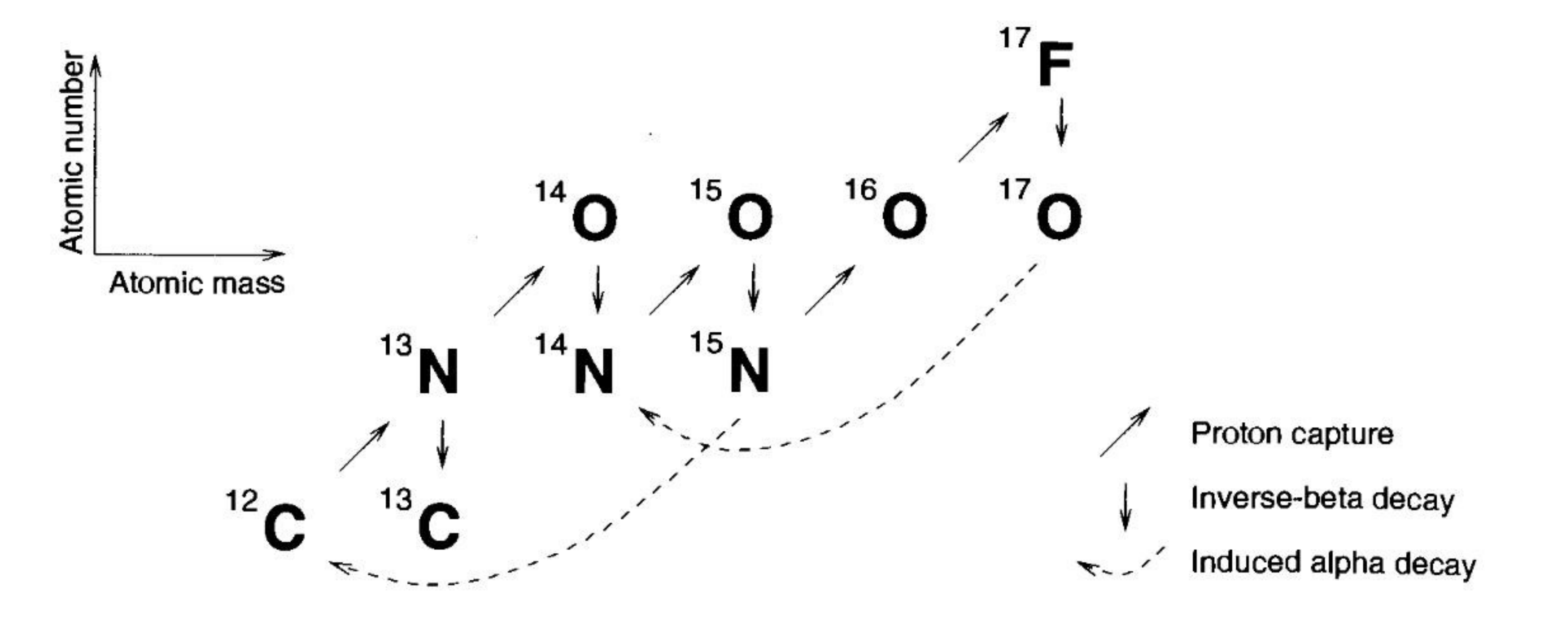} 
\caption[CNO cycle reactions]{CNO cycle reactions \citep{2001cvs..book.....H}}
\label{fig:CNO}
\end{figure}

The Classical Nova outburst occurs due to the accumulation of hydrogen-rich material accreted on the surface of the electron degenerate white dwarf from the companion. This was first suggested by \citet{1962ApJ...135..408K,1964ApJ...139..457K}. The accreted matter increases the density and pressure of the gas which is not allowed to expand and cool down due to degeneracy (degeneration pressure is higher than thermal pressure), therefore the temperature and density increase to the level where proton-proton chain reactions can start. The thermal energy caused by the nuclear reactions does not cause expansion and the temperature increases further causing the rate of nuclear reactions to increase. This effect is called Thermonuclear Runaway (TNR) \nomenclature{TNR}{Thermonuclear Runaway} \citep{2001cvs..book.....H,Petz05} and references therein. The pressure at the  bottom of the accreted layer needs to reach a critical value of  $P_{crit}=\sim 10^{20} \mathrm{dyne cm^{-2}}$ \nomenclature{$P_{crit}$}{Critical Pressure} before the TNR process starts. The ignition mass ($M_{ign}$\nomenclature{$M_{ign}$}{Ignition Mass}) required to be accumulated to get the critical prressure is calculateded using the equation $\colon$

\begin{equation}
M_{ign}=\frac{4\pi P_{crit} R^{4}_\mathrm{{W\-D}}}{G M_\mathrm{{W\-D}}}
\end{equation}

it can be seen that it is inversely proportional to the mass of the white dwarf. For a layer suitable for TNR production to accumulate on the surface of the white Dwarf (WD) \nomenclature{WD}{White Dwarf}, both the mass accretion rate and the WD luminosity have to be sufficiently low. Due to degeneracy, the accreted gas will not expand with increasing temperature. The accreted matter is heated by compression and nuclear reactions (if the temperature reaches $\sim 10^6$K and the proton-proton chain reactions start). Two protons collide and one of them is converted into a neutron and a positron is released. A deutron is formed which later captures a proton and transforms into helium.  When the temperature of the accreted matter reaches $\sim 10^7$K the CNO (Carbon-Nitrogen-Oxygen) \nomenclature{CNO}{Carbon-Nitrogen-Oxygen} cycle starts. Carbon, oxygen and nitrogen nuclei act as catalysts enhancing the transformation of hydrogen into helium as seen in fig.~\ref{fig:CNO}. The temperature increase further enhancing the nuclear reactions (CNO cycle reactions have a rate $\propto  T^{18}$). Due to the high energy at the base of the accreted layer, convection occurs allowing unburned nuclei to be brought to the nuclear burning region. When the temperature reaches $\sim 10^8$K the matter is no longer degenerate and it starts to expand and cool down \citep{2001cvs..book.....H,starrfield2008thermonuclear,2016PASP..128e1001S}.


Novae can be classified depending on the composition of the core of the underlying white dwarf. CO novae occur on white dwarfs with a carbon-oxygen  cores. These novae are characterized by having ejecta with high abundances of carbon and oxygen and they often go through a dust formation phase.  ONe novae occur on oxygen-neon core white dwarfs. They are characterized by showing strong Neon lines in their nebular phase spectra. ONe novae produce stronger explosions which lie within the fast and very fast speed classes \citep{warner2008Properties,gehrz2008ir,2016ApJ...816...26H}. 



\subsection{Dwarf Novae}
Dwarf novae outbursts have magnitude differences in the range from 2 to 5 magnitudes and can reach 8 magnitudes in some rare cases. Dwarf novae are characterized by having several outbursts with intervals ranging from about 10 days to tens of years. Each source has a well defined interval between outbursts. Dwarf nova outbursts occur when the rate of mass transfer in the accretion disc increases temporarily and gravitational energy is released causing the outburst. Dwarf novae have three different sub classes depending on the light curve of the outburst. Z Cam type shows a halt in the light curve at about 0.7 mag below the maximum. SU UMa stars are characterized by having superoutbursts where the outburst lasts for a period $\sim$ 5 times the period of the normal outburst. The third type U Gem does not show the features recognizing Z Cam and SU UMa types.

\subsection{Recurrent Novae}
When a system is classified as classical nova and then another eruption is observed, then it is re classified as a recurrent nova. It is differentiated from dwarf nova spectroscopically where in the case of recurrent novae a shell of gas is ejected at high speeds while for dwarf nova, no shell is ejected away.

Recurrent novae have three subclasses \citep{2003cvs..book.....W}. The first is the T pyx subclass which is characterized by having a short orbital period and long decay time after outburst.The second is the U Sco subclass whose quiescent spectra show that the disc is composed entirely of He II. They have orbital periods of $\sim 1 d$ which suggests that these systems have evolved secondary stars. The third is the T CrB subclass. The outbursts of this subclass rise very quickly and  they are characterized by having giant secondaries of the M spectral type suggesting that they have long orbital periods $\gtrsim 100 d$. 

\subsection{Nova-like}
Nova-like systems are cataclysmic variables with no observed eruptions. It is believed that nova-like systems include pre-nova, post-nova, recurrent nova dwarf nova systems in quiescent states. They are characterized by their low brightness variation and the stable rate of transfer of matter in their disks.
Nova-like systems are subdivided into the following subclasses $\colon$ UX UMa stars whose spectra show broad Balmer absorption lines persistently. RW Tri stars wich are characterized by having emission line dominated spectra which sometimes have absorption cores. It is believed that the differece  between these two subclasses results from the difference in orbital inclination. SW Sex subclass are systems with high orbital inclinations showing large phase shifts in their radial velocity curves.
VY Scl subclass or the anti-dwarf nova are stars whose magnitudes dim for long time intervals (may reach years) and at these times their spectra resemble dwarf novae spectra \citep{2003cvs..book.....W}.

\subsection{Polars}

Polars or AM Herculis stars are cataclysmic variables where the strength of the magnetic fields of the white dwarf are of the order of $\> 10^7 G$. The strong magnetic field forces rotation of the white dwarf to synchronize with the orbital motion. The strength of the magnetic field prevents the formation of a disk and the accreted matter is directed along the field lines and reach the white dwarf from the poles . An accretion column is formed and a shock front is generated due to the interaction between the accretion column and the white dwarf's photosphere emitting hard X-ray photons. These sources provide an opportunity to study the interaction between hot plasma and strong magnetic fields.

\subsection{Intermediate Polars}
Intermediate polars which are also called DQ Her stars are characterized by having a magnetic field which is weaker than that of the polars, therefore it can not force the white dwarf to spin synchronously with orbital motion. The magnetic field disrupts the inner parts of the disk and the motion of accreted matter close to the white dwarf behaves in a similar way to that of the polars. 


\subsection{The White Dwarf}

The White Dwarf is a star whose size is close to that of the Earth while its mass is close to that of the Sun. Most of the hydrogen in the white dwarf has been burnt-out into helium and heavier elements. Therefore, no nuclear reactions occur in its core and it radiates its residual energy and cools down. It is generally believed that white dwarfs originate from the evolution of low and intermediate mass stars ($M \lesssim 8-10 M_{\odot}$ \nomenclature{$M_{\odot}$}{Solar Mass}). However they have an upper limit for their mass which is called the Chandrasekhar Limit ($\sim 1.4 M_{\odot}$). The white dwarf is supported against gravitational collapse by electron degeneracy pressure. In the case of electron degeneracy, high density matter (like in the white dwarf interior) is prevented from further contraction which will lead to having more than one electron with all four quantum numbers equal violating the Pauli Exclusion Principle. Instead electrons are forced to occupy the higher energy states since all the lower energy states are occupied and degeneracy pressure supports the star against further gravitational collapse. This is opposed to the thermal pressure which supports ordinary stars against their gravity.(see \citet{1983bhwd.book.....S,2001cvs..book.....H,Carroll2007,2013sepp.book.....I}). 

Depending on the mass of the progenitor, the core of the white dwarf will be composed of carbon-oxygen (CO) white dwarfs where the nucleosynthesis of elements do not lead to the formation of heavier elements. The masses of CO white dwarfs are typically less than $\sim 1.2 M_{\odot}$. More massive progenitors collapse to form white Dwarfs with oxygen-neon cores (ONe white dwarfs) whose masses are typically larger. The third type of white dwarfs is the Helium core white dwarfs which have the lowest masses $ \lesssim 0.5 M_{\odot}$. They are believed to form from the evolution of close binary systems since they can not have such low masses from single stars within the lifetime of our galaxy \citep{1997ApJ...477..313A}. The atmosphere of the white dwarf is a thin layer composed of lighter elements.


\citet{1972ApJ...175..417N} derived a mass-radius relationship for the white dwarfs in the form $\colon$

\begin{equation}\label{eq:radius}
R_\mathrm{{W\-D}} = 0.78 \times 10^{9}\left[\left(\frac{1.44M_{\odot}}{M_\mathrm{{W\-D}}}\right)^{2/3}-\left(\frac{M_\mathrm{{W\-D}}}{1.44M_{\odot}}\right)^{2/3}\right]^{1/2} (cm)
\end{equation}

\subsection{The Secondary Star}

The secondary star in a CV is a Roche lobe filling star whose main physical characteristics can be known from the Roche lobe properties. The radius of the secondary ($R_2$\nomenclature{$R_2$}{Radius of the Secondary Star}) therefore can be known from the approximate relation:

\begin{equation}\label{radius2}
\frac{R_{2}}{a}= \frac{2}{3^{4/3}}{\left(\frac{q}{1+q}\right)}^{1/3}
\end{equation}
where $a$ \nomenclature{$a$}{Binary Separation} is  the binary separation, $q$ \nomenclature{$q$}{Mass ratio of the two stars $M_2/M_1$} is the mass ratio of the two stars $M_2/M_1$. From Kepler's third law, the orbital period of the binary ($P$\nomenclature{$P$}{Orbital Period of The Binary}) is:
\begin{equation}\label{period}
P^2 = \frac{4\pi^2a^3}{GM}
\end{equation}
where $G$ \nomenclature{$G$}{Universal Gravitational Constant} is the universal gravitational constant and $M=M_1+M_2$. Using the previous two equations, we can get the density of the secondary star, $\overline{\rho}$\nomenclature{$\overline{\rho}$}{The Density of The Secondary Star}
\begin{equation}\label{density}
\overline{\rho}= \frac{3M_2}{4\pi R_2^3} \cong \frac{3^5 \pi}{8 G P^2} \cong 110 \; P_{hr} \; g \; cm^{-3} 
\end{equation}


Absorption lines in the IR spectra of CV secondaries, especially NaI and TiO, show that these stars are late type main sequence stars. Effective temperatures can be estimated using the $\rm{H_{2}O}$ absorption lines at 1.4 $\mu$m and 1.9 $\mu$m. Observing the side of the secondary facing the primary, hence irradiated by the primary star , the boundary layer, the disc and the bright spot can lead to the observation of a spectrum of an earlier spectral type than observed for the unirradiated side \citep{2003cvs..book.....W}and references therein.  

Since these stars are late type main sequence, their mass radius relation can be written in the form 
\begin{equation}\label{mass-radius}
R_2/R_{\odot}=f(M_2/M_{\odot})^{\alpha}
\end{equation}
with $R_{\odot}$ \nomenclature{$R_{\odot}$}{Solar Radius} is the radius of the Sun and $f\simeq \alpha \simeq 1$ \citep{2011arXiv1101.1538K}.  

CV secondary stars tend to be larger than isolated main sequence stars of the same mass, this bloating is due to tidal and rotational distortions. Irradiation may contribute to this swelling to a lesser extent \citep{2011ApJS..194...28K}.
\subsection{Roche lobe}
\begin{figure}
\centering
\includegraphics[height=15cm,width=13cm]{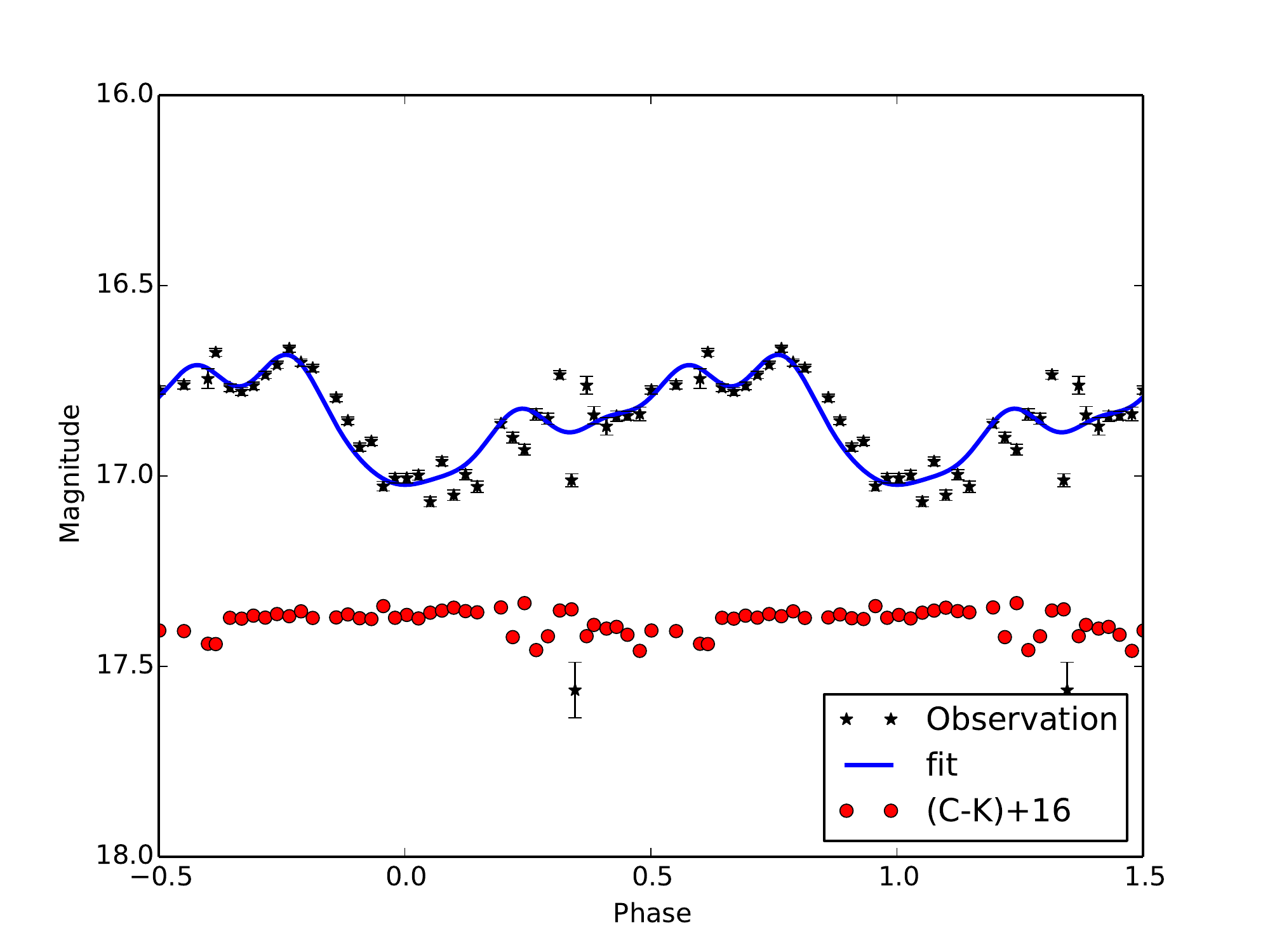} 
\caption[Cross section of the equipotential surfaces in the plane of the binary star orbit]{Cross section of the equipotential surfaces in the plane of the binary star orbit \citep{2002apa..book.....F}}
\label{fig:roche_fig}
\end{figure}

In cataclysmic variables, mass is transferred from the secondary star to the white dwarf via Roche lobe overflow, where the gravitational pull of the white dwarf accretes the outer layers of the secondary. This situation was first studied by the French mathematician Eduoard Roche. The potential of the binary system coming from both the gravitational and centrifugal forces is called the Roche potential ($\Phi_{R}$\nomenclature{$\Phi_{R}$}{Roche Potential}). The equipotential surface has the shape of a peanut and it can be represented by the equation
\begin{equation}\label{roche}
\Phi_{R}(\bold{r})=-\frac {GM_1}{|\bold{r}-\bold{r}_1|}-\frac {GM_2}{|\bold{r}-\bold{r}_2|}-\frac{1}{2}(\omega \times \bold{r})^2
\end{equation}
where $\omega$ \nomenclature{$\omega$}{Angular Frequency of the Binary} is the angular frequency of the binary given by, $\omega = 2\pi/P$ and the last term of the equation comes from the centrifugal force due to the rotation of the binary. The cross section of the equipotential surface in the plane of the orbit is shown in Fig (\ref{fig:roche_fig}). The mass ratio $q=M_2/M_1$ controls the shape of the equipotential. the region surrounding each star is called its Roche lobe. Roche lobe overflow occurs through the inner Lagrangian point $L_1$ where the potential from both stars balance. See \citet{2002apa..book.....F} for a review. 

\subsection{Accretion Disc}
Due to the small size of the white dwarf compared to the binary separation $a$, matter lost from the secondary will not hit the surface of the primary directly. Instead, it will spiral around the primary forming an accretion disc. The matter flowing out of the secondary through the inner Lagrangian point $L_1$ has large specific angular momentum therefore, it does not fall directly towards the surface of the primary but it follows an orbit around the primary. Energy will be dissipated from the gas stream causing the matter to move in a circular orbit in the plane of the primary. The radius of the orbit is smaller than the Roche lobe radius of the primary. A stream of matter will form a ring around the primary and the continuous flow of matter will lead to interaction between gas particles and some of the energy will be dissipated in the form of heat and there will be radiation from the accreted matter. The loss of energy will lead to spiralling of the accreted stream inwards toward the primary losing angular momentum. The ring will then form an accretion disk around the primary. The lost angular momentum will be transferred to the outer parts of the disc causing it to spread outwards. 

The luminosity of the disk \nomenclature{$L_{disc}$}{Luminosity of the Disc} is given by the equation,

\begin{equation}\label{ldisc}
L_{disc} = \frac{GM_\mathrm{{W\-D}}\dot{M}}{2R_\mathrm{{W\-D}}}=\frac{1}{2}L_\mathrm{{acc}}
\end{equation}
where $M_\mathrm{{W\-D}}$ \nomenclature{$M_\mathrm{{W\-D}}$}{Mass of the White Dwarf}, $\dot{M}$ \nomenclature{$\dot{M}$}{Mass Accretion Rate}, $R_\mathrm{{W\-D}}$\nomenclature{$R_\mathrm{{W\-D}}$}{Radius of the White Dwarf} and $L_\mathrm{{acc}}$ \nomenclature{$L_\mathrm{{acc}}$}{Accretion Luminosity} are the mass of the white dwarf, the accretion rate, the radius of the primary and the accretion luminosity respectively. The rest of the accretion luminosity is emitted by the boundary layer between the surface of the star and the accretion disk. And the temperature of the disc at a radius $r$ will be,
\begin{equation}\label{tdisc}
T =\left(\frac{3GM_\mathrm{{W\-D}}\dot{M}}{8\pi\sigma R_\mathrm{{W\-D}}^3}\right) ^{1/4}\left(\frac{R_\mathrm{{W\-D}}}{r}\right)^{3/4}\left(1-\sqrt{R_\mathrm{{W\-D}}/r}\right)^{1/4} 
\end{equation}
where $\sigma$\nomenclature{$\sigma$}{Stefan-Boltzmann constant} is the Stefan-Boltzmann constant \citep{2002apa..book.....F,Carroll2007}


%


\section{Ultraviolet Spectroscopy}

The International Ultraviolet Explorer (IUE\nomenclature{IUE}{International Ultraviolet Explorer}) was launched on Jan $26^{th}$, 1978 and it worked until Sep $30^{th}$, 1996. IUE was a trilateral project between NASA \nomenclature{NASA}{National Aeronautics and Space Administration} (National Aeronautics and Space Administration), ESA \nomenclature{ESA}{European Space Agency} (European Space Agency) and PPARC \nomenclature{PPARC}{Particle Physics and Astronomy Research Council} (Particle Physics and Astronomy Research Council). The IUE had a $45\, \text{cm}$, f/15 Ritchey-Chr\'etien Cassegrain telescope and it was equipped by four cameras, SWP\nomenclature{SWP}{Short-Wavelength Prime} (Short-Wavelength Prime) which worked between $1150-1970\, \text{\AA}$ with a sensitivity of $2 \times 10^{-15}\, erg\, cm^{-2}\,s^{-1}\,\text{\AA}^{-1}$, LWP\nomenclature{LWP}{Long-Wavelength Prime} (Long-Wavelength Prime) which worked between $1750-3300\, \text{\AA}$ with a sensitivity of $1 \times 10^{-15} erg\, cm^{-2}\,s^{-1}\,\text{\AA}^{-1}$, LWR\nomenclature{LWR}{Long-Wavelength Redundant} (Long-Wavelength Redundant) which worked between $1750-3300\, \text{\AA}$ with a sensitivity of $2 \times 10^{-15} erg\, cm^{-2}\,s^{-1}\,\text{\AA}^{-1}$ and the SWR\nomenclature{SWR}{Short-Wavelength Redundant} (Short-Wavelength Redundant) which was never operational. The IUE had two Echelle Spectrographs, the Long-wavelength spectrograph covering the range from  1850 to 3300 $\text{\AA}$ and the short-wavelength spectrograph covering the range from 1150 to 2000 $\text{\AA}$. The IUE Worked for more than 18 years. During the time of its operation it observed 36 classical novae in outburst and 20 old novae.

INES (IUE Newly Extracted Spectra \nomenclature{INES}{IUE Newly Extracted Spectra}) archive is the final archive of the IUE data aiming to ease the access 0f the data. The INES FITS\nomenclature{FITS}{Flexible Image Transport System} (Flexible Image Transport System) files are composed of four columns. The first is the Wavelength in Angstroms $\text{\AA}$, the second column is the flux in $\mathrm{erg\, cm^{-2}\,s^{-1}\,\text{\AA}^{-1}}$, the third column is the sigma is the uncertainty in the flux measurement, the fourth column is the quality of the data.

IRAF \nomenclature{IRAF}{Image Reduction and Analysis Facility} (Image Reduction and Analysis Facility) version 2.16 package distributed by the National Optical Astronomy Observatory (NOAO\nomenclature{NOAO}{National Optical Astronomy Observatory}) was used to normalize the spectra and measure the line fluxes for the objects we present in this work.

In this work, we study the spectral behavior of the three classical novae PW Vul, V1668 Cyg and V1974 Cyg using IUE data along with some optical photometric observations of the three objects. These three novae have different characteristics,  PW Vul and V1668 Cyg are both CO novae one of them is a slow nova while the other is a fast one while V1974 Cyg is an ONe nova.  Therefore studying these three novae would show the effect of the difference in the chemical composition of the central white dwarf and the difference in speed class on the nova outburst and the different emission lines they show on the development of the outburst and the observational properties of the ejecta.

\section{Optical Photometry}

\begin{table}[ht!]                                         
\caption{Known Physical and Orbital Parameters of The Three Novae.}
\label{tab:comp}
\begin{center}
\begin{tabular}{|lrrrr|}\hline\hline                
Parameter      &   PW Vul                                    & V1668 Cyg                    & V1974 Cyg        &            Reference \\  \hline                
$M_{\mathrm{WD}} (\mathrm{M_{\odot}})$      		     &   $0.83 $                    &$0.95 $           &$1.05 $           &             (1)\\
Speed Class         	    &   Slow                                      &fast        &fast       &                    (2)\\
Distance (kpc) 		    &   $2.3 \pm 0.6 $              &$  3.3\pm0.6  $&$ 1.6 \pm 0.2 $ & (3)\\
Orbital Period (d) 		     &   0.213700              &0.138400  &0.081263 & (4)\\
\hline                                                
\end{tabular}
\vspace{+0.5cm}
\caption*{References: (1) \citet{2015ApJ...798...76H} for PW Vul, \citet{2006ApJS..167...59H} for V1668 Cyg and \citep{2005ApJ...631.1094H} for V1974 Cyg. (2) \citet{1988ApJ...329..894G} for PW Vul, \citet{1980A&A....81..157D} for V1668 Cyg and \citep{1993A&A...277..103C} for V1974 Cyg. (3) {\it Gaia} DR2 \citep{2016A&A...595A...1G,2018arXiv180409365G,2018arXiv180409376L,2018arXiv180409366L}\tablefootnote{This work has made use of data from the European Space Agency (ESA) mission
{\it Gaia} (\url{https://www.cosmos.esa.int/gaia}), processed by the {\it Gaia}
Data Processing and Analysis Consortium (DPAC,
\url{https://www.cosmos.esa.int/web/gaia/dpac/consortium}). Funding for the DPAC
has been provided by national institutions, in particular the institutions
participating in the {\it Gaia} Multilateral Agreement.}. (4) \citet{1998A&AS..129...83R} for PW Vul, \citet{1990MNRAS.245..547K} for V1668 Cyg and \citep{1994ApJ...431L..47D} for V1974 Cyg.}                                         
\end{center}                                          
\end{table} 

All the photometric observations presented in this thesis were performed using the CCD\nomenclature{CCD}{Charged-coupled Device} camera attached to the Newtonian focus of the 1.88 m telescope in Kottamia Astronomical Observatory (KAO\nomenclature{KAO}{Kottamia Astronomical Observatory}), Egypt. The Telescope and its instruments are described in detail in \citet{2014arXiv1402.2926A}. 
The limited telescope time, the faintness of the sources along with the challenging observing conditions (weather and light pollution) prevented us from performing complete light curve analysis for the three post-novae.

\subsection{V1668 Cyg\nomenclature{Cyg}{Cygni}}

Photometric observations of classical novae focus mainly on determining their fading rate rather than finding short term variability in their light curves. However, some observations were used for the latter cause. For Nova Cygni 1978 (V1668 Cyg) this started in the first week after the outburst. \citet{1980A&A....85L...4C} detected variations with a 0.4392 d period in the the light curve of this system few days after the outburst. They attribute these variations to perturbations on the envelope caused by the binary orbiting inside it. \citet{1984AcA....34..473P} detected variability with an amplitude of 0.36 mag and a period of 0.07 d in the observations they performed in August 1981. \citet{1990MNRAS.245..547K} observed V1668 Cyg for three nights in June 1989 at Kitt Peak National Observatory using a broadband BV filter. By the time of the observations the system has reached a B magnitude of $\sim$ 20. In these observations there is a clear eclipse at the epoch JD 2447679.848 lasting for about 0.03 d with a period of 0.138400 d between consecutive eclipses. This period is clearly the orbital period of the binary.

\subsection{PW Vul}
Thousands of PW Vul \nomenclature{Vul}{Vulpeculae} (Nova Vul 1984a) observations are available on the American Association of Variable Star Observers (AAVSO\nomenclature{AAVSO}{American Association of Variable Star Observers}) archive. Although, most of these observations were visual, some were done using Johnson V filter. These observations were mainly used to show the long term evolution of the system's brightness rising to and declining from its maximum and they were not used to determine short term variabilities in the system's light curve.

\citet{1987IBVS.2979....1H} observed PW Vul between Sep $30^{th}$ and Oct $10^{th}$, 1986 and found an epoch at JD 2446704.263 and found an orbital period of 0.21372 d.

\subsection{V1974 Cyg}
\citet{1994ApJ...431L..47D} calculated the ephemeris for V1974 Cyg (Nova Cygni 1992) and the epoch in their paper is JD 2449267.562 and the period they calculated was 0.081263 d.
Another period of 0.085 d has been found for V1974 Cyg this was interpreted to be due to the spin of the WD \citep{2002AcA....52..273O} or the superhump phenomenon \citep{1997PASP..109..114S}.

Multiple observations were made for the system aiming to determine the periods in its light curve. Many times of minima and maxima were determined and used to construct an O-C curve to determine the stability of the orbits and it was found that the 0.085 d period was decreasing possibly leading to that the spin period of the white dwarf will eventually synchronize with the orbital period \citep{1995AcA....45..365S,2002AcA....52..273O}.

\chapter{Observations and Data Reduction}

\section{Ultraviolet Spectroscopy}

The study of classical novae in outburst has a great importance in astrophysics since it helps in understanding the nature of white dwarfs, the evolution of binary systems, thermonuclear runaway processes and the hydrodynamics of the explosion where we can determine the mass, surface temperature and core composition of the white dwarf as well as the dynamics and chemical abundance of the ejecta by studying the nova outburst in different phases.

The evolution of the spectrum of a classical nova during the outburst passes through different stages\citep{1994ApJ...421..344S,2001MNRAS.320..103S,2005A&A...439..205C,shore2008evolution,2012BASI...40..185S}. The first one is the "fireball" phase which appears with the explosion as the ejecta is heated by the resulting shock. This phase is characterized by the high optical thickness of the ejecta in both the continuum and the lines. This phase is rarely observed in the classical novae since it lasts for a very short time, typically few days. Fig ~\ref{fig:norm_SWP44030LL} shows the short wavelength ultraviolet spectra of V1974 Cyg in the fireball stage. Then the "iron curtain " phase follows, during which the ejecta cool down and Iron peak elements start to recombine and blanket the spectrum. The third phase, known as  "lifting the iron curtain", is characterized by the decrease in the opacity of the UV lines in the envelope. In this phase, the pseudo-photosphere pulls back towards hotter regions enhancing the ionization and the absorption lines in the UV disappear. \citet{2005A&A...439..205C} found that this phase ends after the visual maximum by about $1.1\, t_3$ in novae occurring in white dwarfs with carbon-oxygen cores (see Figs ~\ref{fig:normswp23675}, ~\ref{fig:normswp02655} and ~\ref{fig:norm_SWP44043LL} for samples of the Fe optically thick spectra). Then the nova spectrum evolves into the "pre-nebular" or "transition" phase, which is characterized by the appearance of semi-forbidden lines and the decrease of the opacity of the ejecta. The emission arises now from low density regions as can be seen in the increasing strength of the C III] line. The maximum flux of the C III] line marks the end of this phase, usually about $2.45\, t_3$ days after the visual maximum (see Figs ~\ref{fig:normswp03134} and ~\ref{fig:norm_SWP44389LL}).  The final phase is the "nebular" phase, during which the spectrum shows (i) strong forbidden emission lines (such as [O I], [N II] and [O III]) and (ii) the emission lines of high ionization states (for example: C IV and N V) reach their maximum flux (Figs ~\ref{fig:normswp26244}, ~\ref{fig:normswp03714} and ~\ref{fig:norm_SWP44762LL}). Then the spectrum returns to its prenova appearance (Figs ~\ref{fig:normswp28461},~\ref{fig:normswp09079} and ~\ref{fig:norm_SWP48220LL}).

The ultraviolet spectra of novae contain a wide range of intercombination, resonance and forbidden lines for many elements, therefore the physical conditions of the ejecta and elemental abundances can be determined accurately \citep{1998ESASP.413..367G}. For example the electron temperature ($T_e$) can be determined from the ratio of the flux of a line produced by dielectronic recombination (e.g. C II 1335) to the flux of a line (of the same element) produced by collisional excitation (e.g. C III] 1909). Knowing the electron temperature, abundances can be determined from the fluxes of lines (see \citealt{1981MNRAS.197..107S}, their equations 5.8, 5.9, 5.11, 5.12 and 6.1). Some of the ultraviolet emission lines are characteristic of some of the phases of the outburst.

\citet{1998ApJ...494..680J} developed a hydrodynamical model to follow the progress of a nova outburst from the accretion stage up to the explosion. The input parameters of the model are the white dwarf (WD) mass (ranging from 0.8 to 1.15 $M_{\odot}$ for carbon-oxygen core (CO\nomenclature{CO}{Carbon-Oxygen} models) and from 1.0 to 1.35 $M_{\odot}$ for oxygen-neon core (ONe\nomenclature{ONe}{Oxygen-Neon} models)) and the mixing ratio of the elements between the core and the envelope (ranging from 25 \% to 75\% ). The mass of the accreted envelope, the mass of the ejected envelope, the ejection speed and the duration of the accretion process are among the main outputs of the model. CO models assume a white dwarf with core composition of $X(^{12}C)=0.495$, $X(^{16}O)=0.495$ and $X(^{22}Ne)=0.01$ accreting solarlike matter from the companion at a rate of $2 \times 10^{-10} \mathrm{M_{\odot} yr^{-1}}$ and the mass of the white dwarf is $0.8 \mathrm{M_{\odot}}$ for CO1 and CO2 models, $1.0 \mathrm{M_{\odot}}$ for CO3 model and $1.15 \mathrm{M_{\odot}}$ for models CO4 to CO7. The degree of mixing between the accreted envelope and the white dwarf's core is $25 \%$ for CO1 and CO2 models, $50 \%$ for CO2, CO3, CO5 and CO7 models and $75 \%$ for the CO6 model. 

ONe models assume white dwarf mass of $1.0 \mathrm{M_{\odot}}$ for the ONe1 model, $1.15 \mathrm{M_{\odot}}$ for the ONe2, ONe3 and ONe4 models, $1.25 \mathrm{M_{\odot}}$ for the ONe5 model and $1.35 \mathrm{M_{\odot}}$ for the ONe6 an ONe7 models. The degree of mixing between the accreted envelope and the white dwarf's core is $25 \%$ for ONe2 model, $50 \%$ for ONe1, ONe3, ONe5 and ONe6 models and $75 \%$ for the ONe4 and ONe7 models. Each model also determines the chemical composition of the ejecta (\citealt{1998ApJ...494..680J} , see their Tables 1-4). The chemical composition and the dynamical outputs of the model can be compared to observations to determine the model best describing each nova outburst.

\citet{2005ApJ...623..398Y} used a lagrangian hydrodynamic code to follow the development of the nova outburst on CO white dwarfs by varying three parameters (the mass of the white dwarf $M_\mathrm{{W\-D}}$, the temperature of the core $T_\mathrm{{W\-D}}$ and the mass transfer rate $\dot{M}_\mathrm{{acc}}$). Among the results of the code are the accreted mass, the ejected mass , the expansion velocity, the maximum luminosity, the chemical composition of the envelope and the recurrence time of the nova outbursts. Although they assumed a CO white dwarf core composition, they argue that the results are applicable to ONe novae as well.

In this chapter, we present ultraviolet observations of three classical novae (PW Vul, V1668 Cyg and V1974 Cyg) obtained with the International Ultraviolet Explorer (IUE) satellite, following their spectral evolution after the nova events. The spectra and data reductions are presented in sections ~\ref{sec:obs1}- ~\ref{sec:obs3}.


International Ultraviolet Explorer low resolution ($\sim 6 \,\AA$ ) short wavelength (1150 to 2000 $\AA$) spectra are used in the present work. All the spectra were taken by the short wavelength prime camera with low dispersion (this mode uses a cross-disperser grating yielding a $\sim 6 \, \AA$ resolution) and large aperture (a 10 $\times$ 20 arcsecond slit). The files were retrieved through the INES (IUE newly extracted spectra) server at \url {http://sdc.cab.inta-csic.es/ines/}. 

\subsection{PW Vul}\label{sec:obs1}
IUE observations of PW Vul are presented in Table ~\ref{tab:journal1}. These observations were taken through the period between 12/08/84 and 09/06/86. Sample spectra in different phases illustrating different emission lines are presented in Figs ~\ref{fig:normswp23675}-~\ref{fig:normswp28461}. Making use of the quality column in the FITS files of the spectra and following the guidance of \citeauthor{1998ESASP.413..715L}(1998, see their Table 1), we inspected each spectrum by eye and discarded those with high noise. Image Reduction and Analysis Facility (IRAF) version 2.16 was used to normalize the spectra and measure the line fluxes. The continuum level was best fitted using a fifth order Chebyshev function, which was then used to normalize the spectrum. The properties of the emission lines were then measured interactively using the splot task of the onedspec package where a gaussian fit was applied to each selected line to determine its properties.

\begin{table}
\caption{Journal of IUE Observations for PW Vul}
\begin{center}
\label{tab:journal1}                                                                                             
\begin{tabular}{|c|c|c|c|}\hline
Spectrum ID   &  JD & Days after Explosion&   Exposure Time  (s) \\
\hline
SWP23577LL    &2445915.0  & 7   & 1500         \\   
SWP23580LL    &2445915.3  & 7   & 1200         \\   
SWP23603LL    &2445917.9$^1$& 10  & 900          \\   
SWP23604LL    &2445917.9  & 10  & 720          \\   
SWP23605LL    &2445918.0  & 10  & 810          \\   
SWP23638LL    &2445920.3  & 12  & 2100         \\   
SWP23675LL    &2445924.9  & 17  & 2519         \\   
SWP23717LL    &2445930.8  & 23  & 99           \\   
SWP23825LL    &2445943.3  & 35  & 1800         \\   
SWP23837LL    &2445944.9  & 37  & 300          \\   
SWP23860LL    &2445947.0  & 39  & 300          \\   
SWP24072LL    &2445972.6  & 65  & 600          \\   
SWP24473LL    &2446018.2$^2$  & 110 &  600         \\
SWP25622LL    &2446163.2$^3$  & 255 &  1200        \\
SWP25623LL    &2446163.3  & 255 &  300         \\
SWP26243LL    &2446241.1  & 333 &  360         \\
SWP26244LL    &2446241.1  & 333 &  1200        \\
SWP26342LL    &2446249.6  & 342 &  2219        \\
SWP26427LL    &2446263.2  & 355 &  600         \\
SWP26470LL    &2446271.6  & 364 &  1080        \\
SWP26997LL    &2446367.8  & 460 &  900         \\
SWP28068LL    &2446521.0$^4$  & 613 &  4199        \\
SWP28461LL    &2446591.0  & 683 &  4199        \\
\hline
\end{tabular}
\end{center}
\caption*{Notes: Times of different phases in the evolution of the outburst. (1) The visual maximum. (2) Start of the pre-nebular phase. (3) Start of the nebular phase. (4) Start of the quiescent phase.}                                                                                                         
\end{table}                                                                                                     

\begin{figure}
\centering
\includegraphics[height=14cm,width=13cm]{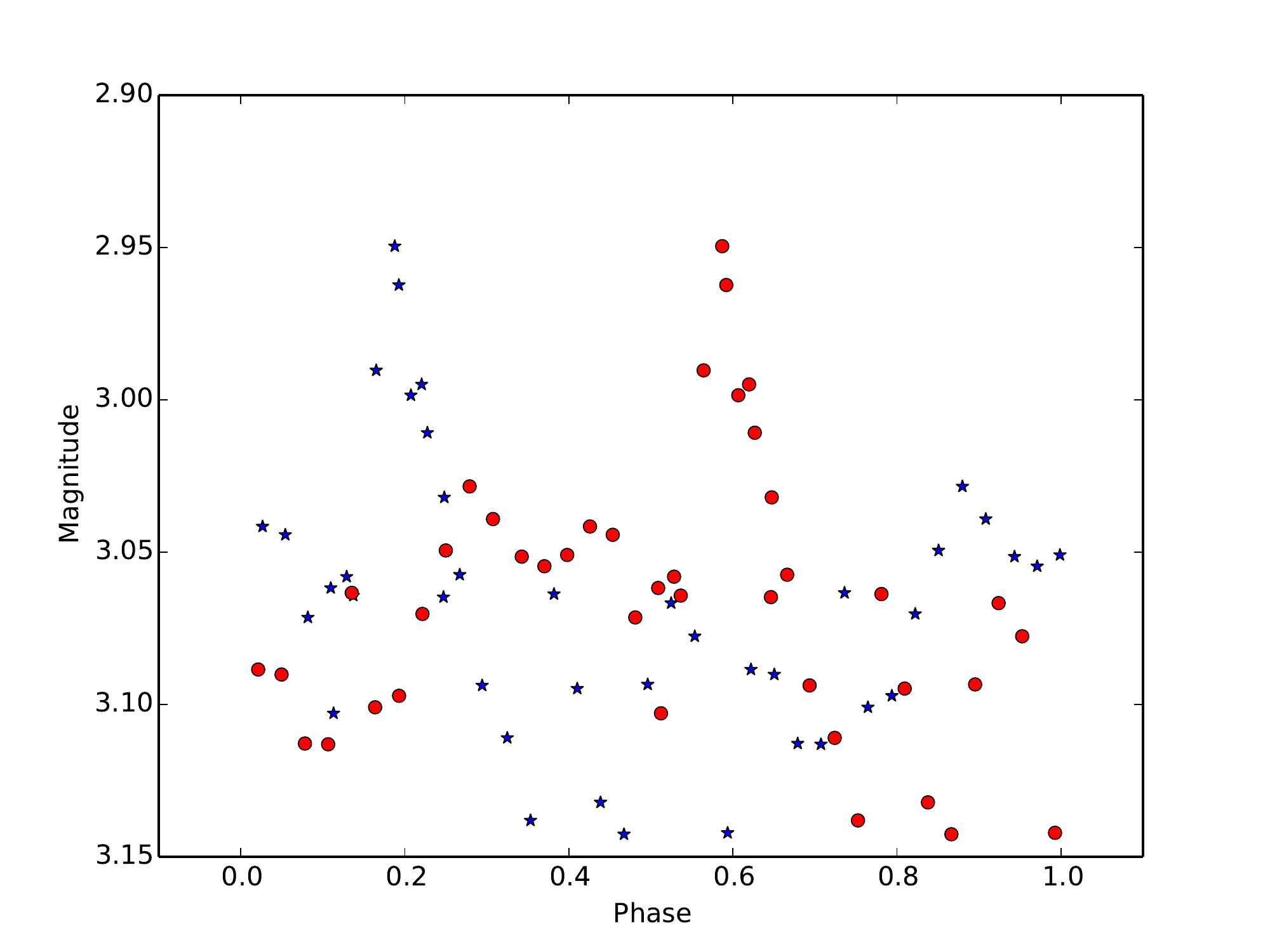}
\caption[PW Vul UV V Magnitude Evolution]{PW Vul V magnitude evolution from AAVSO archives. The dashed line represents the time of visual maximum and the dash-dotted lines represent the end of the different phases of evolution. The dotted line represents the start of the quiescent phase.}
\label{fig:pwvul_aavso}
\end{figure}

\begin{figure}
\centering
\includegraphics[height=14cm,width=13cm]{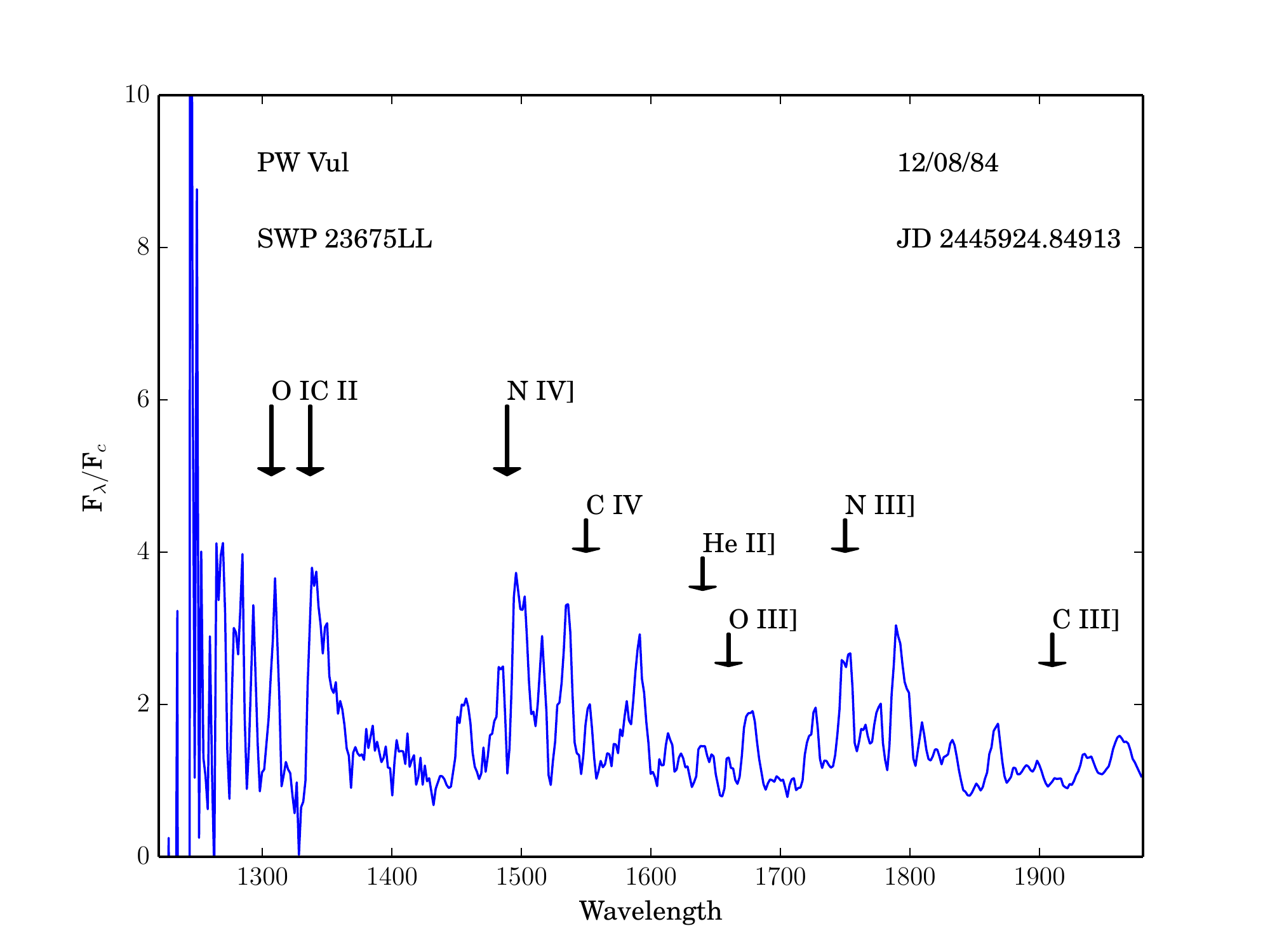}
\caption{PW Vul lifting the iron curtain (initial) Phase.}
\label{fig:normswp23675}
\end{figure}

\begin{figure}
\centering
\includegraphics[height=14cm,width=13cm]{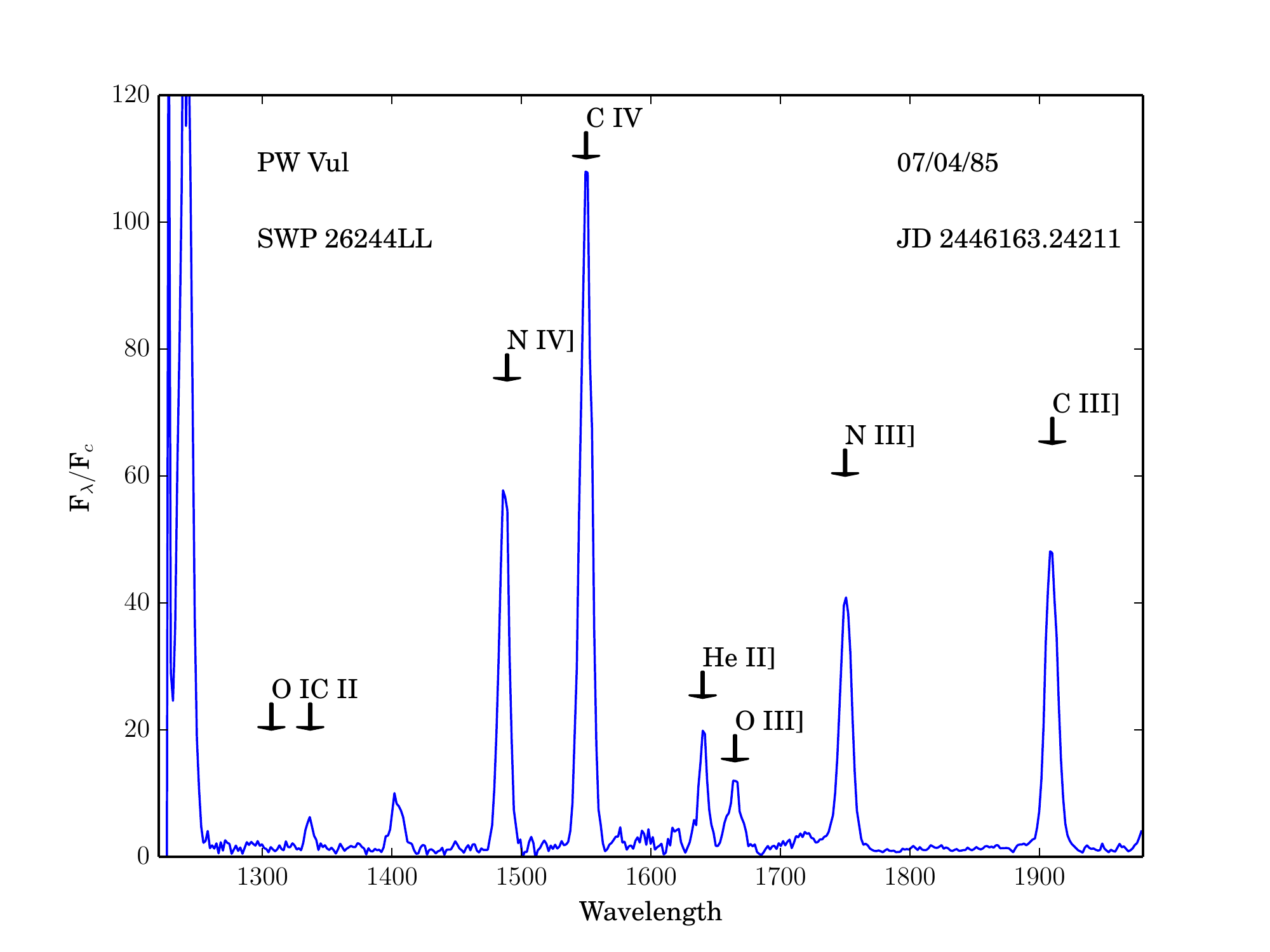}
\caption{PW Vul Nebular Phase.}
\label{fig:normswp26244}
\end{figure}

\begin{figure}
\centering
\includegraphics[height=14cm,width=13cm]{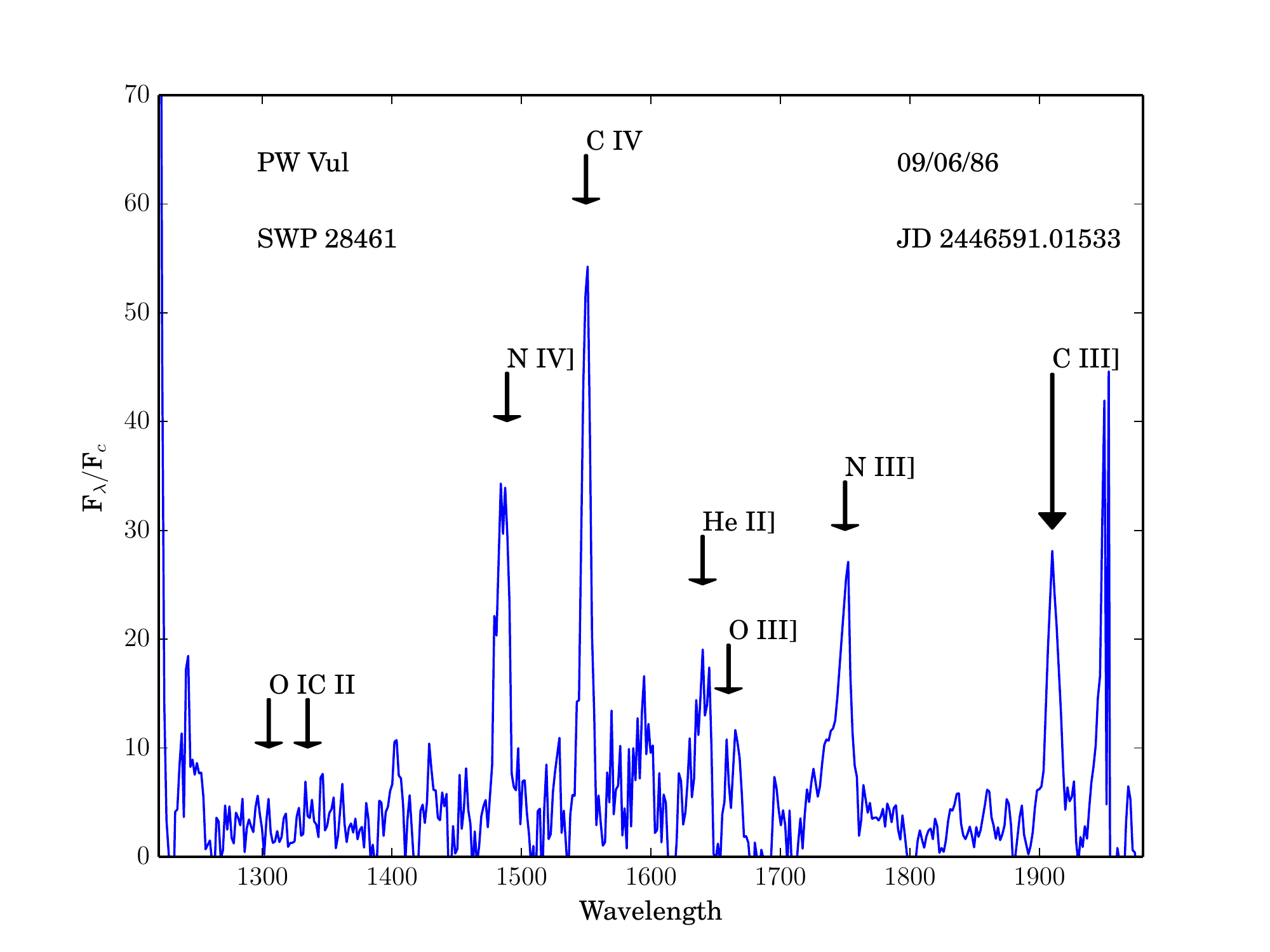}
\caption{PW Vul Quiescent Phase.}
\label{fig:normswp28461}
\end{figure}

\subsection{V1668 Cyg}\label{sec:obs2}

V1668 Cyg observations cover the period between 11/09/78 and 24/12/80. The observations are presented in Table \ref{tab:journal2}. Sample spectra in different phases illustrating different emission lines are presented in Figs ~\ref{fig:normswp02655}-~\ref{fig:normswp09079}.

\vspace{+0.5cm}
\begingroup
\small \renewcommand{\arraystretch}{0.8} \setlength{\tabcolsep}{3pt}
\begin{longtable}{|c|c|c|c|}
\caption{Journal of IUE Observations for V1668 Cyg}
\label{tab:journal2}
\renewcommand{\arraystretch}{0.7}
\cr \hline
Spectrum ID & JD            & Days after Explosion   &  Exposure Time (s)  \\
\hline
  \endfirsthead
  \caption*{Continued. Journal of IUE Observations for V1668 Cyg}\\
  \hline  
Spectrum ID & JD            & Days after Explosion   &  Exposure Time (s)  \\
  \hline
  \endhead
  \hline
  \endfoot
SWP02627LL& 2443763.244$^1$ &4  &  180    \\
SWP02636LL& 2443764.2 &5  & 1500   \\
SWP02641LL& 2443765.0 &6  & 1200    \\
SWP02653LL& 2443767.0 &8  & 1200    \\
SWP02655LL& 2443767.2 &8  & 2099   \\
SWP02666LL& 2443768.6 &10 &  900   \\
SWP02679LL& 2443769.9 &11 & 1200   \\
SWP02680LL& 2443767.0 &11 &  600   \\
SWP02697LL& 2443771.3 &12 &  119   \\
SWP02702LL& 2443771.7 &13 & 1800   \\
SWP02703LL& 2443771.8 &13 & 300    \\
SWP02727LL& 2443773.8 &15 & 1800   \\
SWP02734LL& 2443774.6 &16 & 1380    \\
SWP02742LL& 2443775.5 &16 & 180     \\
SWP02752LL& 2443776.6 &18 & 1200    \\
SWP02795LL& 2443779.9 &21 & 900      \\
SWP02820LL& 2443782.9 &24 & 480     \\
SWP02883LL& 2443789.8 &31 & 119     \\
SWP02884LL& 2443789.9 &31 & 480     \\
SWP02902LL& 2443792.4 &33 & 540      \\
SWP02966LL& 2443795.6 &37 & 96      \\
SWP02990LL& 2443797.4 &38 & 60      \\
SWP03011LL& 2443799.3$^2$ &40 & 300     \\
SWP03089LL& 2443803.4 &44 & 96       \\
SWP03134LL& 2443806.8 &48 & 150     \\
SWP03135LL& 2443806.9 &48  &  150     \\
SWP03136LL& 2443806.9 &48 & 150       \\
SWP03169LL& 2443808.9  &50 & 180      \\
SWP03182LL& 2443810.8  &52 & 180      \\
SWP03190LL& 2443811.4 &52 & 360       \\
SWP03203LL& 2443813.8 &55 & 180       \\
SWP03238LL& 2443819.3 &60 &  540      \\
SWP03274LL& 2443821.6&63  & 150       \\
SWP03362LL& 2443829.1 &70 & 540       \\
SWP03375LL& 2443830.8$^3$ &72 & 292       \\
SWP03526LL& 2443847.2 &88 & 540       \\
SWP03532LL& 2443847.8 &89 & 720       \\
SWP03714LL& 2443869.2 &110 & 720      \\
SWP03886LL& 2443885.2 &126 & 180      \\
SWP03908LL& 2443887.0 &128 & 2400     \\
SWP04505LL& 2443938.7 &180  & 600     \\
SWP04637LL& 2443947.7 &189 & 180      \\
SWP04720LL& 2443954.4 &195 & 1620     \\
SWP04737LL& 2443956.8 &198 & 1500     \\
SWP05755LL& 2444063.5 &305 & 18000    \\
SWP06077LL& 2444090.7$^4$ &332 & 14400    \\
SWP07621LL& 2444248.3 &489 & 14400    \\
SWP09065LL& 2444380.7 &622 & 23700    \\
SWP09079LL& 2444382.7 &624 & 24780    \\
SWP10886LL& 2444598.6 &840 & 21600    \\
\hline                                                                                                       
\caption*{Notes: Times of different phases in the evolution of the outburst. (1) The visual maximum. (2) Start of the pre-nebular phase. (3) Start of the nebular phase. (4) Start of the quiescent phase.}
\vspace{-1.18cm}                                                                                                    
\end{longtable}                                                  


\begin{figure}
\centering
\includegraphics[height=14cm,width=13cm]{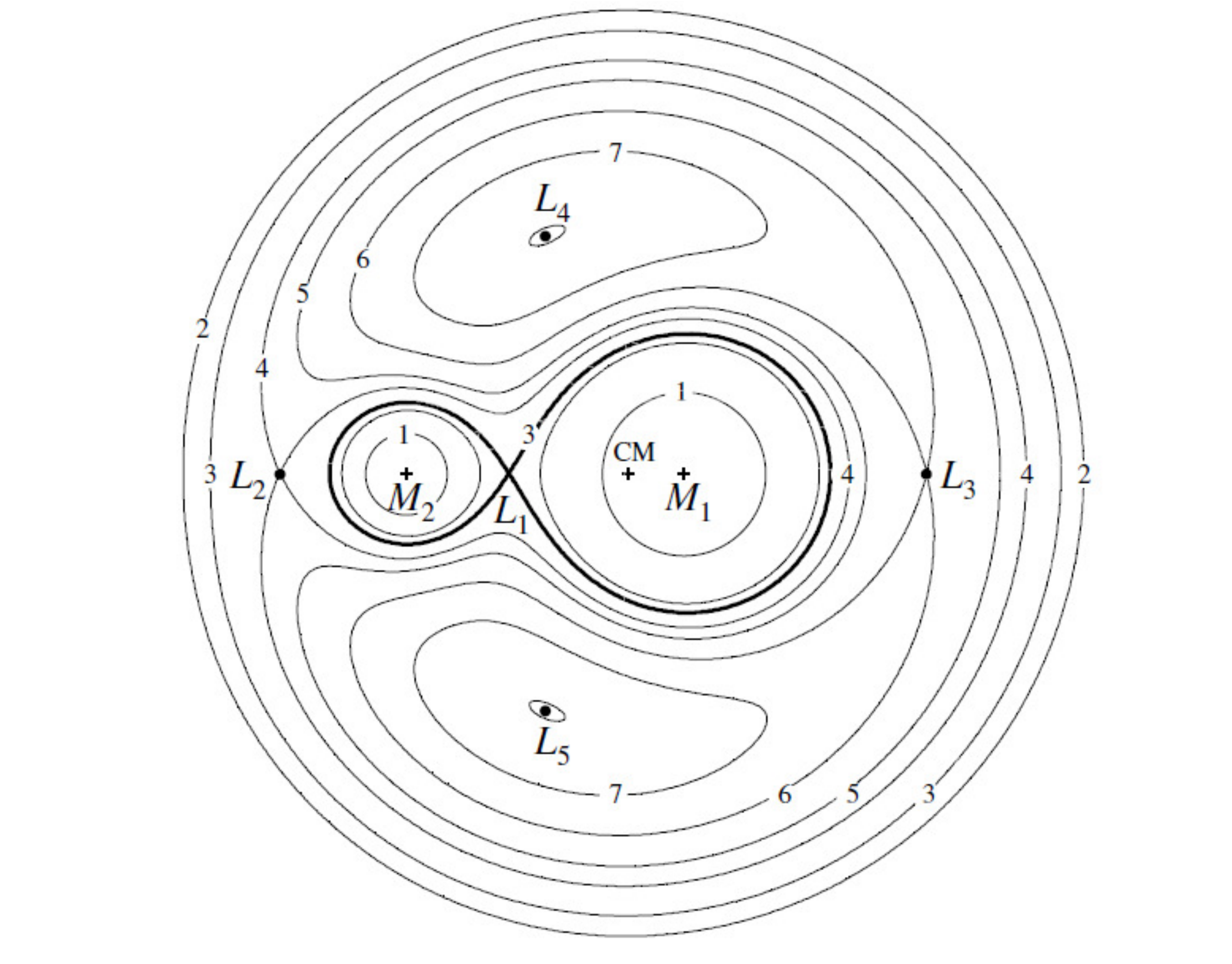}
\caption[V1668 Cyg V Magnitude Evolution]{V1668 Cyg V magnitude evolution from AAVSO archives. The dashed line represents the time of visual maximum and the dash-dotted lines represent the end of the different phases of evolution. The dotted line represents the start of the quiescent phase.}
\label{fig:v1668cyg_aavso}
\end{figure}

\begin{figure}[hb!]
\centering
\includegraphics[height=14cm,width=13cm]{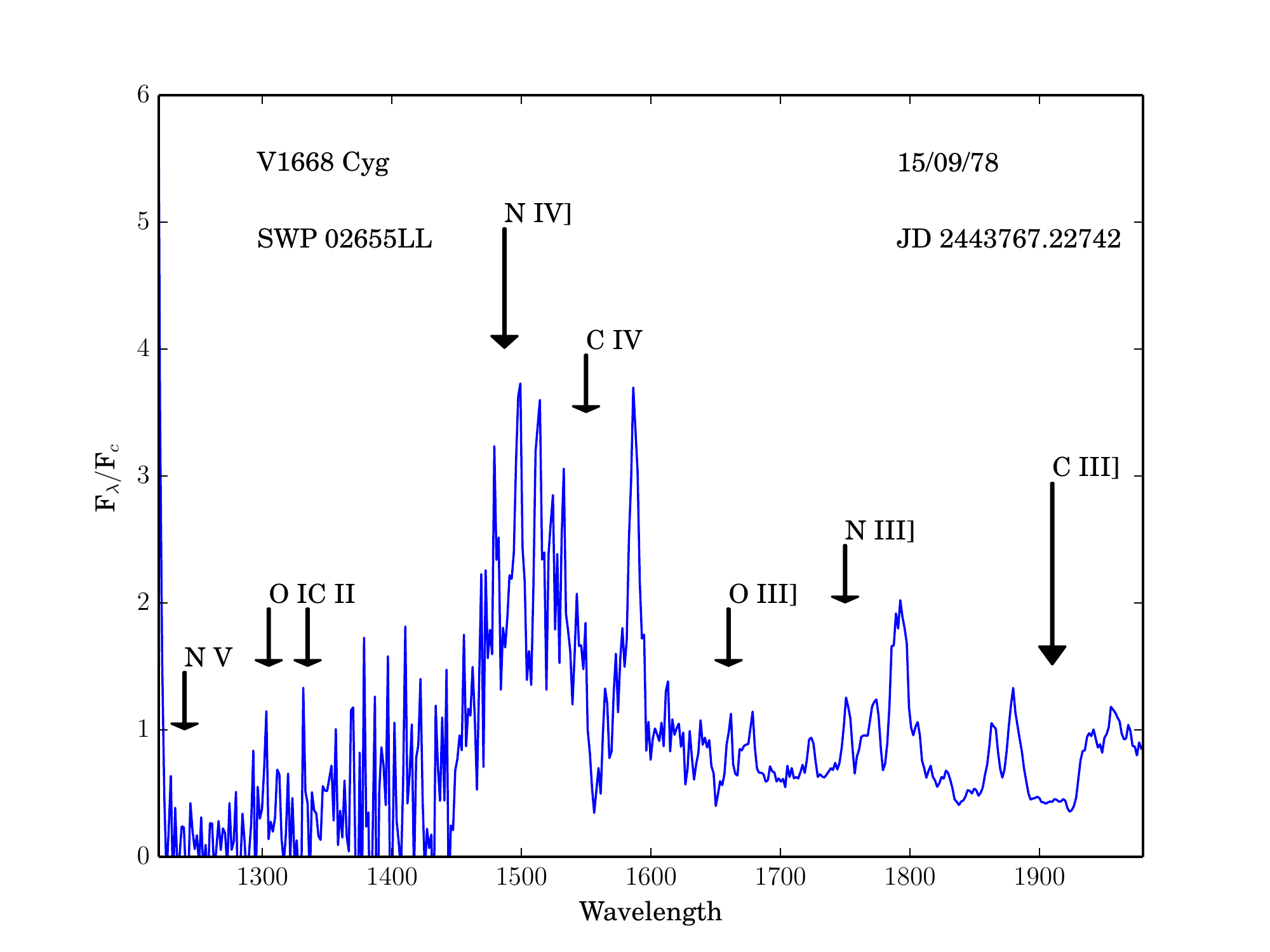}
\caption{V1668 Cyg lifting the iron curtain (initial) Phase.}
\label{fig:normswp02655}
\end{figure}

\begin{figure}
\centering
\includegraphics[height=14cm,width=13cm]{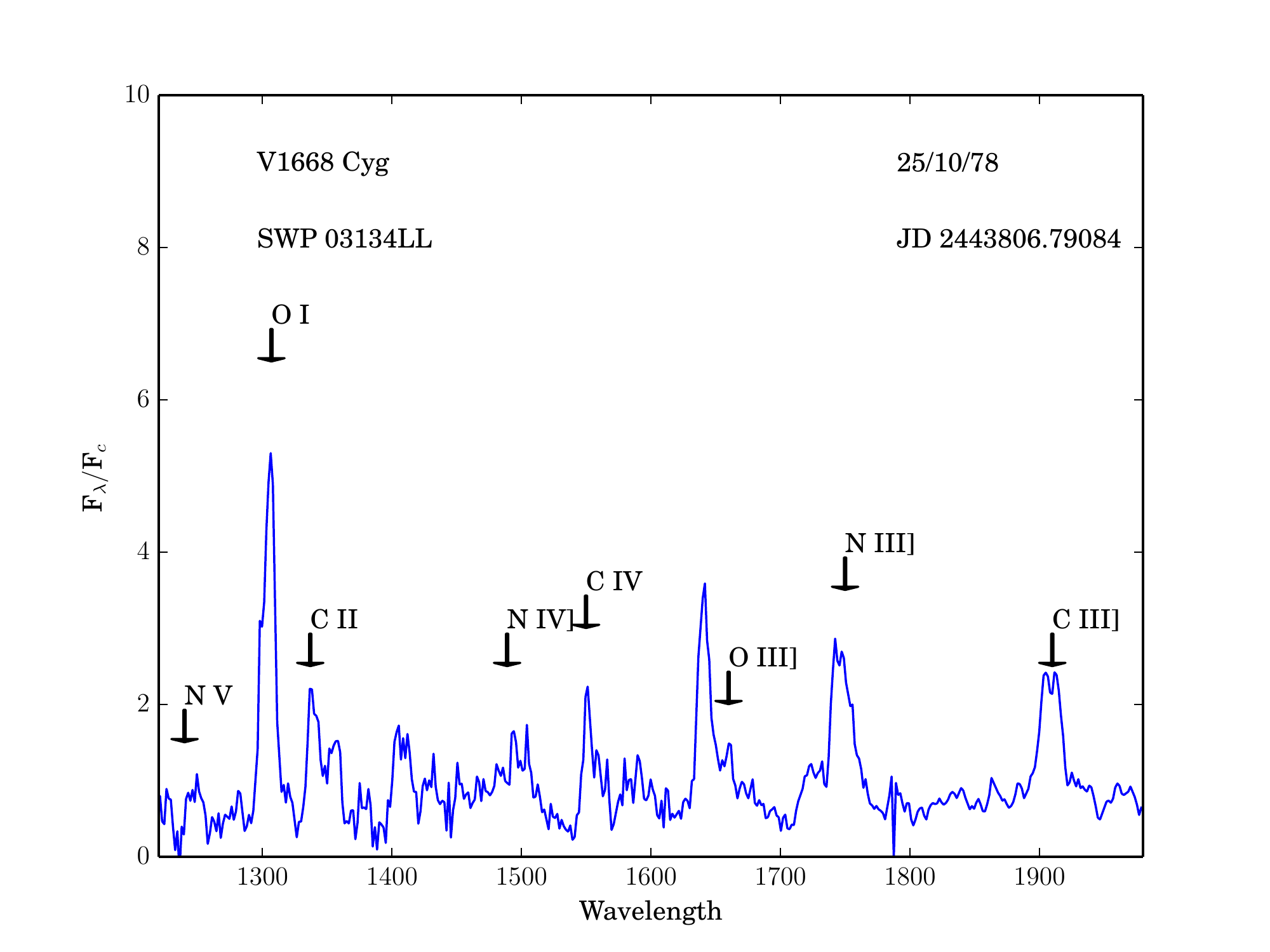}
\caption{V1668 Cyg Pre-nebular Phase.}
\label{fig:normswp03134}
\end{figure}

\begin{figure}
\centering
\includegraphics[height=14cm,width=13cm]{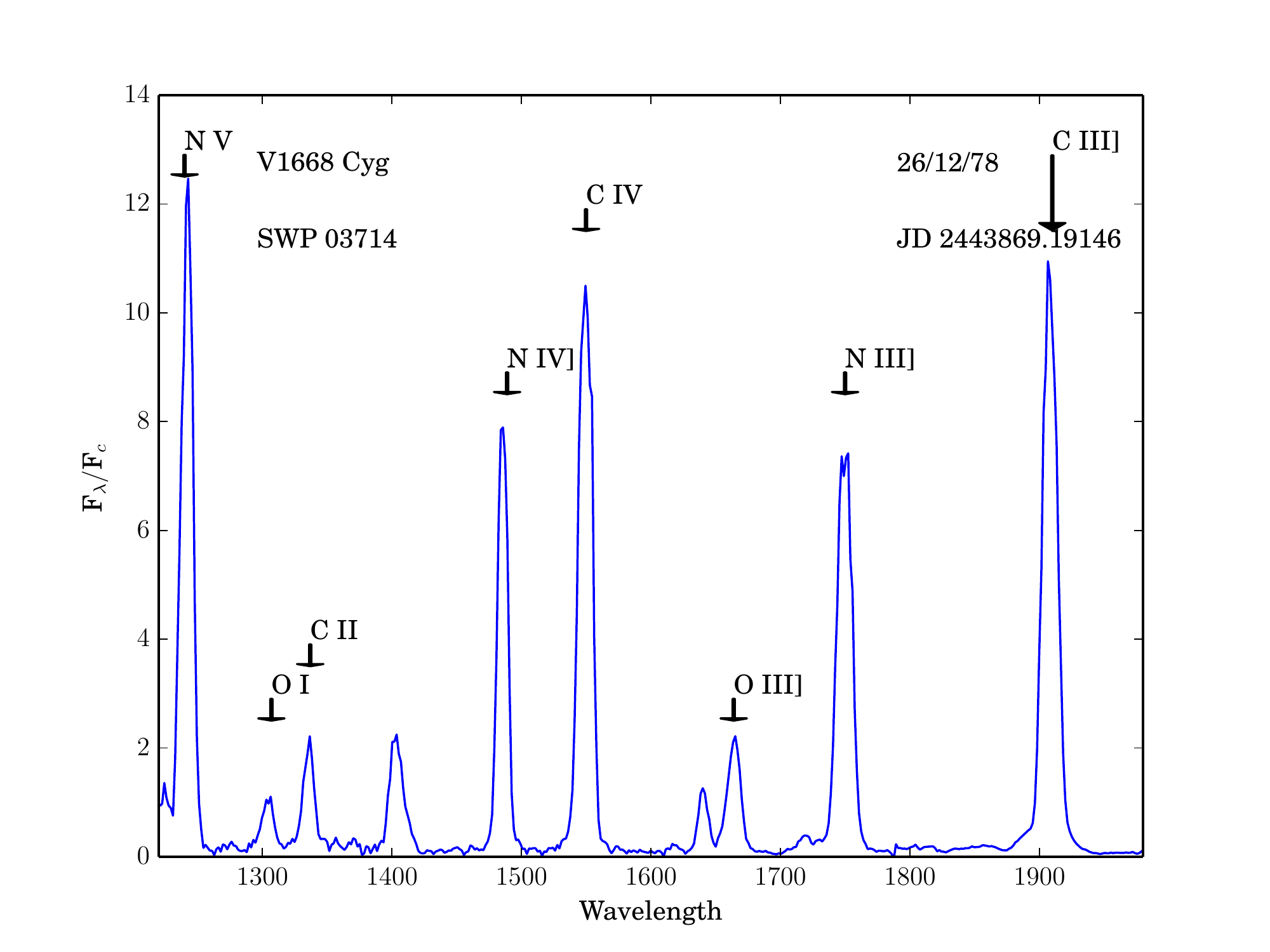}
\caption{V1668 Cyg Nebular Phase.}
\label{fig:normswp03714}
\end{figure}

\begin{figure}
\centering
\includegraphics[height=14cm,width=13cm]{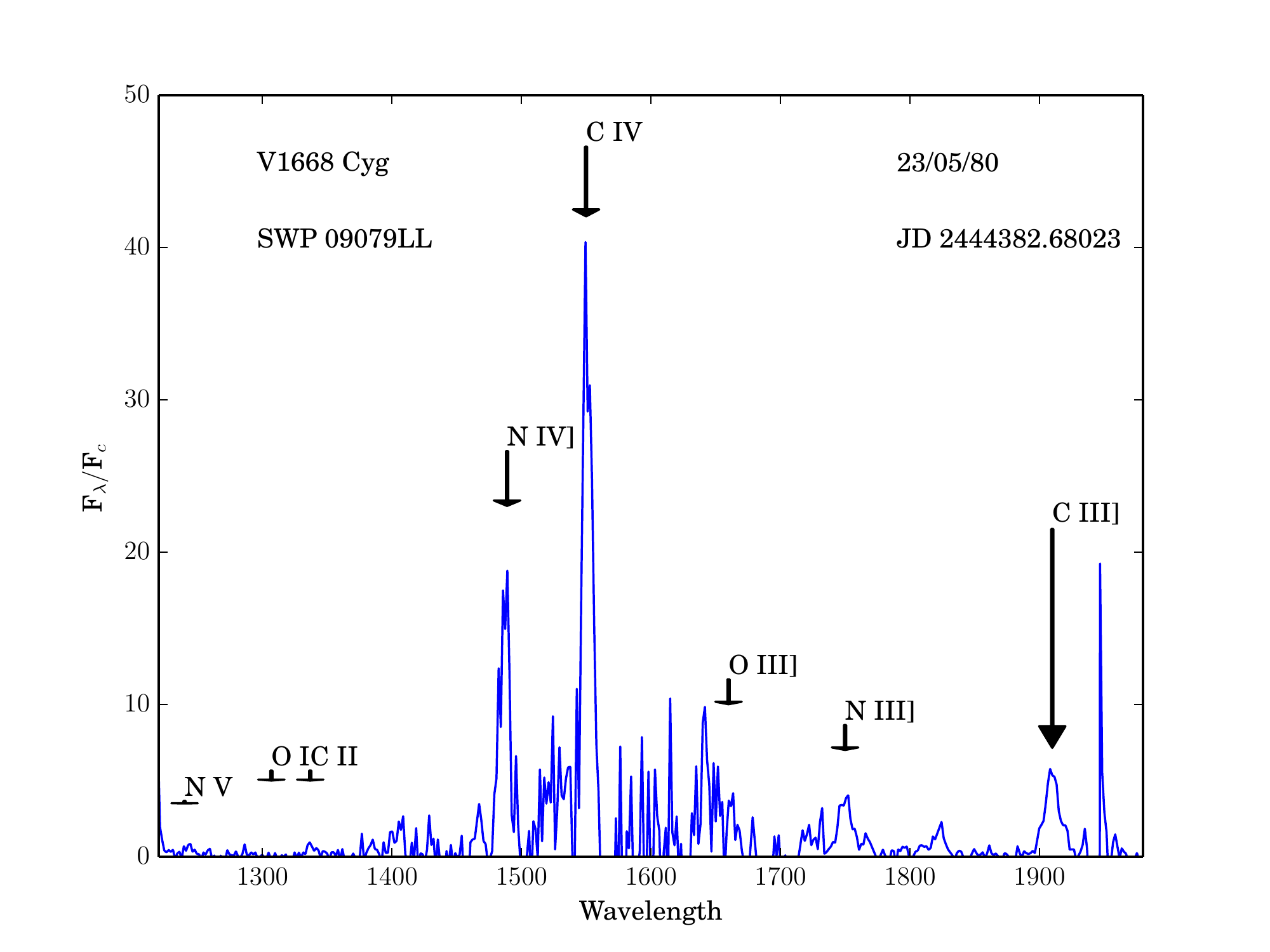}
\caption{V1668 Cyg Quiescent Phase.}
\label{fig:normswp09079}
\end{figure}

\clearpage

\subsection{V1974 Cyg}\label{sec:obs3}


V1974 Cyg observations cover the period between 20/02/92 and 20/11/94. Table ~\ref{tab:journal3} contains the journal of the used spectra. Sample normalized spectra in different stages illustrating different emission lines are presented in Figs ~\ref{fig:norm_SWP44030LL}-~\ref{fig:norm_SWP48220LL}.

\begingroup
\small \renewcommand{\arraystretch}{0.8} \setlength{\tabcolsep}{3pt}
\begin{longtable}{|c|c|c|c|}
\caption{Journal of IUE Observations for V1974 Cyg.}
\label{tab:journal3}
\renewcommand{\arraystretch}{0.7}
\cr \hline
Spectrum ID & JD            & Days after Explosion   &  Exposure Time (s)  \\
\hline
  \endfirsthead
  \caption*{Continued. Journal of IUE Observations for V1974 Cyg.}\\
  \hline
Spectrum ID & JD            & Days after Explosion   &  Exposure Time (s)  \\
  \hline
  \endhead
  \hline
  \endfoot
SWP44030LL & 2448673.4 &  4   & 9                \\ 
SWP44031LL & 2448673.5 &  4   & 44               \\ 
SWP44040LL & 2448674.5$^1$ &  5   & 419              \\ 
SWP44043LL & 2448675.2$^2$ &  6   & 479              \\ 
SWP44044LL & 2448675.3 &  6   & 2099             \\ 
SWP44050LL & 2448676.4 &  7   & 179              \\ 
SWP44051LL & 2448676.7 &  8   & 59               \\ 
SWP44055LL & 2448677.6 &  9   & 29               \\ 
SWP44056LL & 2448677.7 &  9   & 119              \\ 
SWP44060LL & 2448678.6 &  10  & 49               \\ 
SWP44062LL & 2448678.7 &  10  & 179              \\ 
SWP44064LL & 2448678.9 &  10  & 59               \\ 
SWP44073LL & 2448680.4 &  11  & 39               \\ 
SWP44086LL & 2448682.8 &  14  & 49               \\ 
SWP44102LL & 2448684.5 &  15  & 19               \\ 
SWP44115LL & 2448686.3 &  17  & 34               \\ 
SWP44130LL & 2448688.8 &  20  & 34               \\ 
SWP44155LL & 2448693.5 &  25  & 34               \\ 
SWP44174LL & 2448695.7 &  27  & 34               \\ 
SWP44193LL & 2448700.5 &  31  & 34               \\ 
SWP44209LL & 2448703.9 &  35  & 34               \\ 
SWP44233LL & 2448707.5 &  39  & 29               \\ 
SWP44268LL & 2448711.9 &  43  & 29               \\ 
SWP44305LL & 2448715.3 &  46  & 29               \\ 
SWP44338LL & 2448717.8$^3$ &  49  & 29               \\ 
SWP44377LL & 2448723.4 &  54  & 29               \\      
SWP44389LL & 2448725.8 &  57  & 29               \\            
SWP44439LL & 2448732.8 &  64  & 39               \\            
SWP44632LL & 2448752.3$^4$ &  83  & 39               \\            
SWP44634LL & 2448752.4 &  83  & 19               \\            
SWP44717LL & 2448761.7 &  93  & 29               \\            
SWP44761LL & 2448767.7 &  99  & 29               \\            
SWP44762LL & 2448767.7 &  99  & 44               \\            
SWP44790LL & 2448770.3 &  101 & 29               \\            
SWP44808LL & 2448772.6 &  104 & 29               \\            
SWP44901LL & 2448783.3 &  114 & 34               \\            
SWP44937LL & 2448790.2 &  121 & 37               \\            
SWP44970LL & 2448794.4 &  125 & 34               \\
SWP44973LL & 2448795.6 &  127 & 34               \\
SWP45030LL & 2448802.6 &  134 & 37               \\
SWP45059LL & 2448807.2 &  138 & 41               \\
SWP45061LL & 2448807.3 &  138 & 39               \\
SWP45135LL & 2448818.2 &  149 & 50               \\
SWP45244LL & 2448833.1 &  164 & 59               \\
SWP45310LL & 2448845.0 &  176 & 79               \\
SWP45359LL & 2448851.4 &  182 & 59               \\
SWP45469LL & 2448864.4 &  195 & 69               \\
SWP45547LL & 2448873.0 &  204 & 109              \\
SWP45548LL & 2448873.0 &  204 & 99               \\
SWP45670LL & 2448883.3 &  214 & 109              \\
SWP46047LL & 2448919.1 &  250 & 169              \\
SWP46064LL & 2448921.6 &  253 & 114              \\
SWP46404LL & 2448960.9 &  292 & 129              \\
SWP47027LL & 2449041.7 &  373 & 269              \\
SWP47278LL & 2449061.5 &  392 & 269              \\
SWP47397LL & 2449078.3 &  409 & 269              \\
SWP47416LL & 2449082.3 &  413 & 269              \\
SWP47417LL & 2449082.3 &  413 & 599              \\
SWP48026LL & 2449171.0 &  502 & 1139             \\
SWP48027LL & 2449171.1 &  502 & 3599             \\
SWP48028LL & 2449171.2 &  502 & 1139             \\
SWP48218LL & 2449192.4 &  523 & 1799             \\
SWP48219LL & 2449192.5 &  523 & 3599             \\
SWP48220LL & 2449192.6 &  524 & 6599             \\
SWP48221LL & 2449192.8 &  524 & 1379             \\
SWP48222LL & 2449192.8 &  524 & 1499             \\
SWP48638LL & 2449246.2$^5$ &  577 & 1799         \\
SWP48639LL & 2449246.3 &  577 & 4499             \\
SWP49320LL & 2449317.0 &  648 & 3599             \\
SWP49321LL & 2449317.2 &  648 & 7199             \\
SWP49322LL & 2449317.3 &  648 & 3299             \\
SWP50494LL & 2449449.7 &  781 & 16499            \\
SWP50941LL & 2449503.6 &  835 & 20399            \\
SWP51387LL & 2449543.8 &  875 & 20695            \\
SWP51983LL & 2449594.4 &  925 & 25199            \\
SWP52846LL & 2449677.5 &  1008& 24599            \\
\hline
\caption*{Notes: Times of different phases in the evolution of the outburst. (1) Start of the iron curtain phase. (2) The visual maximum. (3) Start of the pre-nebular phase. (4) Start of the nebular phase. (5) Start of the quiescent phase.}
\vspace{-1.7cm} 
\end{longtable}                                                 

\begin{figure}
\centering
\includegraphics[height=14cm,width=13cm]{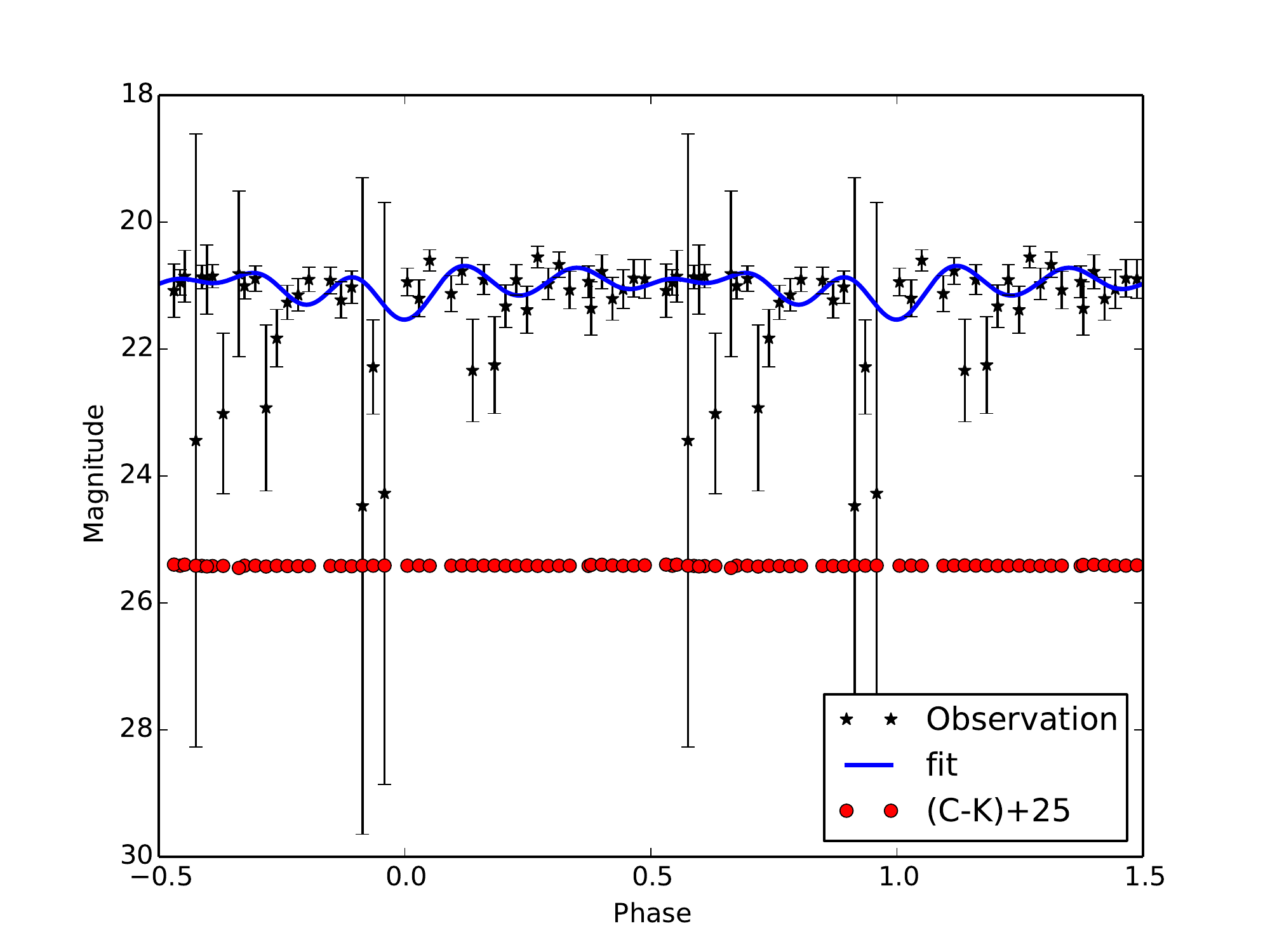}
\caption[V1974 Cyg V Magnitude Evolution]{V1974 Cyg V magnitude evolution from AAVSO archives. The numbers 1,2,3,4,and 5 correspond to the fireball, Fe optically thick, transition, nebular and quiescent phases, respectively.}
\label{fig:v1974cyg_aavso}
\end{figure}

\begin{figure}
\includegraphics[height=14cm,width=13cm]{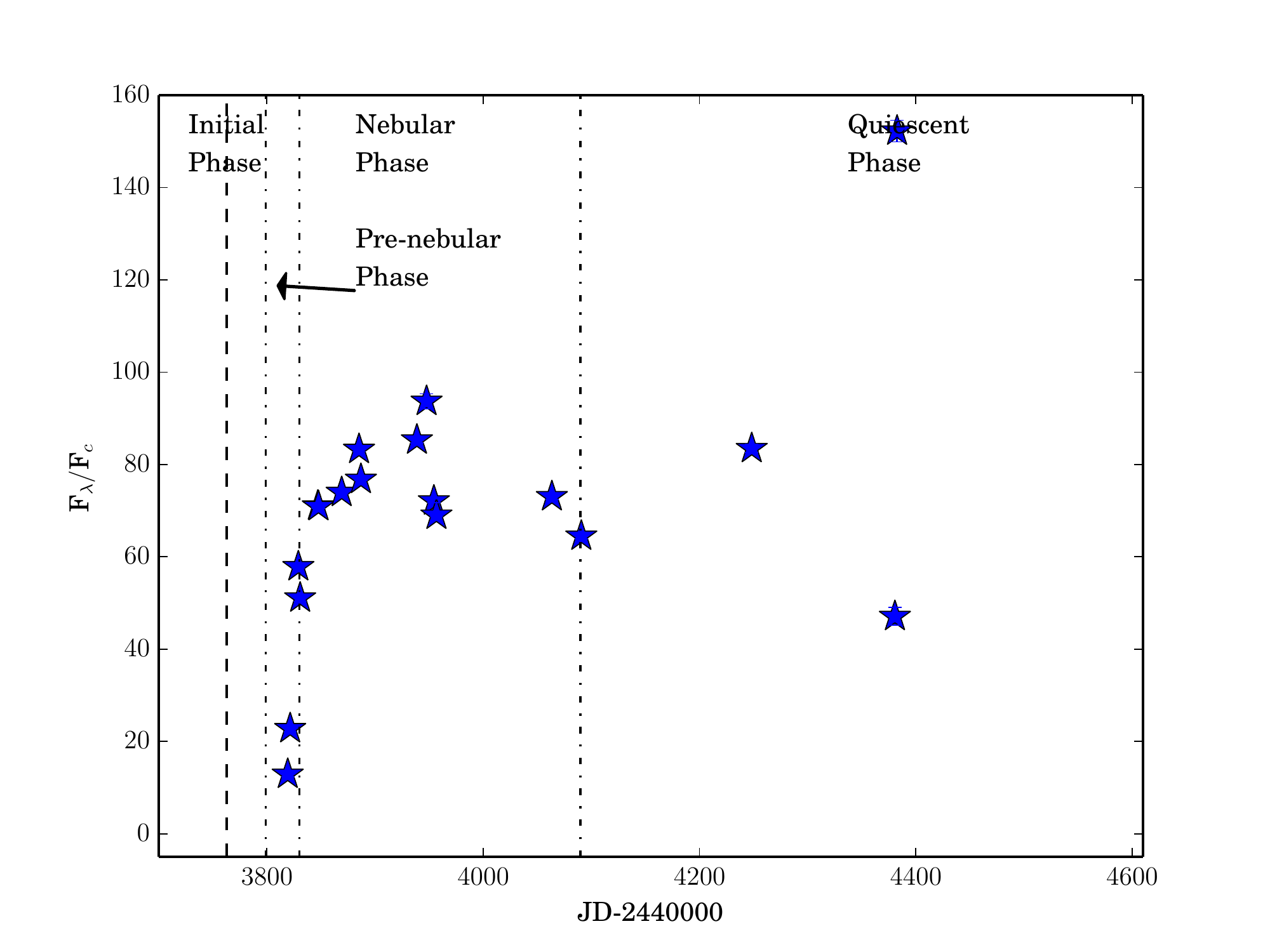}
\caption{V1974 Cyg Fireball Spectrum}
\label{fig:norm_SWP44030LL}
\end{figure}

\begin{figure}
\includegraphics[height=14cm,width=13cm]{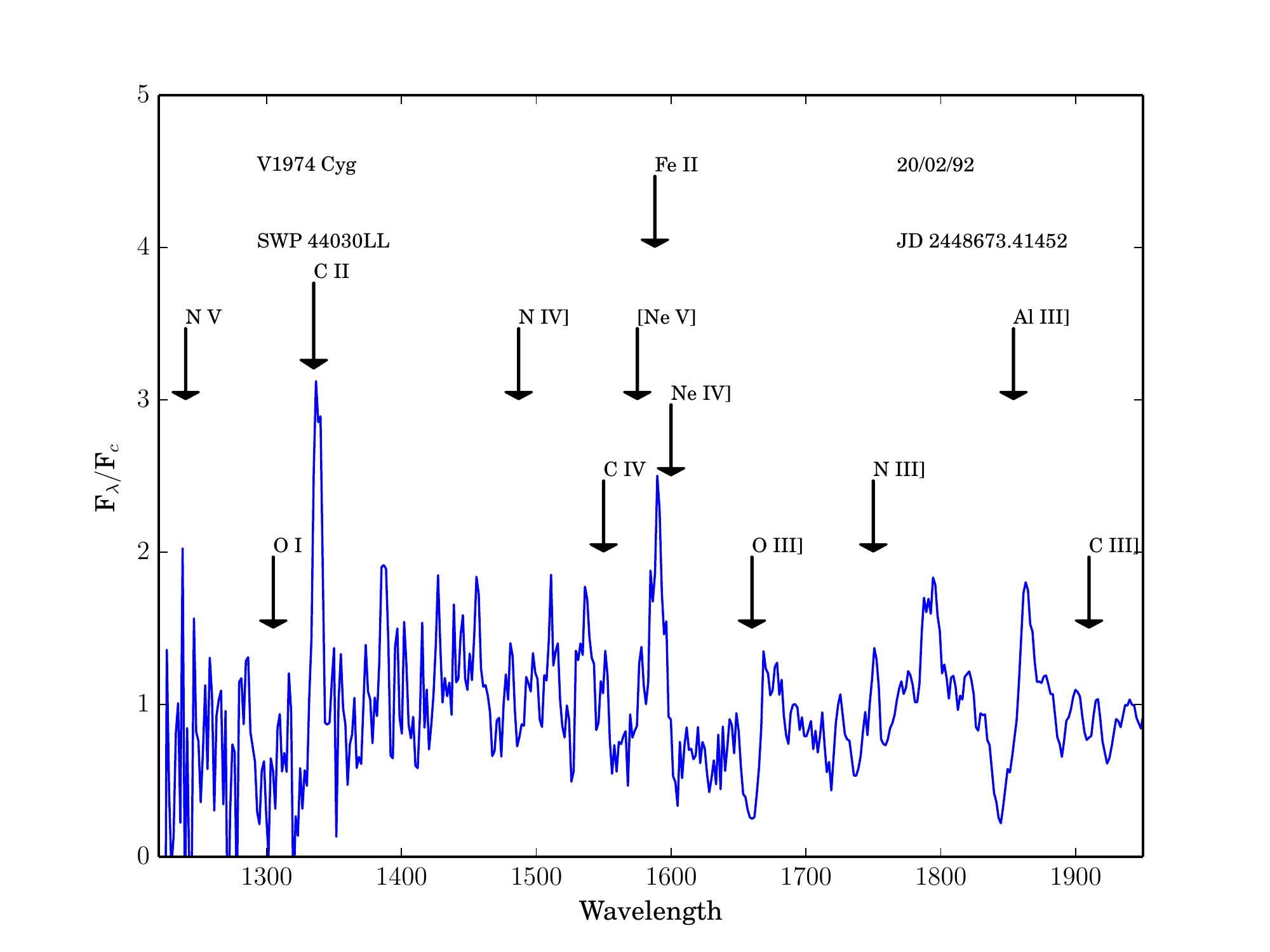}
\caption{V1974 Cyg Fe Optically Thick Phase.}
\label{fig:norm_SWP44043LL}
\end{figure}

\begin{figure}
\includegraphics[height=14cm,width=13cm]{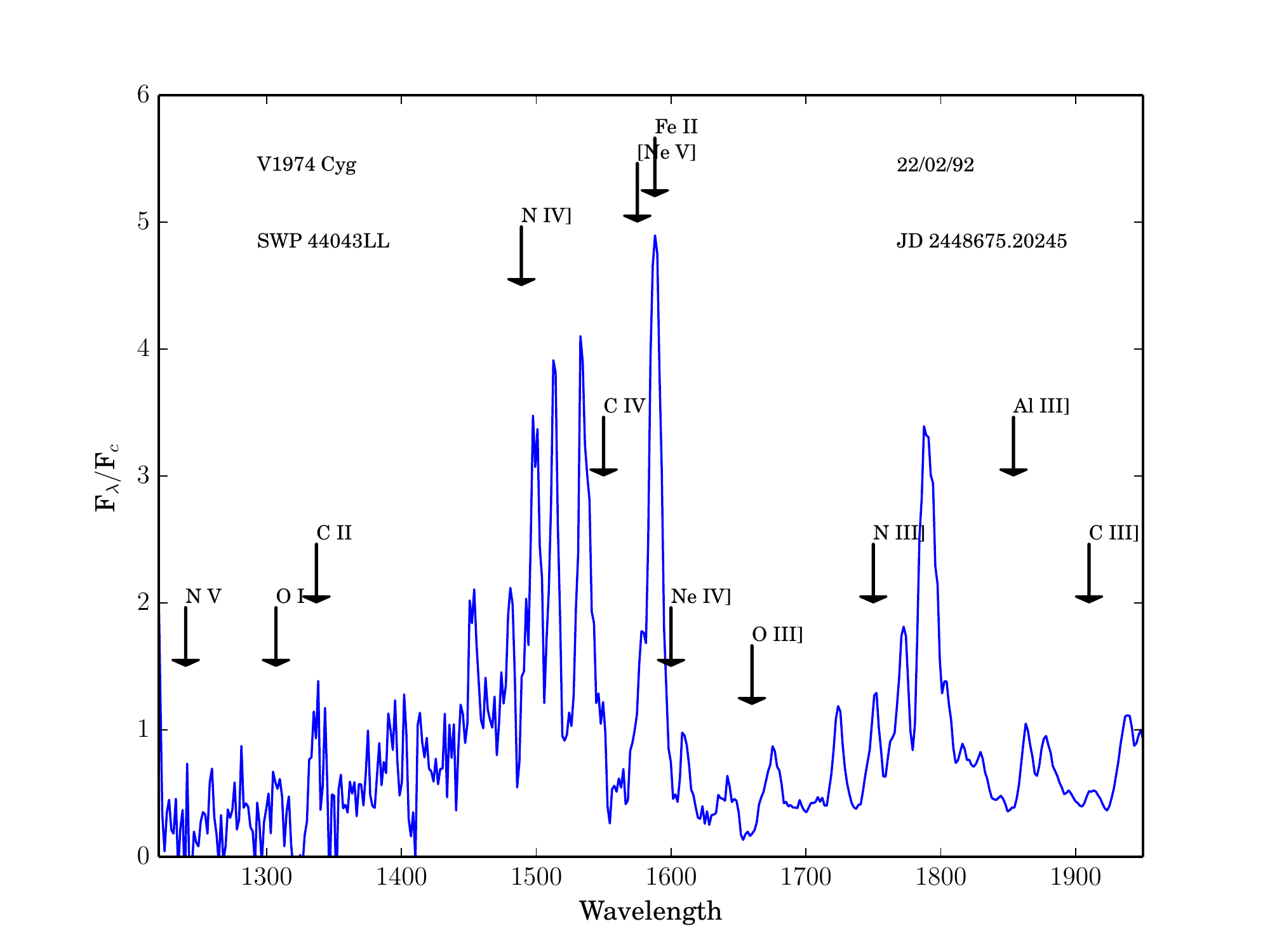}
\caption{V1974 Cyg Transition Phase.}
\label{fig:norm_SWP44389LL}
\end{figure}
\begin{figure}
\includegraphics[height=14cm,width=13cm]{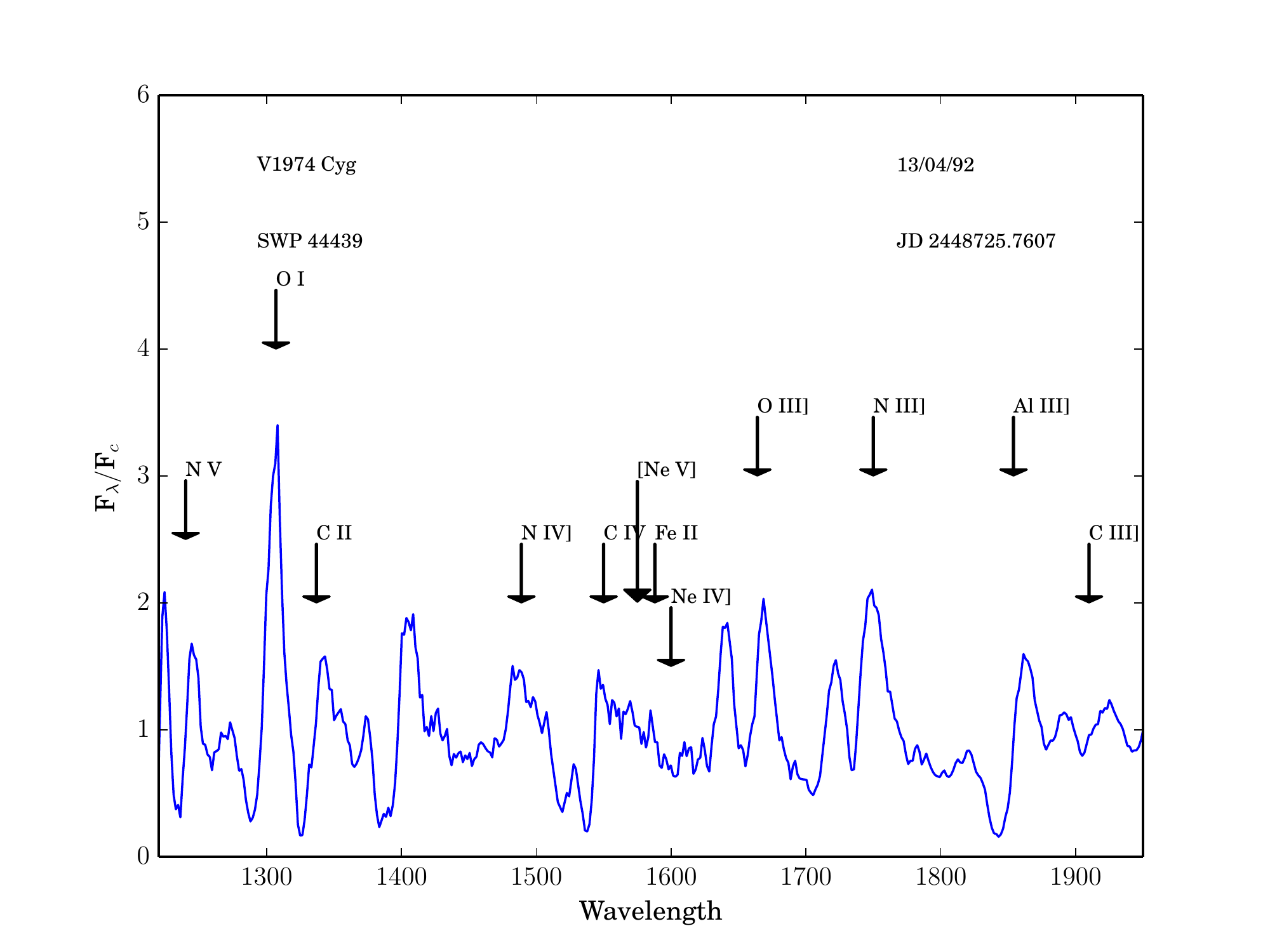}
\caption{V1974 Cyg Nebular Phase.}
\label{fig:norm_SWP44762LL}
\end{figure}

\begin{figure}
\includegraphics[height=14cm,width=13cm]{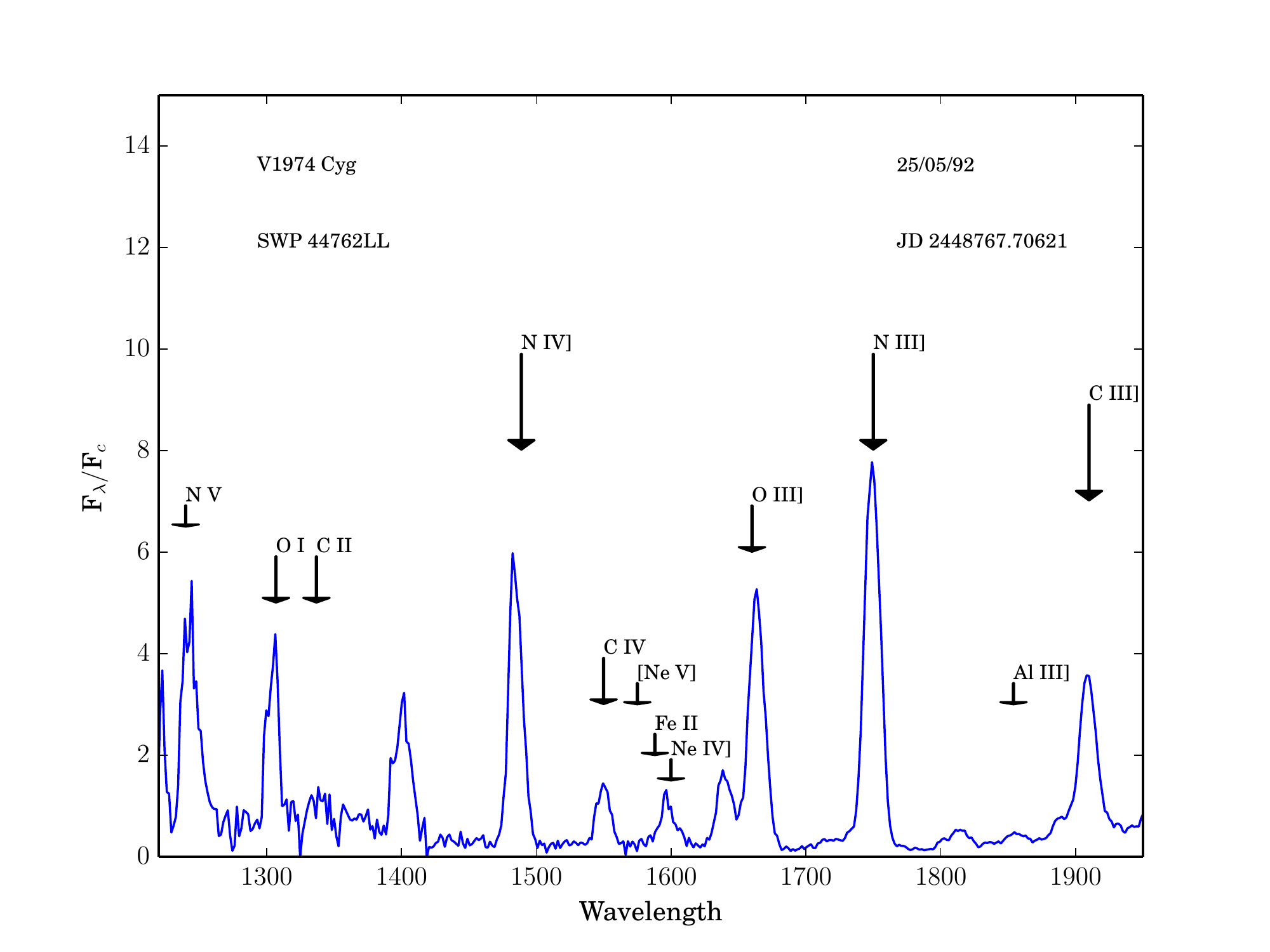}
\caption{V1974 Cyg Quiescent Phase.}
\label{fig:norm_SWP48220LL}
\end{figure}

\clearpage
\section{Optical Photometry}\label{sec:phot_obs}

\subsection{V1668 Cyg}

V1668 Cyg  was observed for 3 nights in the V, $\mathrm{R_c}$ and $\mathrm{I_c}$ filters through the time interval from July $31^{st}$, 2016 and October $4^{th}$, 2016. The exposure times were 240 s in V filter and 300 s in both $\mathrm{R_c}$ and $\mathrm{I_c}$ filters. the total number of 125 images in all filters were taken. All the data reductions were made using the ccdred package of the Image Reduction and Analysis Facility (IRAF) version 2.16 \citep{1993ASPC...52..173T}. The differential photometry was performed using Muniwin program (\url{http://c-munipack.sourceforge.net/} which is based on the DAOPHOT package. The comparison star used was 2MASS 21422902+4400161 and the check star was 2MASS 21423459+4359462. A finding chart showing the three stars is presented in Fig~\ref{fig:v1668cyg}. Table ~\ref{tab:v1668candk} contains the variable, comparison and check stars' parameters.

\begin{figure}
\centering
\includegraphics[height=14cm,width=13cm]{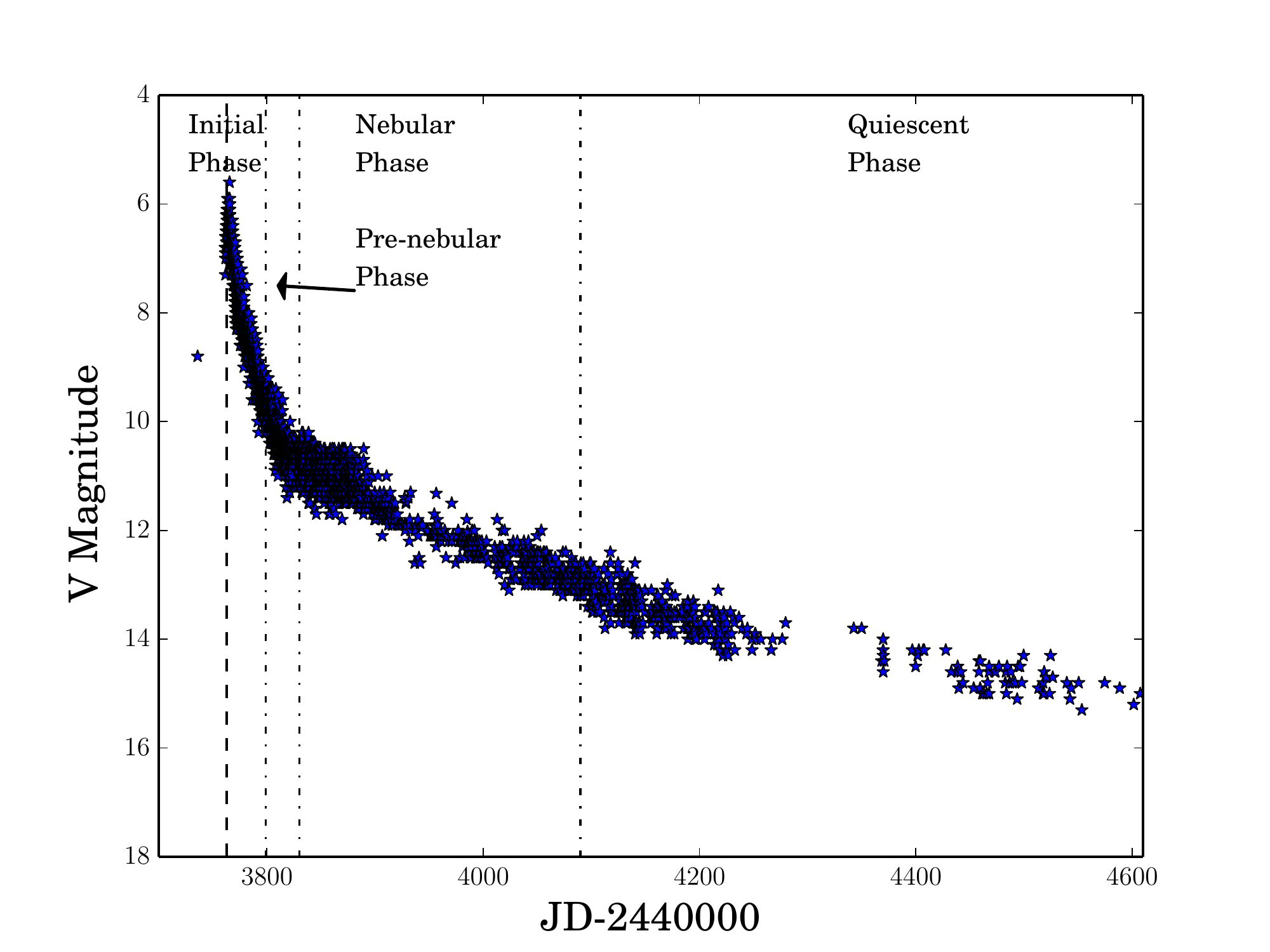}
\caption[V1668 Cyg finding chart]{V1668 Cyg finding chart showing the variable (V), comparison (C) and check (K) stars.}
\label{fig:v1668cyg}
\end{figure}

\begin{table}[h]
\caption{V1668 Cyg Variable, comparison and check stars' parameters}
\begin{center}
\label{tab:v1668candk}
\resizebox{\textwidth}{!}{%
\begin{tabular}{|l|c|c|c|c|c|c|c|}\hline\hline
Star&Identifier&Right Ascension&Declination&B&V&$\mathrm{\mathrm{R_c}}$&$\mathrm{\mathrm{I_c}}$\\
\hline
Variable & V1668 Cyg                 & 21 42 35.22      &  +44 01 54.9     &  20    & 6-20   &      &          \\
Comparison &  2MASS 21422902+4400161 & 21 42 29.024     &  +44 00 17.15    & 16.000 & 15.057 &14.384&  13.994  \\
Check     &  2MASS 21423459+4359462  & 21 42 34.670     &  +43 59 45.83    & 15.298 & 14.594 &14.097&  13.707  \\
\hline
\end{tabular}}
\end{center}
\end{table}
  
The magnitudes of the comparison and check stars were obtained from \citet{2016yCat.2336....0H} converting the APASS r' and i' magnitudes to $\mathrm{R_c}$ and $\mathrm{I_c}$ magnitudes using the conversion formulas of \citet{2005AJ....130..873J} since APASS uses Sloan r' and i' filters.

We clearly detect the eclipse found by \citet{1990MNRAS.245..547K} especially in our $\mathrm{R_c}$ filter observations. Using the Kwee-van Woerden method \citep{1956BAN....12..327K}, we found a new epoch at JD 2457665.281560 from our $\mathrm{R_c}$ filter observations. This was done using AVE (An\`{a}lisis de Variabilidad Estelar) \nomenclature{AVE}{An\`{a}lisis de Variabilidad Estelar}program (\url {http://astrogea.org/soft/ave/aveint.htm}) distributed by the Grup d'Estudis Astron\`{o}mics. If we use this new epoch to plot the phase magnitude diagram, there will be  a phase shift of 0.0864 d. Figure ~\ref{fig:v1668bothepochs} shows an $\mathrm{R_c}$ filter phase-differential magnitude diagram showing both epochs. 

\begin{figure}
\centering
\includegraphics[height=14cm,width=13cm]{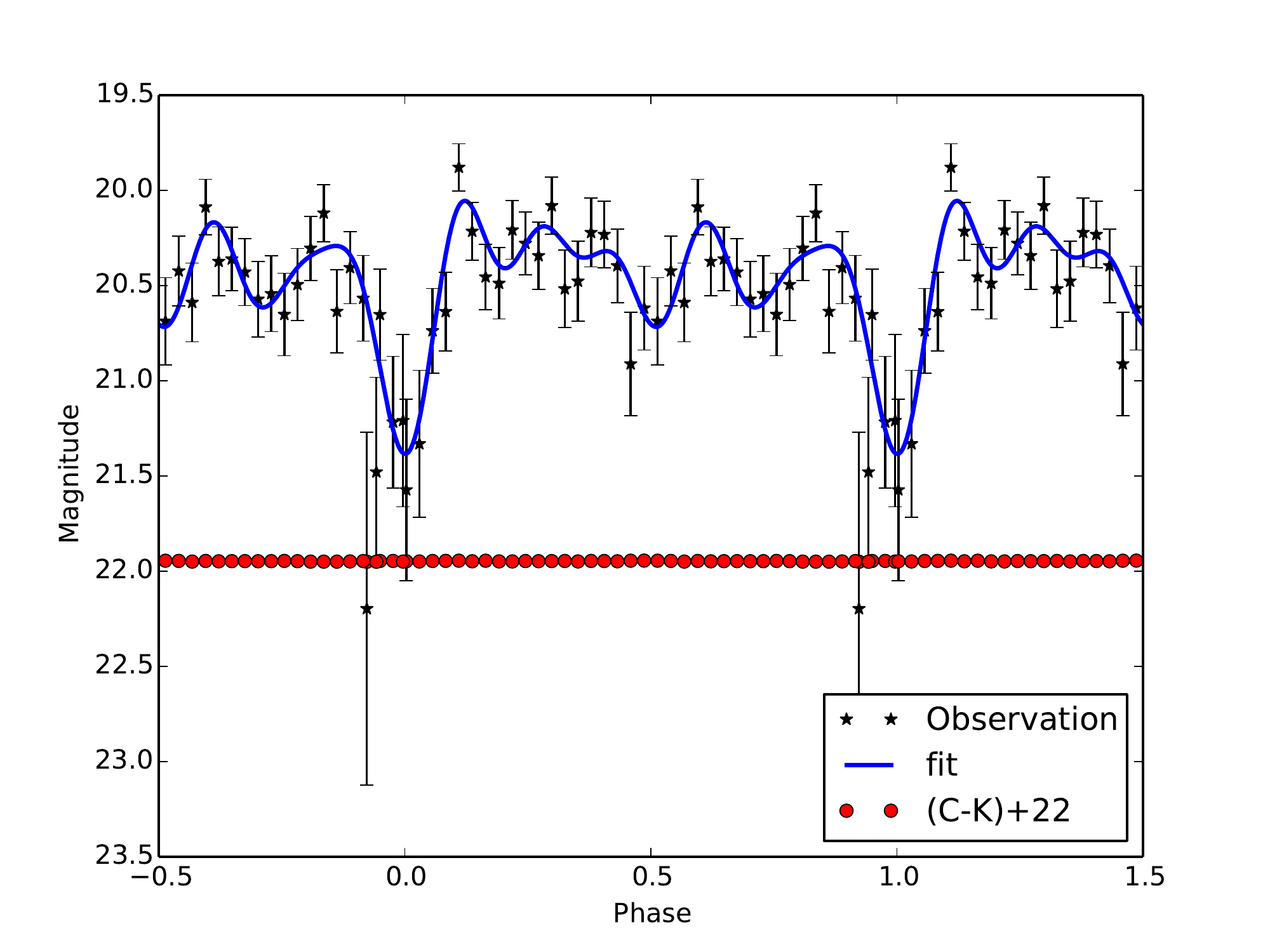}
\caption[V1668 Cyg $\mathrm{R_c}$ phase-differential magnitude diagram ]{V1668 Cyg $\mathrm{R_c}$ phase-differential magnitude diagram for the ephemeris $HJD_{min}=2457665.281560 + 0.138400 \times E$ (dots) and the ephemeris $HJD_{min}=2447679.848 + 0.138400 \times E$ (stars).}
\label{fig:v1668bothepochs}
\end{figure}

Based on the new epoch, the ephemeris will be

\begin{eqnarray}~\label{eq:v1668cyg}
HJD_{min}&=&2457665.281560 + 0.138400 \times E\\
          &&\pm 0.000562596 \nonumber
\end{eqnarray}

\clearpage
\subsection{PW Vul}

We observed PW Vul for 4 nights between October $13^{th}$, 2015 and October $5^{th}$, 2016 using V, $\mathrm{R_c}$ and $\mathrm{I_c}$ filters . The exposure times were 420 s in all filters. A total number of 162 images in all filters were taken.

The comparison star used was 2mass 19260413+2721363 and the check star was 2mass 19255867+2723282. A finding chart for the field of PW Vul is shown in Fig~\ref{fig:pwvul}. Table ~\ref{tab:pwvulcandk} contains the variable, comparison and check stars' parameters.

\begin{figure}
\centering
\includegraphics[height=14cm,width=13cm]{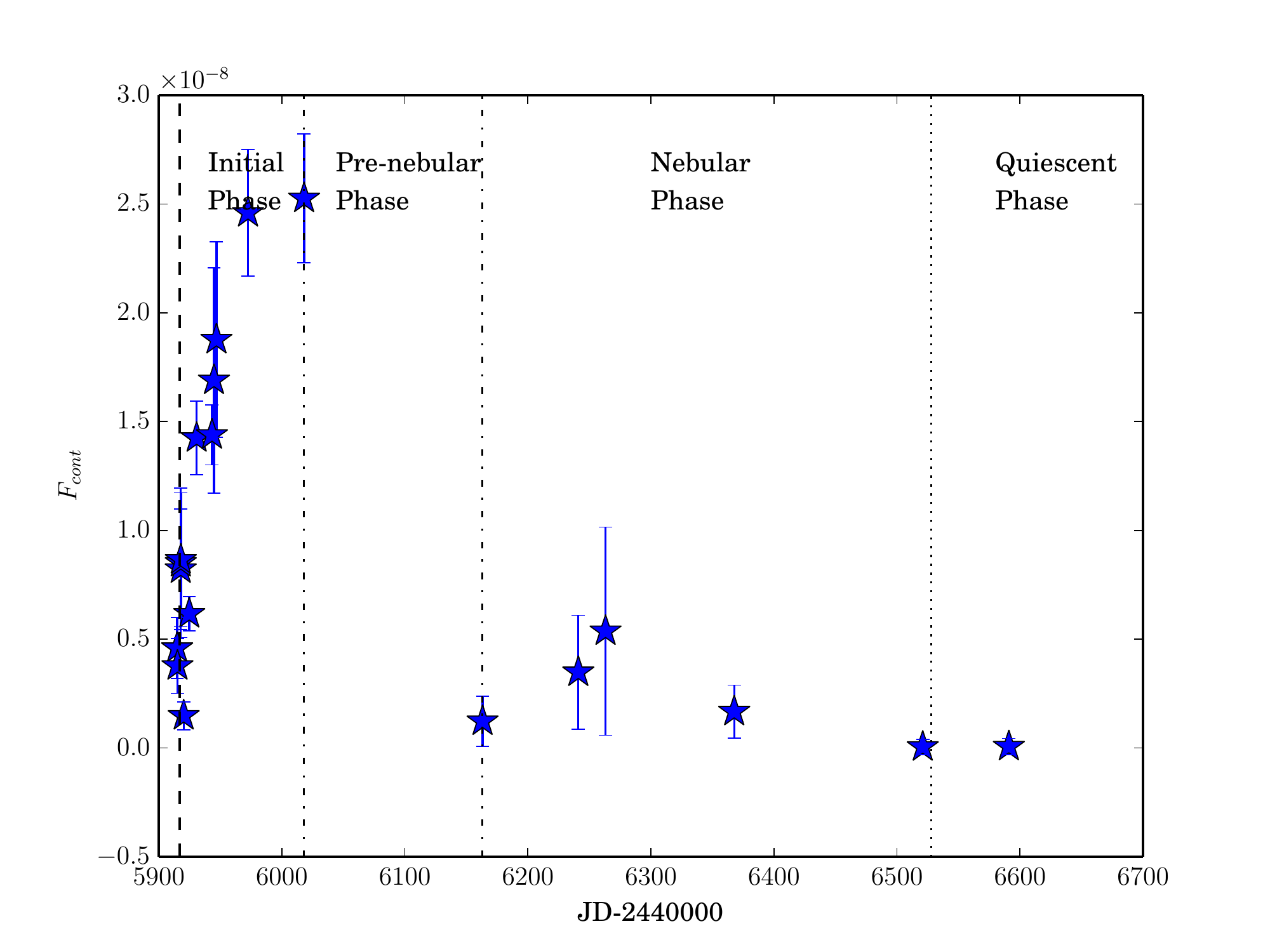}
\caption[PW Vul finding chart]{PW Vul finding chart showing the variable (V), comparison (C) and check (K) stars.}
\label{fig:pwvul}
\end{figure}

\begin{table}
\caption{PW Vul Variable, comparison and check stars' parameters}
\begin{center}
\label{tab:pwvulcandk}
\resizebox{\textwidth}{!}{%
\begin{tabular}{|l|c|c|c|c|c|c|c|}\hline\hline
Star&Identifier&Right Ascension&Declination&B&V&$\mathrm{\mathrm{R_c}}$&$\mathrm{\mathrm{I_c}}$\\
\hline
Variable   & PW Vul                  & 19 26 05.06      & +27 21 58.8      &        &6.4 - 16.9 &    &          \\                  
Comparison &  2MASS 19260413+2721363 & 19 26 04.143     &  +27 21 36.71    & 17.930 & 16.120 &16.070&  13.6584  \\
Check     &  2MASS 19255867+2723282 & 19 25 58.669     &  +27 23 28.26    & 15.880 & 15.120 &14.320&  14.0864  \\
\hline
\end{tabular}}
\end{center}
\end{table}

The magnitudes of the comparison and check stars were obtained from \citet{2005yCat.1297....0Z} and the $\mathrm{\mathrm{I_c}}$ magnitude was calculated using the formula

\begin{equation}\label{eq:iconversion}
I=B-(2.36\times (B-V))
\end{equation}

of \citet{1994A&A...289..756N}.

We found a new epoch in the $\mathrm{I_c}$ light curve observed on Oct $13^{th}$, 2015  at JD 2457309.229200. We found another epoch in the V light curve of our observations at JD 2457568.482263. If we use these two epochs to plot the phase magnitude diagram, we will get a phase shift of 0.1594 d from the epoch at JD 2457309.229200 and a phase shift of 0.1089 d from the epoch at JD 2457568.48226.

In figure ~\ref{fig:pwvlallepochs} we plot an $\mathrm{I_c}$ filter phase-differential magnitude diagram showing all three epochs.
\begin{figure}
\centering
\includegraphics[height=14cm,width=13cm]{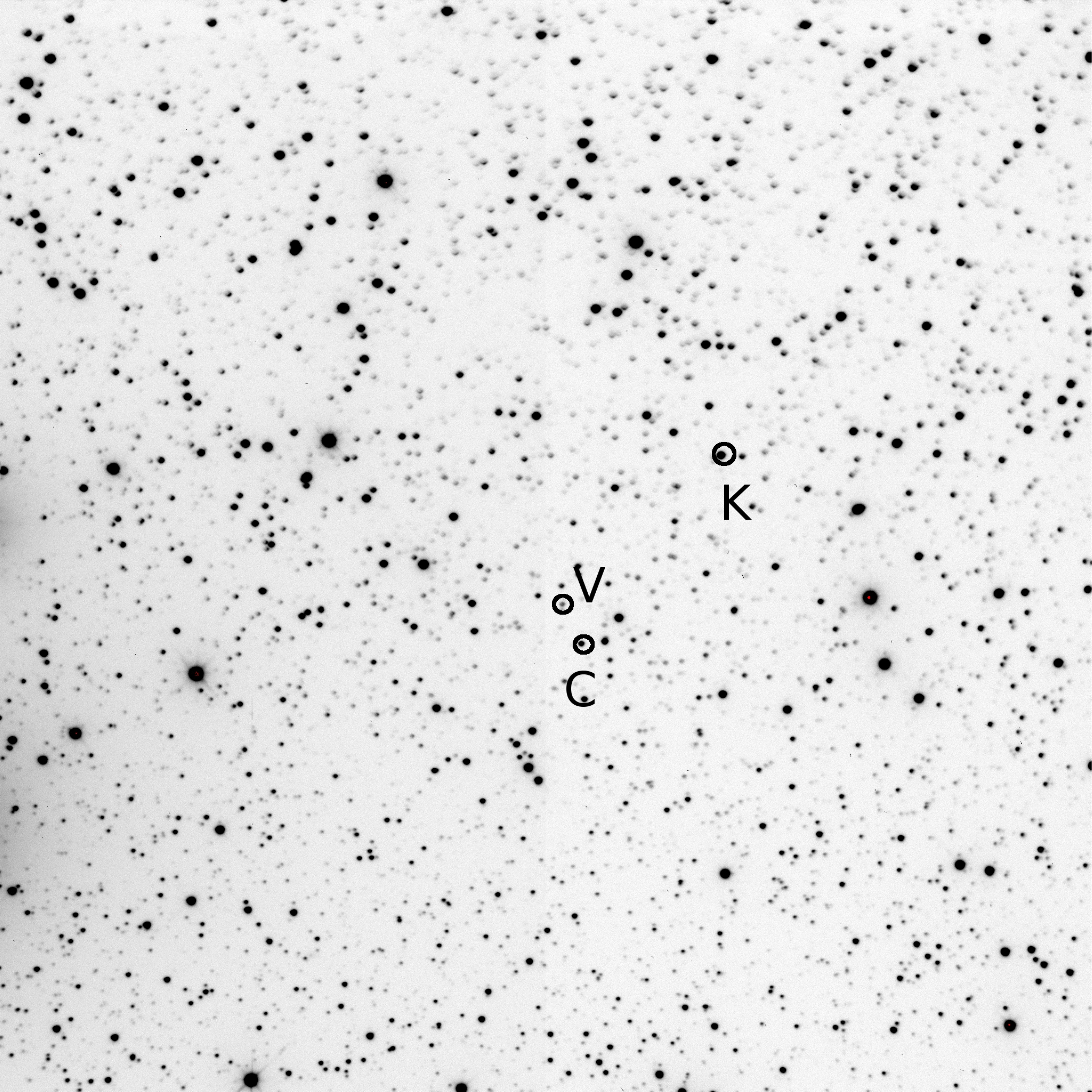}
\caption[PW Vul $\mathrm{I_c}$ phase-differential magnitude diagram]{PW Vul $\mathrm{I_c}$ phase-differential magnitude diagram for the ephemeris $HJD_{min}=2457309.229200 + 0.213700 \times E$ (dots),\\ $HJD_{min}=2457568.482263 + 0.213700 \times E$ (x signs) and \\$HJD_{min}=2446704.263 + 0.213700 \times E$ (stars) .}
\label{fig:pwvlallepochs}
\end{figure}

Based on the time of minimum of 13-10-2015 observations, the ephemeris will be

\begin{eqnarray}~\label{eq:pwvul}
HJD_{min}&=&2457309.229200 + 0.213700 \times E\\
         &&\pm 0.0015508 \;\;\;\;\;\; \pm 0.00091 \nonumber
\end{eqnarray}

\clearpage
\subsection{V1974 Cyg}

We present observations for V1974 Cyg taken over four nights between October $20^{th}$, 2015 and July $31^{st}$, 2016. A total of 148 images were taken in V, $\mathrm{R_c}$ and $\mathrm{I_c}$ filters. The exposure times were 120-180 s for the V filter, 180-480 s for the $\mathrm{R_c}$ filter and 180 s for the $\mathrm{I_c}$ filter. The comparison star used was 2MASS 20303767+5238093 and the check star was 2MASS 20302890+5237123. A finding chart showing the three stars is presented in Fig~\ref{fig:v1974cyg}. Table ~\ref{tab:v1974cygcandk} contains the variable, comparison and check stars' parameters.

\begin{figure}
\centering
\includegraphics[height=14cm,width=13cm]{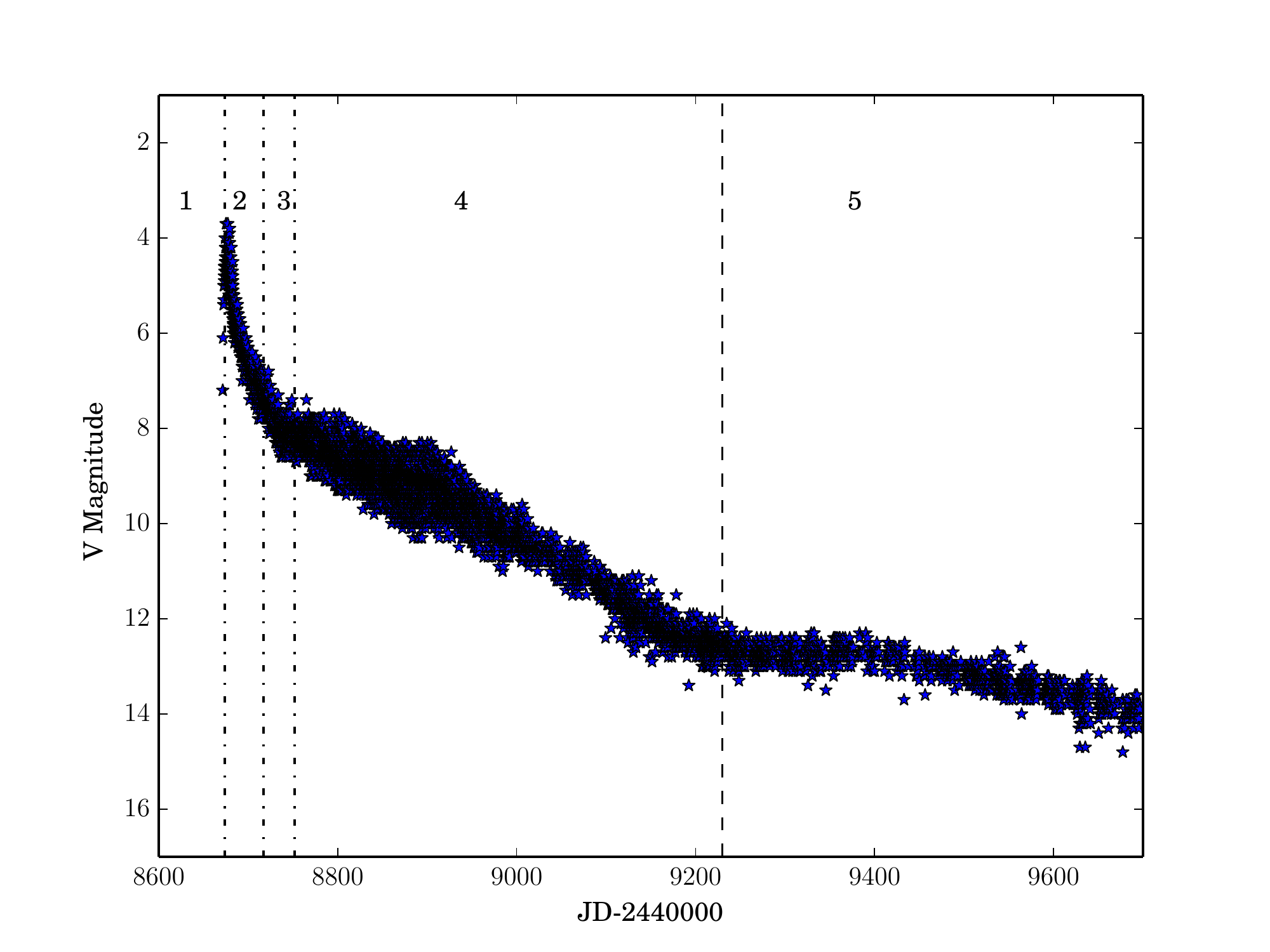}
\caption[V1974 Cyg finding chart]{V1974 Cyg finding chart showing the variable (V), comparison (C) and check (K) stars.}
\label{fig:v1974cyg}
\end{figure}

\begin{table}
\caption{V1974 Cyg Variable, comparison and check stars' parameters}
\begin{center}
\label{tab:v1974cygcandk}
\resizebox{\textwidth}{!}{%
\begin{tabular}{|l|c|c|c|c|c|c|c|}\hline\hline
Star&Identifier&Right Ascension&Declination&B&V&$\mathrm{\mathrm{R_c}}$&$\mathrm{\mathrm{I_c}}$\\
\hline
Variable   & V1974 Cyg              & 20 30 31.61   &  +52 37 51.3      &  18.5     &  4.2 - 17.5 & 17.9  &             \\
Comparison & 2MASS 20303767+5238093 & 20 30 37.652  &  +52 38 09.83     &  14.797   &  14.101 &   13.735  &   13.345    \\
Check      & 2MASS 20302890+5237123 & 20 30 28.900  &  +52 37 12.47     &  15.972   &  14.489 &   13.697  &   13.307    \\
\hline
\end{tabular}}
\end{center}
\end{table}

We found an epoch in the 20-10-2015 $\mathrm{I_c}$ light curve of our observations at JD 2457316.385920. We used these two epochs to plot a phase-differential magnitude diagram and we find a phase shift of 0.399066 d.

\begin{eqnarray}~\label{eq:v1974cyg}
HJD_{min}&=&2457316.385920 + 0.081263 \times E\\
        &&\pm 0.00512738 \;\;\;\;\;\ \pm 0.000003 \nonumber
\end{eqnarray}

\begin{figure}
\centering
\includegraphics[height=14cm,width=13cm]{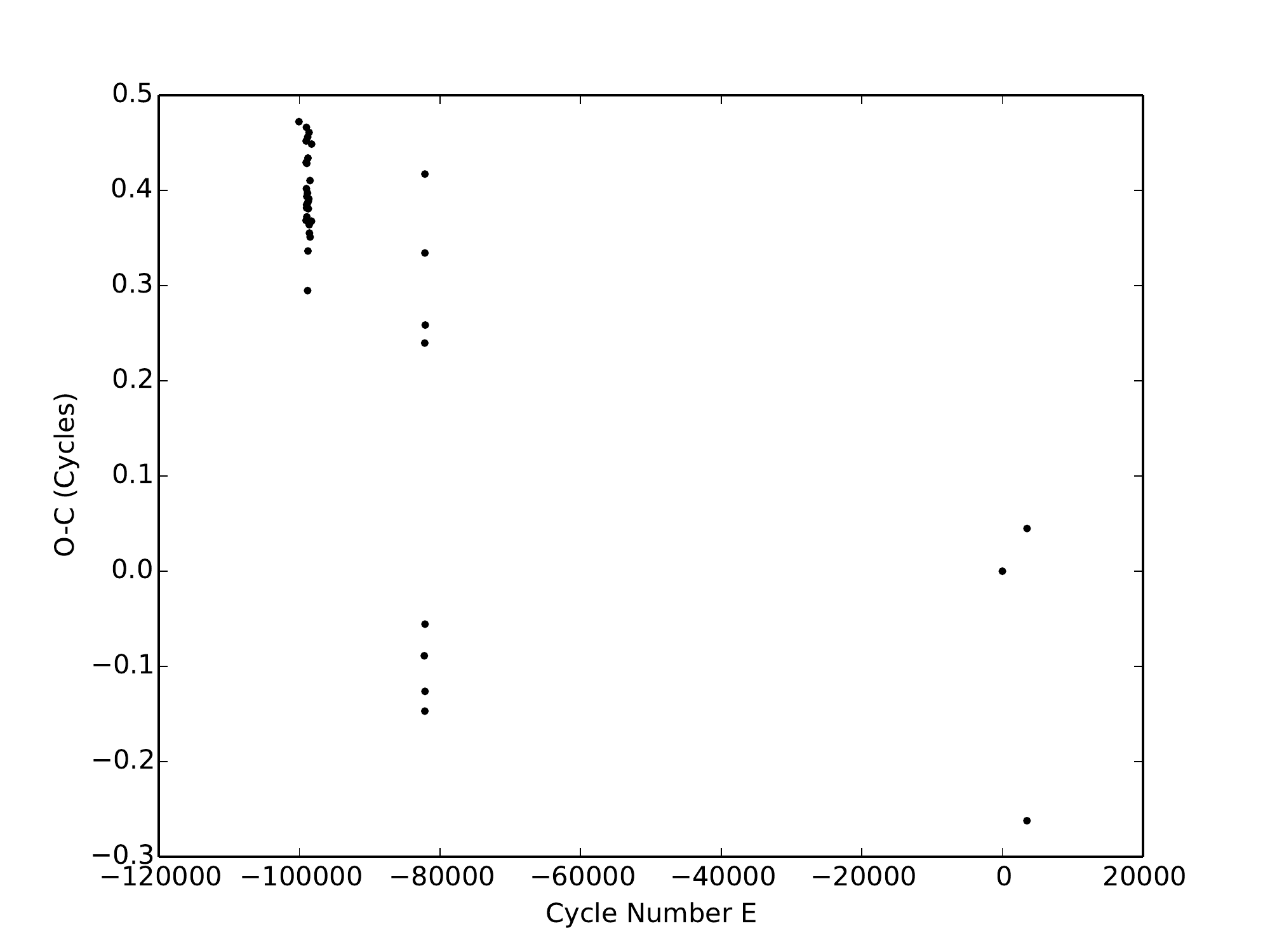}
\caption[V1974 V phase-differential magnitude diagram]{V1974 V phase-differential magnitude diagram for the ephemeris $HJD_{min}=2449267.562 + 0.081263 \times E$ (stars) and\\ $HJD_{min}=2457316.385920 + 0.081263 \times E$ (circles)}
\label{fig:v1974cygtwoepochs}
\end{figure}

\chapter{Results and Discussion}

\section{Ultraviolet Spectroscopy Results}\label{sec:res}

In this section, we present the results of our analysis for the reslts presented in the previous chapter. Section ~\ref{sec:res} summarizes our main results: from the analysis of the evolution of the normalized fluxes of selected emission lines, we calculate the ultraviolet luminosities which, in turn, provide the mass accretion rates of the three novae in quiescence. A discussion on the results is presented in section ~\ref{sec:dis} while the main conclusions are listed in section ~\ref{sec:conc}.

\subsection{PW Vul}\label{subsec:pwvul}
PW Vul (Nova Vul 1984a) was discovered in outburst on July $28^{th}$ 1984 (JD\nomenclature{JD}{Julian Day} 2445910) and it reached its visual maximum seven days later \citep{2005A&A...439..205C,1988ApJ...329..894G,1984IAUC.3963....2K}. Its V magnitude at discovery was 9.2 mag and at maximum it reached  a magnitude of 6.3 mag \citep{1991A&A...244..111A}. \citet{2014ApJ...785...97H} assumed the outburst day was JD 2445908.0 (UT 1984 July 26.5). \citet{1988ApJ...329..894G} report a $t_3$ time of 100 days for PW Vul making it a slow nova. \citet{2000AJ....120.2007D} used expansion parallax method to obtain a distance of $1.8 \pm 0.05 \,\mathrm{kpc}$. We adopted a value of $2.3 \pm 0.6 \,\mathrm{kpc}$ obtained from {\it Gaia} parallax \citep{2018arXiv180409366L,2018arXiv180409365G} with E(B-V) = 0.55 mag \citep{2015ApJ...798...76H,2000AJ....120.2007D}. 

The emission lines seen in the UV spectra of PW Vul cover a wide range of ionization states and ionization potentials from 13.60 eV for O I to 47.98 eV for C IV. In this thesis we studied the C IV 1550 \AA \, resonance emission doublet, the O I 1306 \AA \, collisionally excited resonance triplet pumped by Hydrogen Lyman $\beta$ and the N IV] 1487 \AA \, and C III] 1909 \AA \, inter-combination lines.  
The normalized line fluxes are plotted as a function of Julian date in Figs ~\ref{fig:oi} -~\ref{fig:civ}. We used these fluxes to calculate line luminosities of the spectral lines using the equation
\begin{equation}\label{eq:uv_luminosity}
L_{\lambda} = 4 \pi F_{\lambda} d^2
\end{equation}
where $F_{\lambda}$ \nomenclature{$F_{\lambda}$}{Integrated Flux} is the integrated flux corrected for interstellar extinction using the equation
\begin{equation}\label{eq:extinction}
F =  F_o 10^{0.4X(\lambda)E(B-V)}
\end{equation}
from \citet{1981MNRAS.197..107S}, where $F_o$ is the observed unnormalised flux and $X(\lambda)$ is the analytic function fitted by \citet{1979MNRAS.187P..73S} for the interstellar extinction in the UV. Maximum, intermediate and minimum luminosity values are listed in tables ~\ref{tab:calc1} and ~\ref{tab:calc2}, where the maximum is the peak UV flux and the minimum is the lowest value we calculated for the UV luminosity.
We consider that the system has entered the quiescent stage for the last two spectra (SWP28068LL ans SWP28461), when the flux of the continuum and emission lines have decreased to 0.2\% and less than 3\% of the maximum values, respectively.

Our assumption is consistent with the model of \citet{2015ApJ...798...76H} where they assumed that the wind stopped about 620 days after the outburst (marked by the dotted line in figs ~\ref{fig:oi}-~\ref{fig:civ} which lies between the last two spectra). We then used the quiescent luminosity to calculate the mass accretion rate onto the white dwarf $\dot{M}_\mathrm{{acc}}$ using the equation

\begin{equation}\label{eq:mass_acc2}
\dot{M}_\mathrm{{acc}} = \frac{2{L}R_\mathrm{{W\-D}}}{GM_\mathrm{{W\-D}}}
\end{equation}


from \citet{2002apa..book.....F}, where $G$ is the universal gravitational constant, $M_\mathrm{{W\-D}}$ is the mass of the white dwarf (= $0.83 \mathrm{M_{\odot}}$ as derived by \citealp{2015ApJ...798...76H}). $R_\mathrm{{W\-D}}$ is the radius of the white dwarf calculated using equation ~\ref{eq:radius} and it was found to be $0.0097 R_{\odot}$.

Equation ~\ref{eq:mass_acc2} is used to calculate the mass accretion rate on non-magnetic white dwarfs. We used it to calculate the mass accretion rate on PW Vul and we have not found any observations reported in the literature detecting the presence of a magnetic field in PW Vul, therefore we assumed it is a non-magnetic novae.
 
The normalized flux of all emission lines start generally at relatively low values and increase gradually to reach maxima on JD 2446018, 101 days after the visual maximum  for O I line. The C III], C IV, and  N IV] lines reach maxima on JD 2446163, 246 days after the visual maximum. This behavior can be seen in Figs ~\ref{fig:oi} -~\ref{fig:civ} 

\begin{figure}[hb!]
\centering                                                                                                   
\includegraphics[height=14cm,width=15cm]{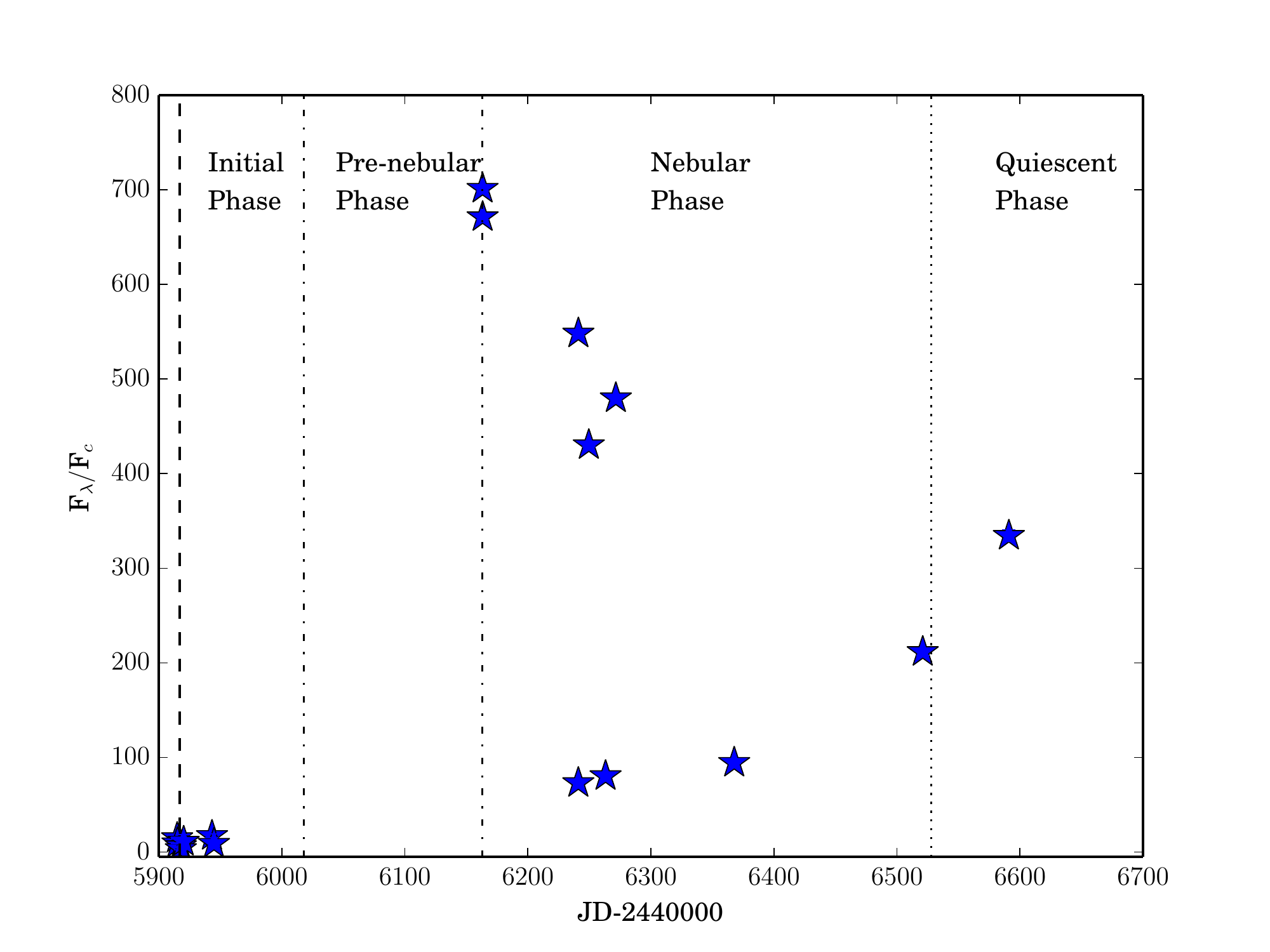}                                                            
\caption[PW Vul O I line spectral evolution]{PW Vul O I line spectral evolution. The dashed line represents the time of visual maximum and the dash-dotted lines represent the end of the different phases of evolution. The dotted line represents the start of the quiescent phase.}
\label{fig:oi}
\end{figure}

\begin{figure}
\centering 
\includegraphics[height=14cm,width=13cm]{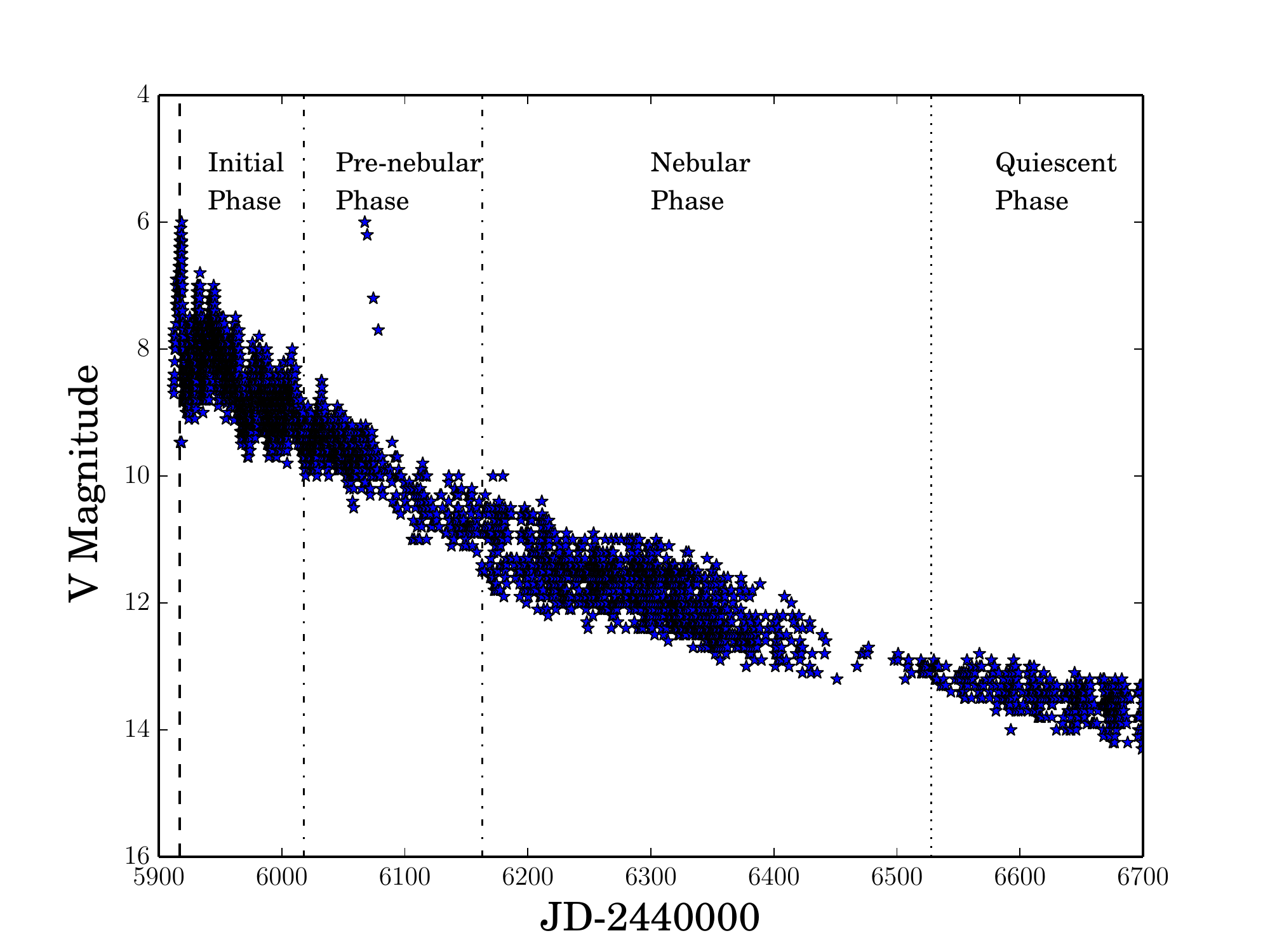}
\caption[PW Vul C III{]} line spectral evolution]{PW Vul C III] line spectral evolution. The dashed line represents the time of visual maximum and the dash-dotted lines represent the end of the different phases of evolution. The dotted line represents the start of the quiescent phase.}
\label{fig:ciii}
\end{figure}

\begin{figure}                                                                                                  
\centering                                                                                                      
\includegraphics[height=14cm,width=13cm]{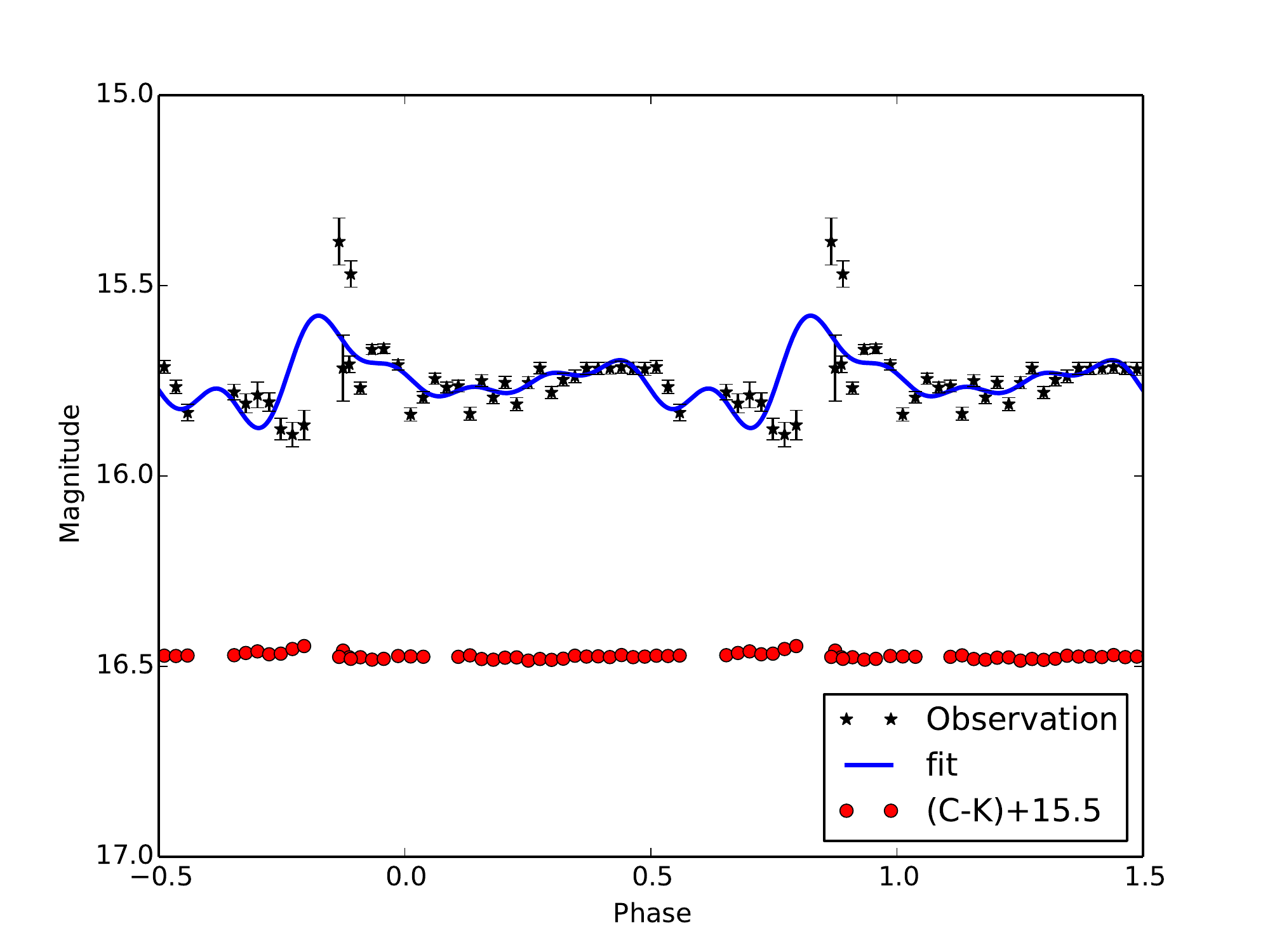}
\caption[PW Vul N IV{]} line spectral evolution]{PW Vul N IV] line spectral evolution. The dashed line represents the time of visual maximum and the dash-dotted lines represent the end of the different phases of evolution. The dotted line represents the start of the quiescent phase.}
\label{fig:niv}
\end{figure}

\begin{figure}
\centering 
\includegraphics[height=14cm,width=13cm]{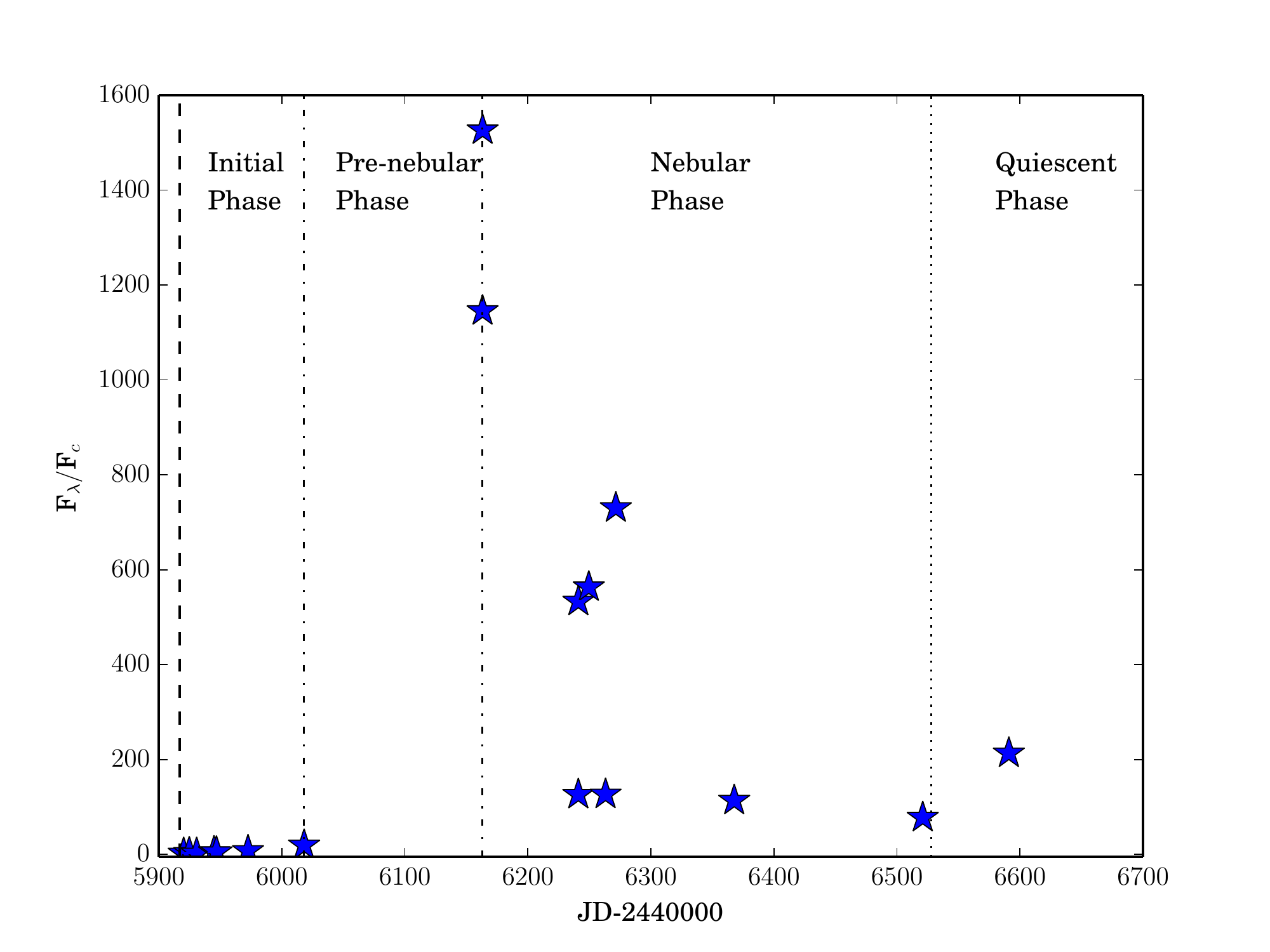}
\caption[PW Vul C IV line spectral evolution]{PW Vul C IV line spectral evolution. The dashed line represents the time of visual maximum and the dash-dotted lines represent the end of the different phases of evolution. The dotted line represents the start of the quiescent phase.}
\label{fig:civ}
\end{figure}

\begin{figure}
\centering
\includegraphics[height=14cm,width=13cm]{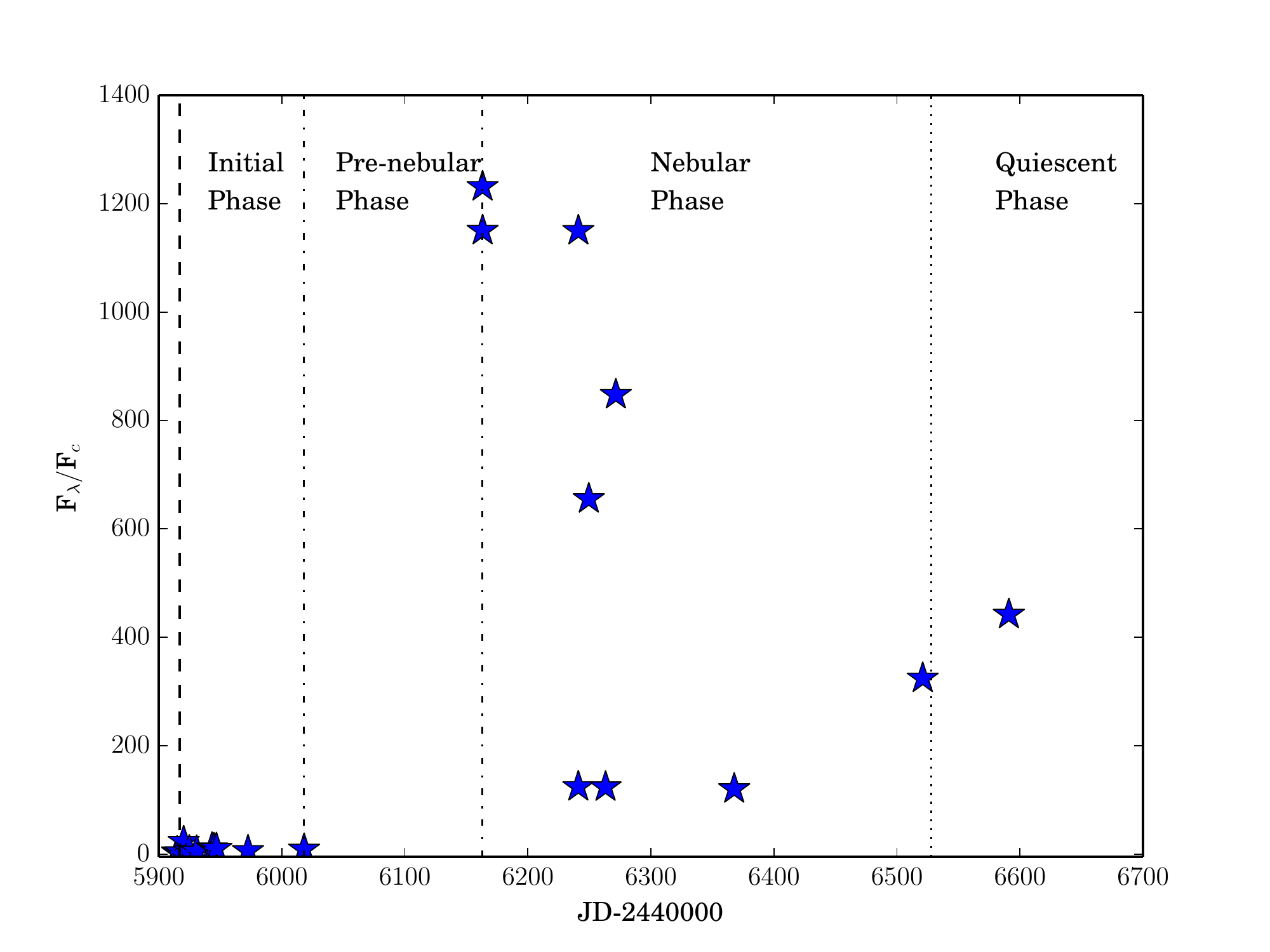}
\caption[PW Vul UV shortwavelength continuum evolution]{PW Vul UV shortwavelength continuum evolution. The dashed line represents the time of visual maximum and the dash-dotted lines represent the end of the different phases of evolution. The dotted line represents the start of the quiescent phase.}
\label{fig:pwvul_continuum}
\end{figure}

\begin{table}[t]                       
\caption[PW Vul O I and C III{]} emission line parameters]{PW Vul O I and C III] emission line parameters. $(F_{\lambda}/F_c)$ \nomenclature{$F_{\lambda}/F_c$}{Normalized Flux} is the normalized flux, $F_{\lambda}$ is the absolute flux and L is the Luminosity. Intermediate values are the mean of all the values calculated for a specific parameter.}
\label{tab:calc1}                     
\begin{center}
\resizebox{\textwidth}{!}{%
\begin{tabular}{|lcc|}\hline\hline
Value 											&         O I        &  C III]           \\ \hline 
$(F_{\lambda}/F_c)_{\mathrm{max}}                                                       $&$ 40 \pm 1        $&$ 1527   \pm 5    $\\
$F_{\lambda(\mathrm{max})} (  10^{-09} \mathrm{erg\, cm^{-2}\,s^{-1}\,\text{\AA}^{-1}}) $&$ 1.92\pm 0.05    $&$ 1.265 \pm 0.002 $\\
$L_{\mathrm{max}}(  10^{35}\mathrm{erg\, s^{-1}})                                       $&$ 12 \pm 5     $&$ 8 \pm 3     $\\
$(F_{\lambda}/F_c)_{\mathrm{mid}}                                                       $&$ 9 \pm 1         $&$ 113.7 \pm 0.8   $\\
$F_{\lambda(\mathrm{mid})} (  10^{-11} \mathrm{erg\, cm^{-2}\,s^{-1}\,\text{\AA}^{-1}}) $&$ 2.0 \pm 0.1     $&$ 22 \pm 3        $\\
$L_{\mathrm{mid}}   (  10^{34} \mathrm{erg\, s^{-1}})                                   $&$ 1.3 \pm 0.4 $&$ 14 \pm 5         $\\
$(F_{\lambda}/F_c)_{\mathrm{min}}                                                       $&$ 2 \pm 1         $&$ 2.3 \pm 0.9     $\\
$F_{\lambda(\mathrm{min})} (  10^{-12} \mathrm{erg\, cm^{-2}\,s^{-1}\,\text{\AA}^{-1}}) $&$ 2.9 \pm 0.4     $&$ 8 \pm 4         $\\
$L_{\mathrm{min}}  (  10^{33} \mathrm{erg\, s^{-1}})                                    $&$ 1.8 \pm 0.7     $&$ 5 \pm 3         $\\
\hline                                                
\end{tabular}}
\end{center}                                          
\end{table}   

\begin{table}[t]                       
\caption[PW Vul N IV{]} and C IV emission line parameters]{PW Vul N IV] and C IV emission line parameters. $(F_{\lambda}/F_c)$ is the normalized flux, $F_{\lambda}$ is the absolute flux and L is the Luminosity. Intermediate values are the mean of all the values calculated for a specific parameter.}
\label{tab:calc2}                     
\begin{center}
\resizebox{\textwidth}{!}{%
\begin{tabular}{|lcc|}\hline\hline
Value 											 &     N IV]              & C IV            \\ \hline 
$(F_{\lambda}/F_c)_{\mathrm{max}}                                                       $&$ 701 \pm 2           $&$ 1231 \pm 4     $\\
$F_{\lambda(\mathrm{max})} (  10^{-09} \mathrm{erg\, cm^{-2}\,s^{-1}\,\text{\AA}^{-1}}) $&$ 1.39 \pm 0.01       $&$ 2.099 \pm 0.006$\\
$L_{\mathrm{max}}(  10^{35}\mathrm{erg\, s^{-1}})                                       $&$ 8 \pm 3         $&$ 13.2 \pm 0.5    $\\
$(F_{\lambda}/F_c)_{\mathrm{mid}}                                                       $&$ 80 \pm 1            $&$ 11 \pm 1       $\\
$F_{\lambda(\mathrm{mid})} (  10^{-11} \mathrm{erg\, cm^{-2}\,s^{-1}\,\text{\AA}^{-1}}) $&$ 15 \pm 2            $&$ 15 \pm 1       $\\
$L_{\mathrm{mid}}   (  10^{34} \mathrm{erg\, s^{-1}})                                   $&$ 9 \pm 4         $&$ 9 \pm 4    $\\
$(F_{\lambda}/F_c)_{\mathrm{min}}                                                       $&$ 3 \pm 1             $&$ 2 \pm 1        $\\
$F_{\lambda(\mathrm{min})} (  10^{-12} \mathrm{erg\, cm^{-2}\,s^{-1}\,\text{\AA}^{-1}}) $&$ 0.8 \pm 0.5         $&$ 7 \pm 3        $\\
$L_{\mathrm{min}}  (  10^{33} \mathrm{erg\, s^{-1}})                                    $&$ 0.5 \pm 0.4         $&$ 4 \pm 2        $\\
\hline                                                
\end{tabular}}
\end{center}                                          
\end{table}
\clearpage

\subsection{V1668 Cyg}\label{subsec:v1668cyg}

V1668 Cyg (Nova Cygni 1978) was discovered in outburst on JD 2443762 with a visual magnitude of 6.2 \citep{1978IAUC.3264....3M}. At maximum, it reached a visual magnitude of 6.0 mag on JD 2443764 \citep{1978IAUC.3268....1A}. \citet{1980ApJ...239..570G} estimated the outburst day to be JD 2443759.0. \citet{1980A&A....81..157D} reported a $t_3$ time of 23 days making it a fast nova. 

For V1668 Cyg, we studied the C IV 1550 \AA \,, N IV] 1487 \AA \,, O I 1306 \AA \, lines as well as the N III] 1750 \AA \, inter-combination line, the N V 1240 \AA \, resonance line and the C II 1336 \AA \,  resonance doublet. The emission line with highest ionization potential was the N V resonance line at 1240 \AA \, with 77.47 eV. The normalized line fluxes for all the studied emission lines are plotted as a function of Julian date in figs ~\ref{fig:cii_1668cyg} -~\ref{fig:nv_v1668cyg}.

We calculated the line luminosities for V1668 Cyg using equation ~\ref{eq:uv_luminosity} and assuming a distance of $3.3\pm 0.6 \,\mathrm{kpc}$ from {\it Gaia} archives \citep{2018arXiv180409366L,2018arXiv180409365G} and E(B-V) = 0.38 mag, as derived by \citet{1979Natur.277..114S}. Maximum, intermediate and minimum luminosity values are listed in tables ~\ref{tab:calc3} -~\ref{tab:calc5}.

The normalized fluxes of different emission lines show similar behaviors, starting generally at relatively low values and then increasing gradually towards the maximum as can be seen in Figures ~\ref{fig:cii_1668cyg} -~\ref{fig:nv_v1668cyg}. The maximum value is reached at different days after the visual maximum for different ionized species: 59 days (JD 2443821) for both C II and O I, 85 days (JD 2443847) for N III], 123 days (JD 2443885) for N V, 176 days (JD 2443938) for C IV and 185 days (JD 2443947) for N IV]. However, in the quiescent phase (on JD 2444382), we measured a value for the C IV normalized flux higher than that we measured on JD 2443938. This may be attributed to the very low value of the continuum flux on that day. See Fig. ~\ref{fig:v1668cyg_continuum}.

\begin{figure}
\includegraphics[height=14cm,width=13cm]{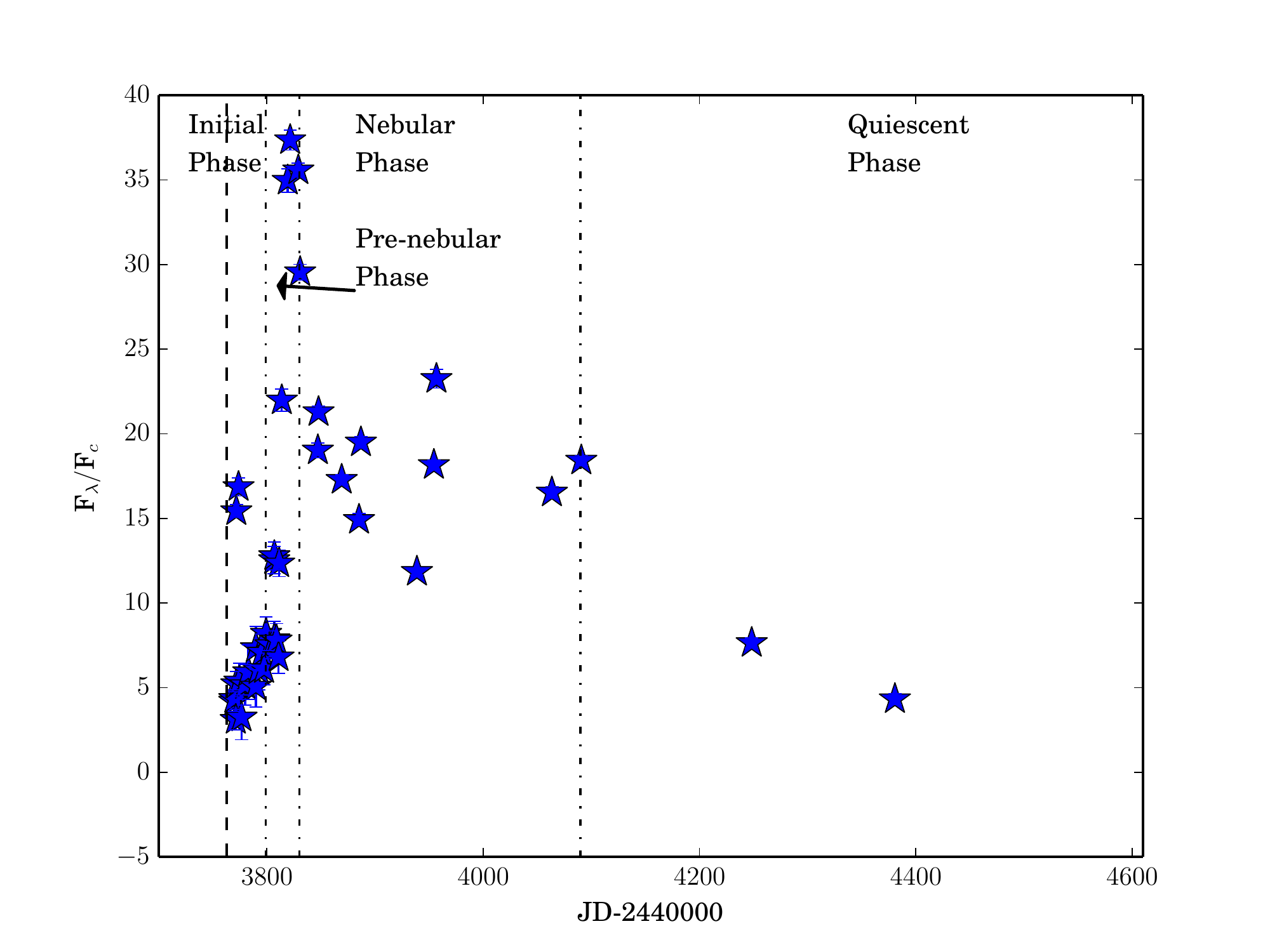}
\caption[V1668 Cyg C II line spectral evolution.]{V1668 Cyg C II line spectral evolution. The dashed line represents the time of visual maximum and the dash-dotted lines represent the end of the different phases of evolution. The dotted line represents the start of the quiescent phase.}
\label{fig:cii_1668cyg}
\end{figure}

\begin{figure}
\centering 
\includegraphics[height=14cm,width=13cm]{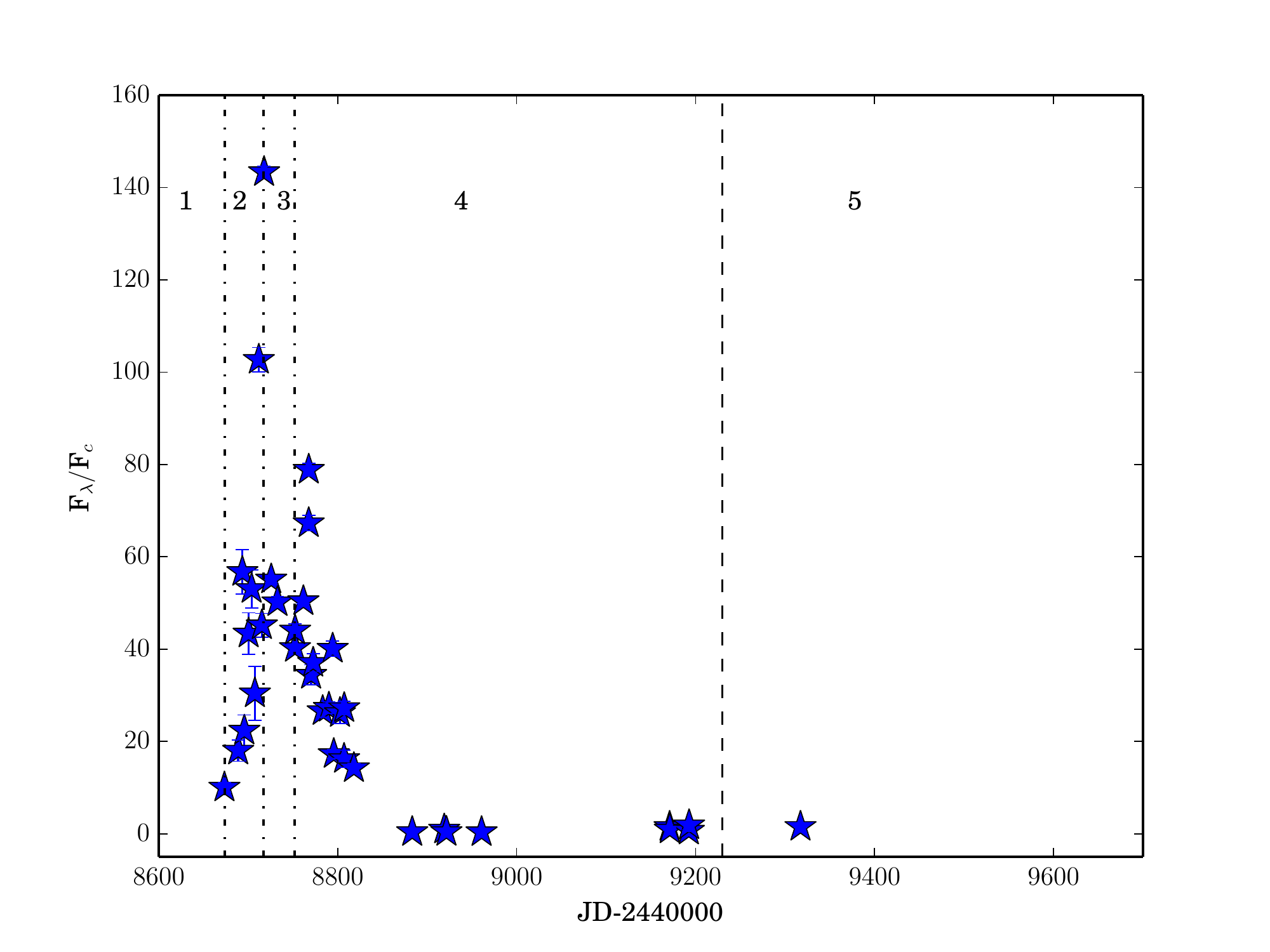}
\caption[V1668 Cyg O I line spectral evolution.]{V1668 Cyg O I line spectral evolution. The dashed line represents the time of visual maximum and the dash-dotted lines represent the end of the different phases of evolution. The dotted line represents the start of the quiescent phase.}
\label{fig:oi_v1668cyg}
\end{figure}

\begin{figure}
\centering
\includegraphics[height=14cm,width=13cm]{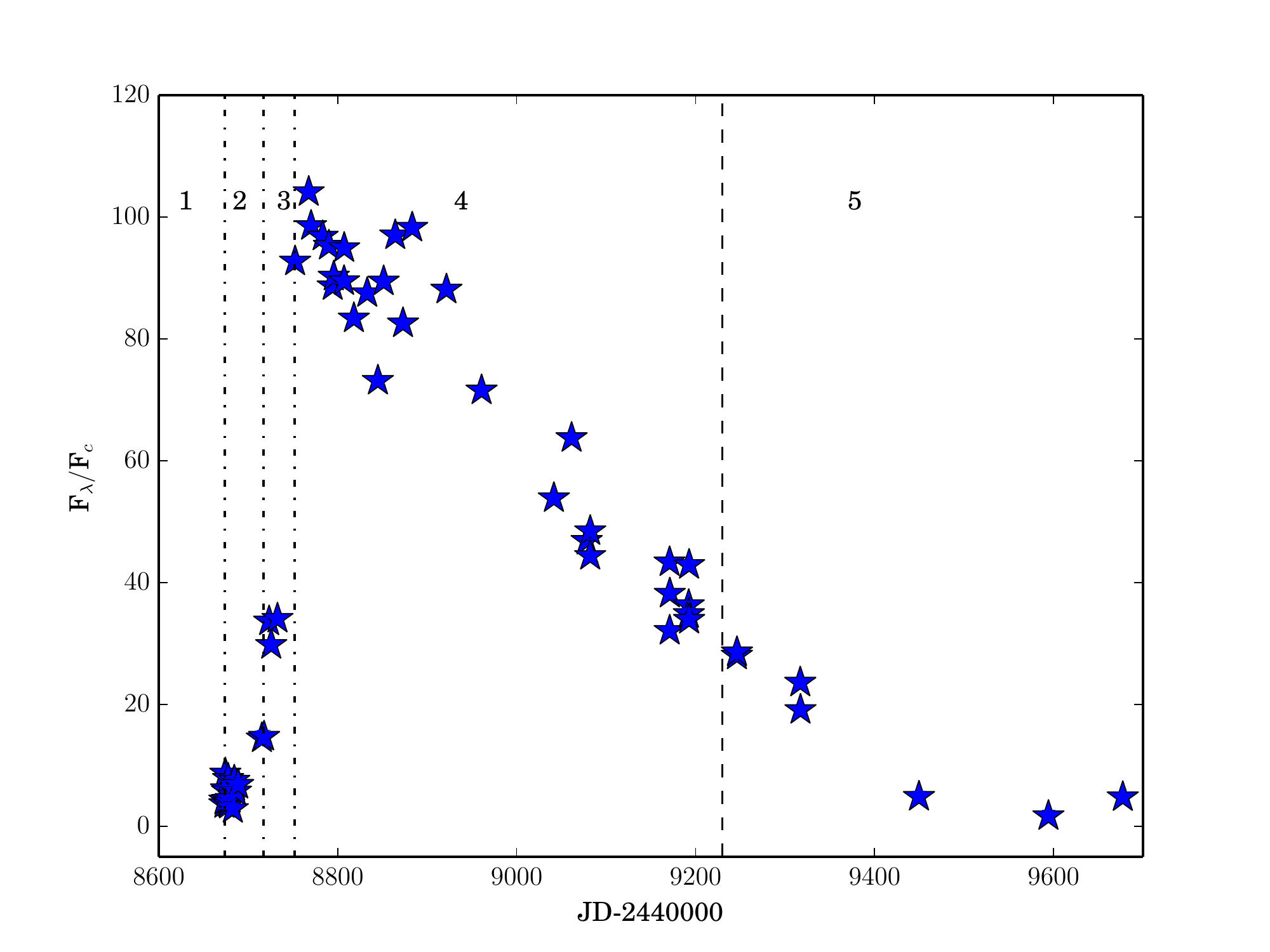}
\caption[V1668 Cyg N III{]} line spectral evolution.]{V1668 Cyg N III] line spectral evolution. The dashed line represents the time of visual maximum and the dash-dotted lines represent the end of the different phases of evolution. The dotted line represents the start of the quiescent phase.}
\label{fig:niii_v1668cyg}
\end{figure}

\begin{figure}
\includegraphics[height=14cm,width=13cm]{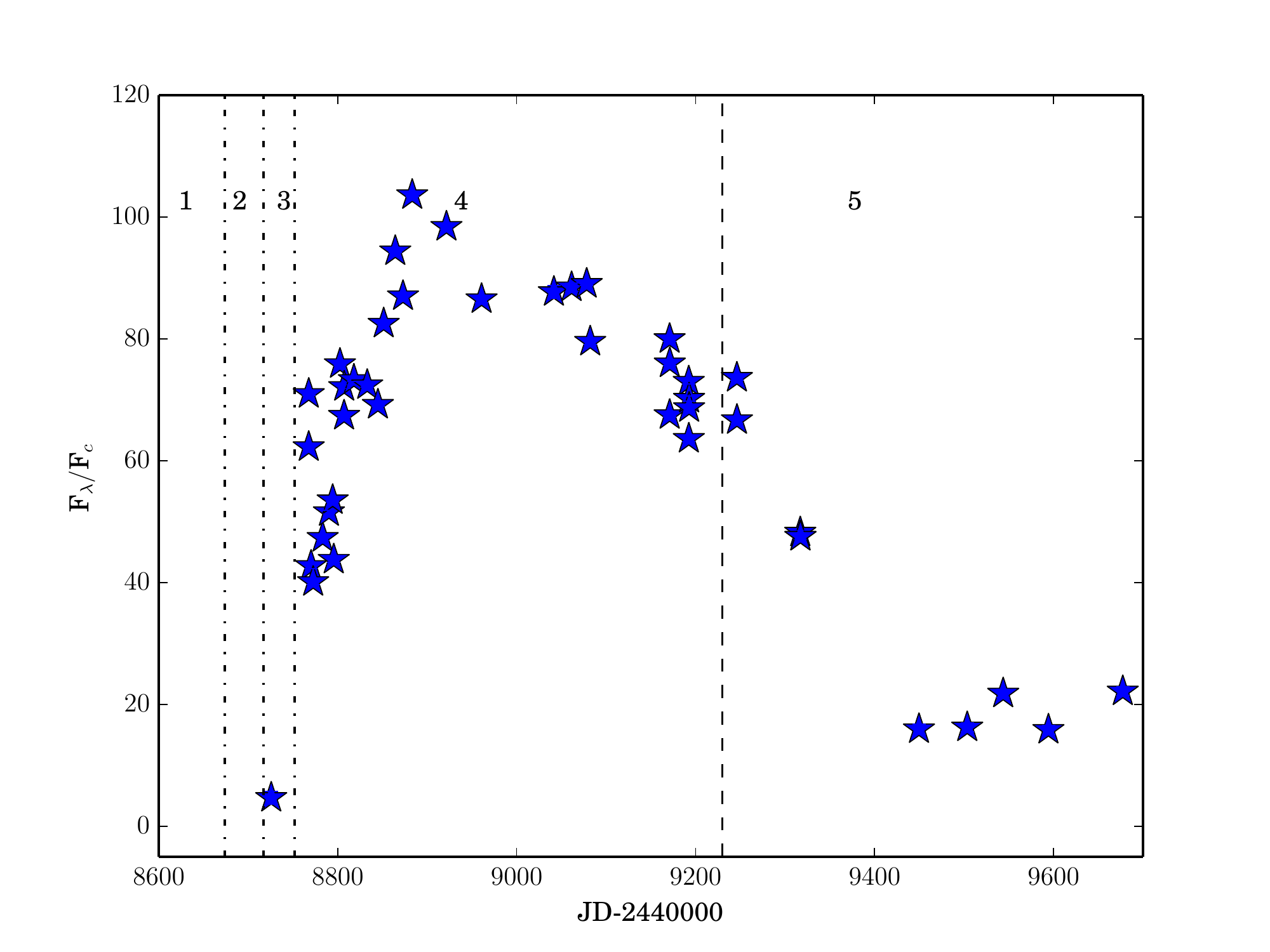}
\caption[V1668 Cyg N IV{]} line spectral evolution.]{V1668 Cyg N IV] line spectral evolution. The dashed line represents the time of visual maximum and the dash-dotted lines represent the end of the different phases of evolution. The dotted line represents the start of the quiescent phase.}
\label{fig:niv_1668cyg}
\end{figure}

\begin{figure}
\centering
\includegraphics[height=14cm,width=13cm]{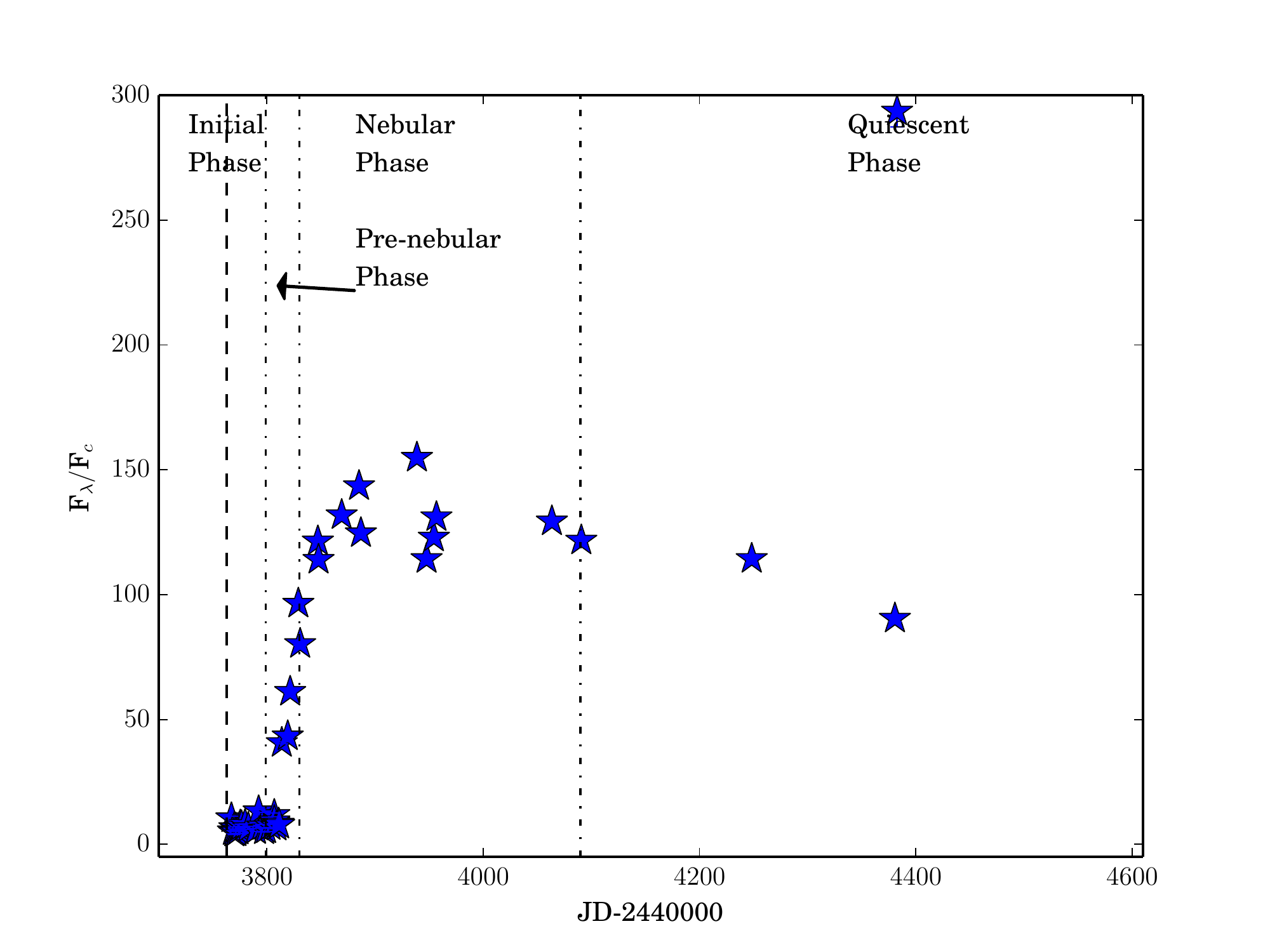}
\caption[V1668 Cyg C IV line spectral evolution.]{V1668 Cyg C IV line spectral evolution. The dashed line represents the time of visual maximum and the dash-dotted lines represent the end of the different phases of evolution. The dotted line represents the start of the quiescent phase.}
\label{fig:civ_v1668cyg}
\end{figure}

\begin{figure}
\centering
\includegraphics[height=14cm,width=13cm]{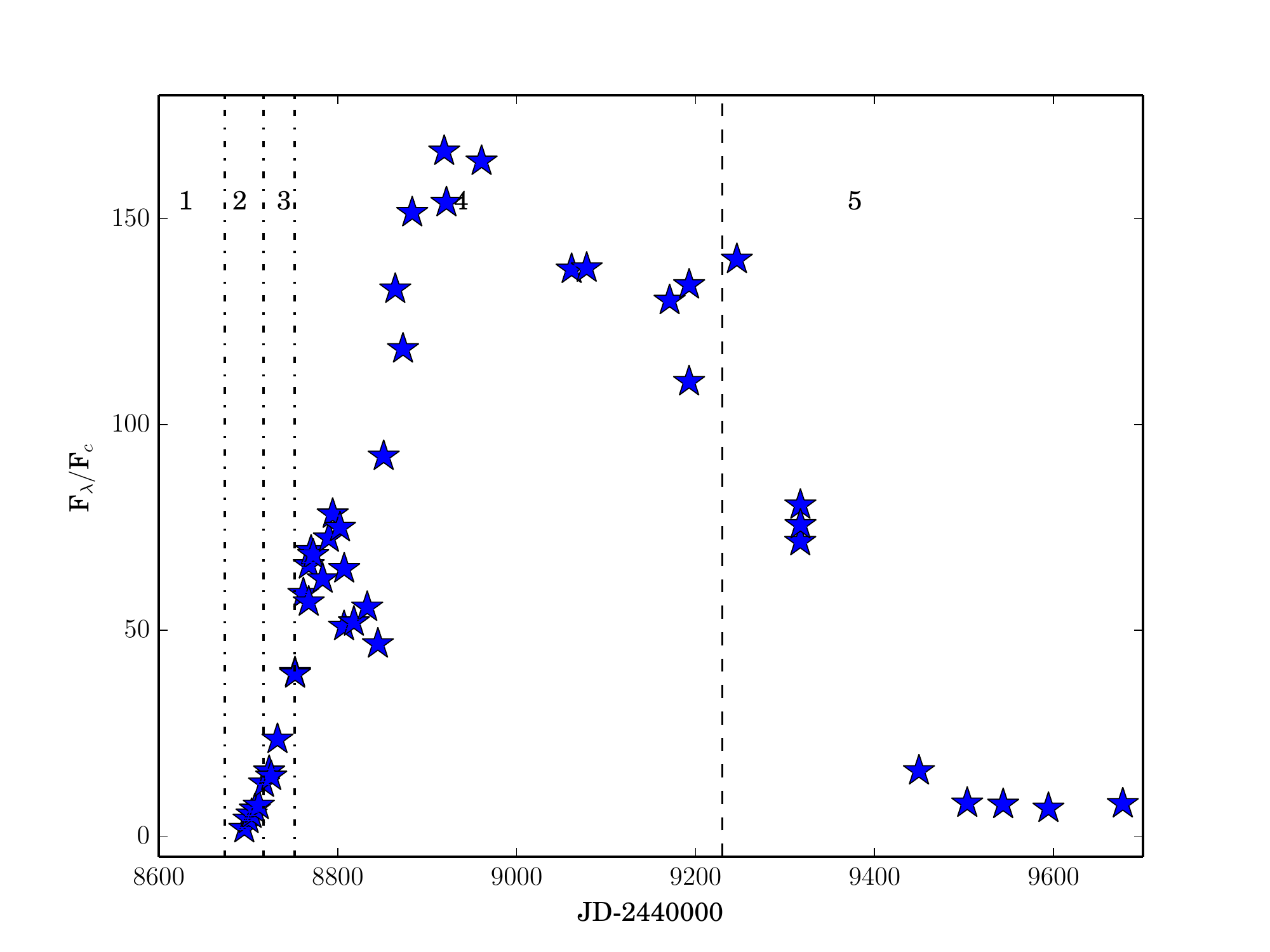}
\caption[V1668 Cyg N V line spectral evolution.]{V1668 Cyg N V line spectral evolution. The dashed line represents the time of visual maximum and the dash-dotted lines represent the end of the different phases of evolution. The dotted line represents the start of the quiescent phase.}
\label{fig:nv_v1668cyg}
\end{figure}

\begin{figure}
\centering
\includegraphics[height=14cm,width=13cm]{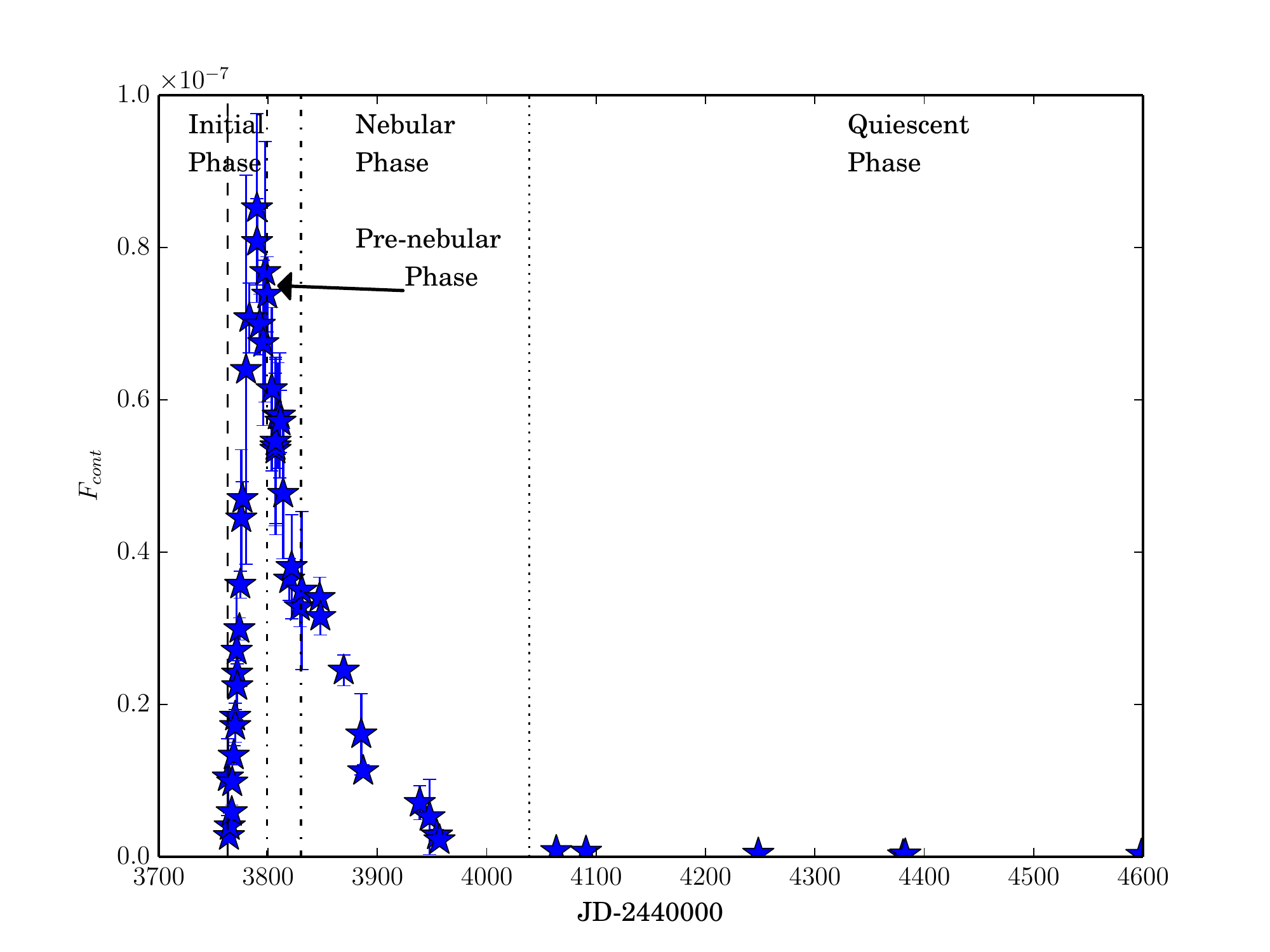}
\caption[V1668 Cyg UV shortwavelength continuum evolution]{V1668 Cyg UV shortwavelength continuum evolution. The dashed line represents the time of visual maximum and the dash-dotted lines represent the end of the different phases of evolution. The dotted line represents the start of the quiescent phase.}
\label{fig:v1668cyg_continuum}
\end{figure}


\begin{table}                       
\caption[V1668 Cyg O I and N III{]} emission line parameters]{V1668 Cyg O I and N III] emission line parameters. $(F_{\lambda}/F_c)$ is the normalized flux, $F_{\lambda}$ is the absolute flux and L is the Luminosity. Intermediate values are the mean of all the values calculated for a specific parameter.}
\label{tab:calc3}                     
\begin{center}
\resizebox{\textwidth}{!}{%
\begin{tabular}{|lcc|}\hline\hline
Value										        &O I                 &        N III]    \\ \hline 
$(F_{\lambda}/F_c)_{\mathrm{max}}                                                      $&$ 55.5 \pm 0.3     $&$ 107.5 \pm 0.2  $\\      
$F_{\lambda(\mathrm{max})}(  10^{-09} \mathrm{erg\, cm^{-2}\,s^{-1}\,\text{\AA}^{-1}}) $&$ 1.70 \pm 0.04    $&$ 1.78 \pm 0.02  $\\
$L_{\mathrm{max}} (  10^{36} \mathrm{erg\, s^{-1}})                                    $&$ 2.2 \pm 0.6    $&$ 2.3 \pm 0.6  $\\
$(F_{\lambda}/F_c)_{\mathrm{mid}}                                                      $&$ 32.4 \pm 0.6     $&$ 43.6 \pm 0.6   $\\      
$F_{\lambda(\mathrm{mid})}(  10^{-10} \mathrm{erg\, cm^{-2}\,s^{-1}\,\text{\AA}^{-1}}) $&$ 7.9  \pm 0.3     $&$ 1.4 \pm 0.2    $\\
$L_{\mathrm{mid}} (  10^{35} \mathrm{erg\, s^{-1}})                                    $&$ 10 \pm 3     $&$ 1.8 \pm 0.5    $\\
$(F_{\lambda}/F_c)_{\mathrm{min}}                                                      $&$ 1.2  \pm 0.2     $&$ 2 \pm 1        $\\      
$F_{\lambda(\mathrm{min})}(  10^{-13} \mathrm{erg\, cm^{-2}\,s^{-1}\,\text{\AA}^{-1}}) $&$ 2    \pm 1       $&$ 3.4 \pm 0.3    $\\
$L_{\mathrm{min}}  (  10^{32} \mathrm{erg\, s^{-1}})                                   $&$ 2    \pm 1       $&$ 4 \pm 1    $\\
\hline
\end{tabular}}
\end{center}
\end{table}

\begin{table}                       
\caption[V1668 Cyg C IV and N V emission line parameters]{V1668 Cyg C IV and N V emission line parameters. $(F_{\lambda}/F_c)$ is the normalized flux, $F_{\lambda}$ is the absolute flux and L is the Luminosity. Intermediate values are the mean of all the values calculated for a specific parameter.}
\label{tab:calc4}                     
\begin{center}
\resizebox{\textwidth}{!}{%
\begin{tabular}{|lcc|}\hline\hline
Value										        & C IV             &N V               \\ \hline 
$(F_{\lambda}/F_c)_{\mathrm{max}}                                                      $&$ 294  \pm 6     $&$ 142  \pm 1     $\\      
$F_{\lambda(\mathrm{max})}(  10^{-09} \mathrm{erg\, cm^{-2}\,s^{-1}\,\text{\AA}^{-1}}) $&$ 1.68 \pm 0.01  $&$ 0.896 \pm 0.007$\\
$L_{\mathrm{max}} (  10^{36} \mathrm{erg\, s^{-1}})                                    $&$ 2.2 \pm 0.6  $&$ 1.2 \pm 0.3  $\\
$(F_{\lambda}/F_c)_{\mathrm{mid}}                                                      $&$ 8.7  \pm 0.8   $&$ 20.5 \pm 0.2   $\\      
$F_{\lambda(\mathrm{mid})}(  10^{-10} \mathrm{erg\, cm^{-2}\,s^{-1}\,\text{\AA}^{-1}}) $&$ 1.4  \pm 0.2   $&$ 2.9  \pm 0.2   $\\
$L_{\mathrm{mid}} (  10^{35} \mathrm{erg\, s^{-1}})                                    $&$ 1.9  \pm 0.5   $&$ 4  \pm 1   $\\
$(F_{\lambda}/F_c)_{\mathrm{min}}                                                      $&$ 4.7  \pm 0.8   $&$ 2.4  \pm 0.1   $\\      
$F_{\lambda(\mathrm{min})}(  10^{-13} \mathrm{erg\, cm^{-2}\,s^{-1}\,\text{\AA}^{-1}}) $&$ 17.8 \pm 0.3   $&$ 17  \pm 2      $\\
$L_{\mathrm{min}}  (  10^{32} \mathrm{erg\, s^{-1}})                                   $&$ 23 \pm 6   $&$ 22  \pm 6      $\\
\hline
\end{tabular}}
\end{center}
\end{table}

\begin{table}                       
\caption[V1668 Cyg C II and N IV{]} emission line parameters]{V1668 C II and N IV] Cyg emission line parameters. $(F_{\lambda}/F_c)$ is the normalized flux, $F_{\lambda}$ is the absolute flux and L is the Luminosity. Intermediate values are the mean of all the values calculated for a specific parameter.}
\label{tab:calc5}                     
\begin{center}
\resizebox{\textwidth}{!}{%
\begin{tabular}{|lcc|}\hline\hline
Value 											 &     C II                 &  N IV]                \\ \hline 
$(F_{\lambda}/F_c)_{\mathrm{max}}                                                       $&$ 37.4 \pm 0.6           $&$ 152 \pm 2           $\\
$F_{\lambda(\mathrm{max})} (  10^{-09} \mathrm{erg\, cm^{-2}\,s^{-1}\,\text{\AA}^{-1}}) $&$0.61 \pm 0.02           $&$ 0.98 \pm 0.01       $\\
$L_{\mathrm{max}}(  10^{35}\mathrm{erg\, s^{-1}})                                       $&$ 8 \pm 2            $&$ 13 \pm 3        $\\
$(F_{\lambda}/F_c)_{\mathrm{mid}}                                                       $&$ 8   \pm 1              $&$ 71.0 \pm 0.3        $\\
$F_{\lambda(\mathrm{mid})} (  10^{-11} \mathrm{erg\, cm^{-2}\,s^{-1}\,\text{\AA}^{-1}}) $&$ 15.3 \pm 0.9           $&$ 10.5 \pm 0.1        $\\
$L_{\mathrm{mid}}   (  10^{34} \mathrm{erg\, s^{-1}})                                   $&$ 20  \pm 5          $&$ 14 \pm 3       $\\
$(F_{\lambda}/F_c)_{\mathrm{min}}                                                       $&$ 3.1 \pm 0.6            $&$ 12.9 \pm 0.6        $\\
$F_{\lambda(\mathrm{min})} (  10^{-12} \mathrm{erg\, cm^{-2}\,s^{-1}\,\text{\AA}^{-1}}) $&$ 0.34  \pm 0.08         $&$ 1.11 \pm 0.04       $\\
$L_{\mathrm{min}}  (  10^{33} \mathrm{erg\, s^{-1}})                                    $&$ 0.4 \pm 0.2            $&$ 1.5 \pm 0.4       $\\
\hline                                                
\end{tabular}}
\end{center}
\end{table}

\subsection{V1974 Cyg}\label{subsec:v1974_cyg}
The outburst of V1974 Cyg (Nova Cygni 1992) was discovered on Feb $19^{th}$ 1992 (JD 2448672.3) by \citet{1992IAUC.5454....1C} at a visual magnitude of 6.8 mag. It reached a maximum visual magnitude of 4.4 mag three days later. Its visual magnitude declined by 3 magnitudes 42 days after maximum making it a fast nova \citep{1993A&A...277..103C}. \citet{2004A&A...420..571C} estimated the outburst day to be approximately JD 2448669.0. It is a well studied classical nova with multiple observations in different bands. There are some estimates of the V1974 Cyg distance a value of $1.8 \pm 0.1 \mathrm{kpc}$ was determined  determined from the MMRD relations and expansion parallax method \citep{1997A&A...318..908C,2000AJ....120.2007D,2005ApJ...631.1094H}. In our investigation we adopt a value of $1.6 \pm 0.2 \mathrm{kpc}$ measured by {\it Gaia} \citep{2018arXiv180409366L,2018arXiv180409365G}. We adopt an E(B-V) value of 0.35 mag \citep{1995A&A...299..823P}.

The study of the UV spectrum originating from the expanding ejecta provides important information about the evolution of the outburst. In this thesis we studied a number of emission lines with a wide range of ionization potentials along with the short wavelength ultraviolet continuum. We studied the C II 1336 \AA \,,  O I 1306 \AA \,, N III] 1750 \AA \,, N IV] 1487 \AA \, and N V 1240 \AA \, lines along with the Fe II 1588 \AA \ emission line, the Al III 1854 \AA \, emission doublet and the [Ne V] 1575 \AA \, forbidden emission line. The time evolution for all the studied emission lines are shown in Figs ~\ref{fig:feii} -~\ref{fig:nev}.

We have calculated the ultraviolet luminosity of the emitting regions using  equation ~\ref{eq:uv_luminosity}

We assumed that V1974 Cyg has reached quiescent state at about JD 2449230 nearly 560 days after the discovery (the dashed line in Figs ~\ref{fig:feii} - ~\ref{fig:contflux}). This happens when the flux of the continuum and emission lines have decreased to $\sim$ 0.3\% and $\sim$ 1.5\% of the maximum values, respectively. This is consistent with the stopping of the hydrogen burning according to the model of \citet{2005ApJ...631.1094H}. We used the ultraviolet luminosity after this date to calculate the mass accretion rate on the white dwarf $\dot{M}_\mathrm{{acc}}$ using equation

\begin{equation}\label{eq:mass_acc}
\dot{M}_\mathrm{{acc}} = \frac{L_\mathrm{{acc}}R_\mathrm{{W\-D}}}{GM_\mathrm{{W\-D}}}
\end{equation}

adopting a value of $1.05 \mathrm{M_{\odot}}$ for the mass of the white dwarf \citep{2005ApJ...631.1094H}. We have calculated the radius of the white dwarf using equation ~\ref{eq:radius} and it was found to be $0.0073 \mathrm{R_{\odot}}$. 


\begin{figure}
\includegraphics[height=14cm,width=13cm]{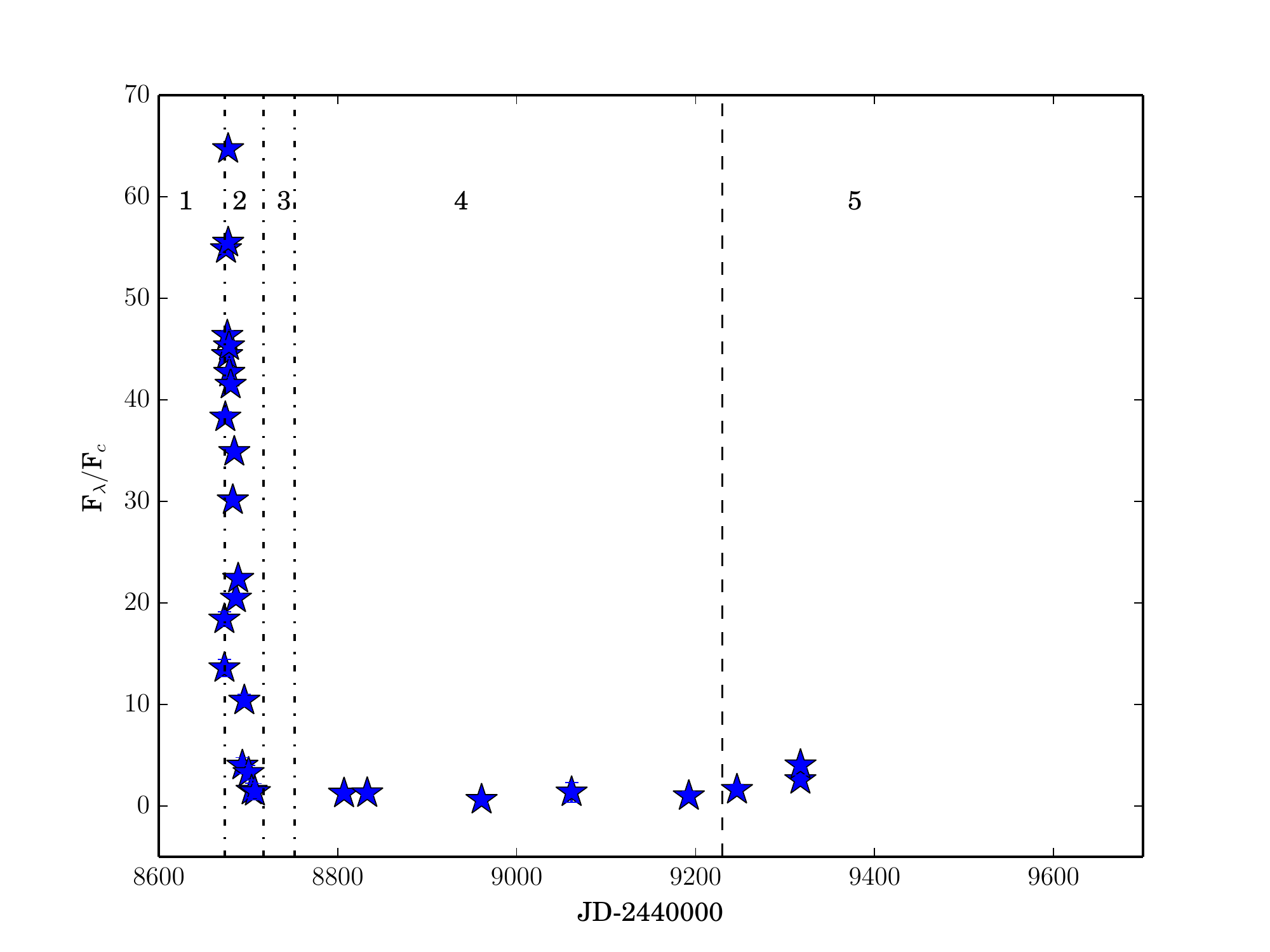}
\caption[V1974 Cyg Fe II line spectral evolution]{V1974 Cyg Fe II line spectral evolution. The numbers 1,2,3,4,and 5 correspond to the fireball, Fe optically thick, transition, nebular and quiescent phases, respectively.}
\label{fig:feii}
\end{figure}

\begin{figure}
\includegraphics[height=14cm,width=13cm]{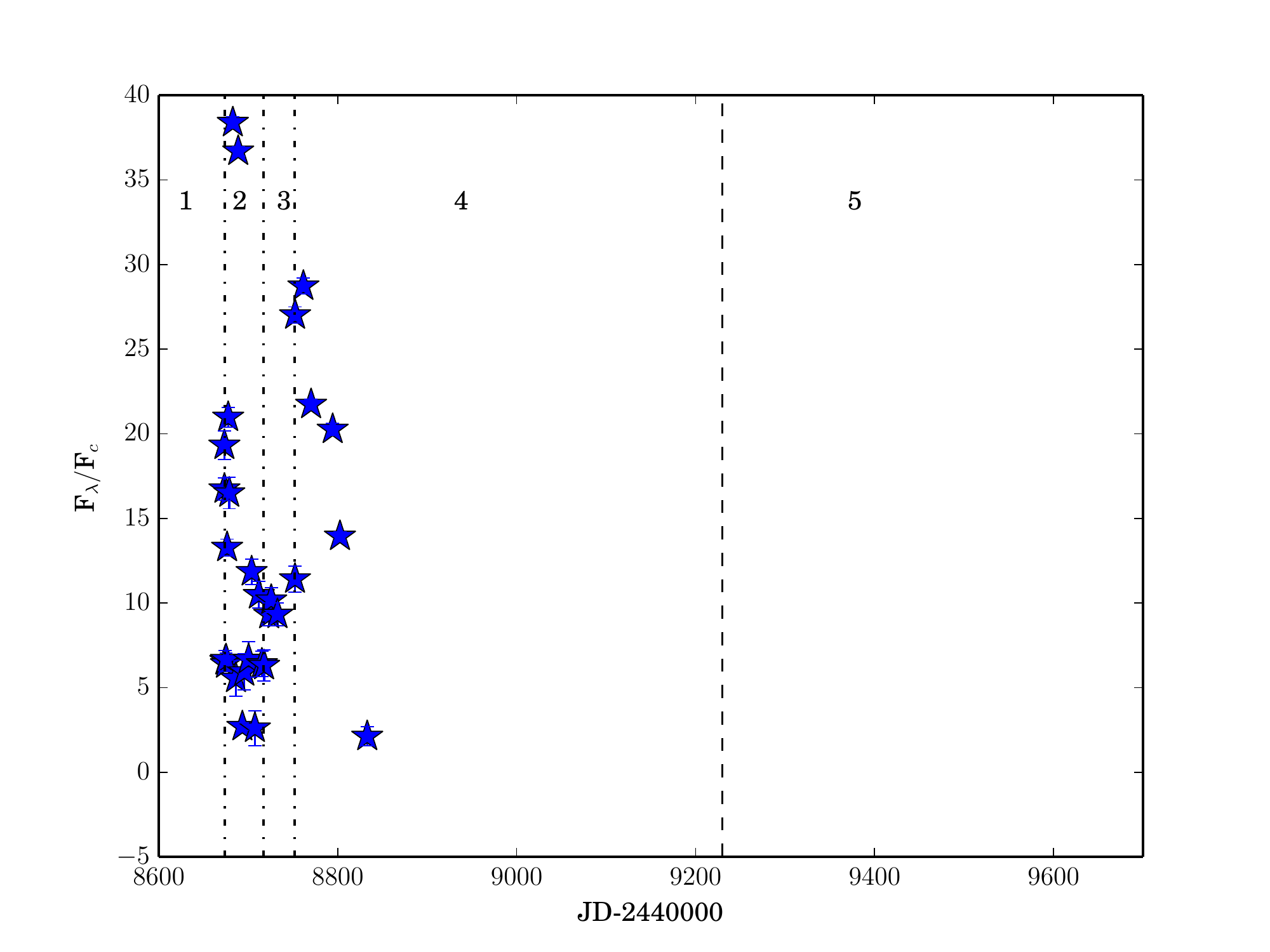}
\caption[V1974 Cyg C II line spectral evolution]{V1974 Cyg C II line spectral evolution. The numbers 1,2,3,4,and 5 correspond to the fireball, Fe optically thick, transition, nebular and quiescent phases, respectively.}
\label{fig:cii_1974}
\end{figure}

\begin{figure}
\includegraphics[height=14cm,width=13cm]{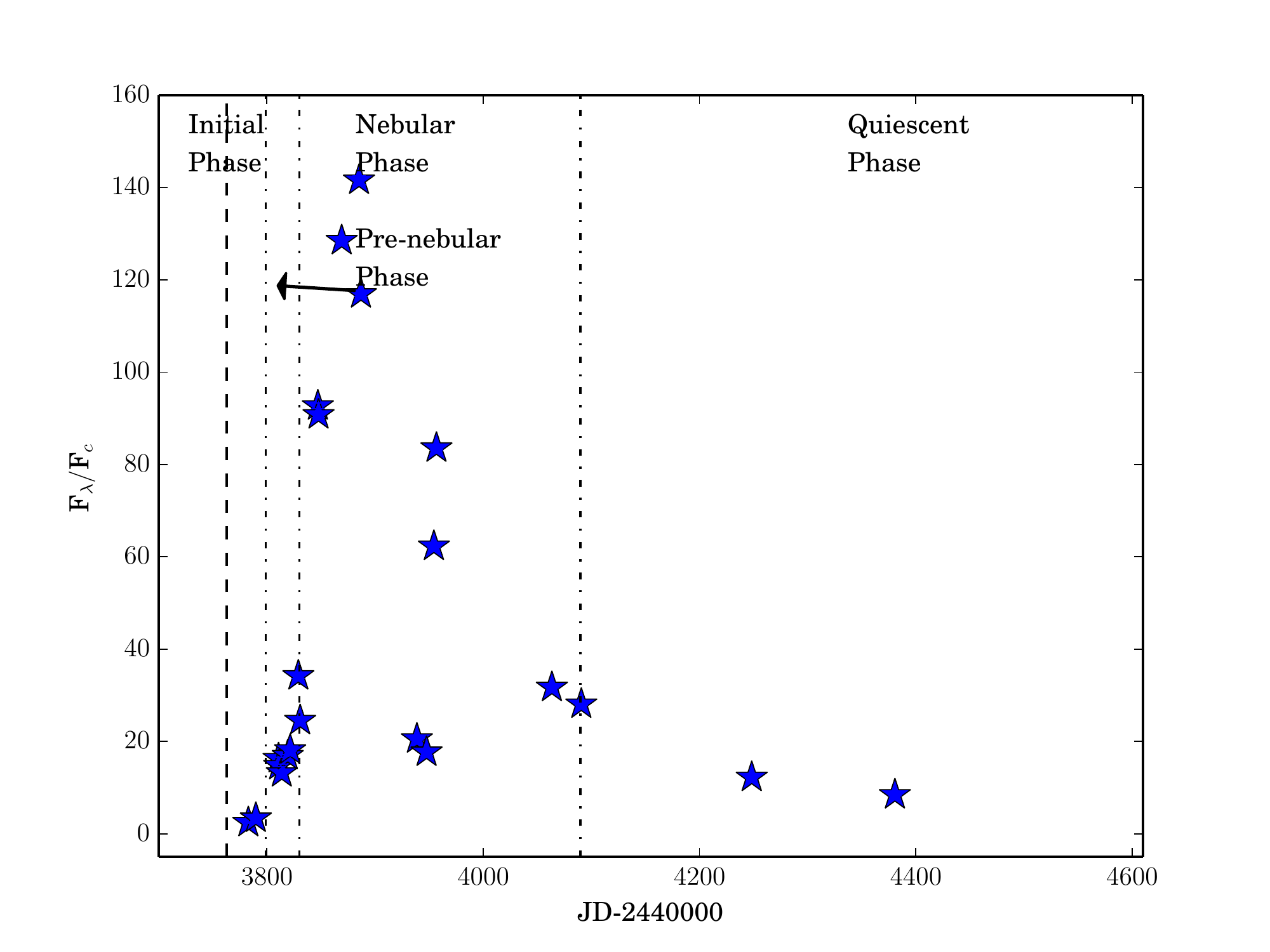}
\caption[V1974 Cyg O I line spectral evolution]{V1974 Cyg O I line spectral evolution. The numbers 1,2,3,4,and 5 correspond to the fireball, Fe optically thick, transition, nebular and quiescent phases, respectively.}
\label{fig:oi_1974}
\end{figure}

\begin{figure}
\includegraphics[height=14cm,width=13cm]{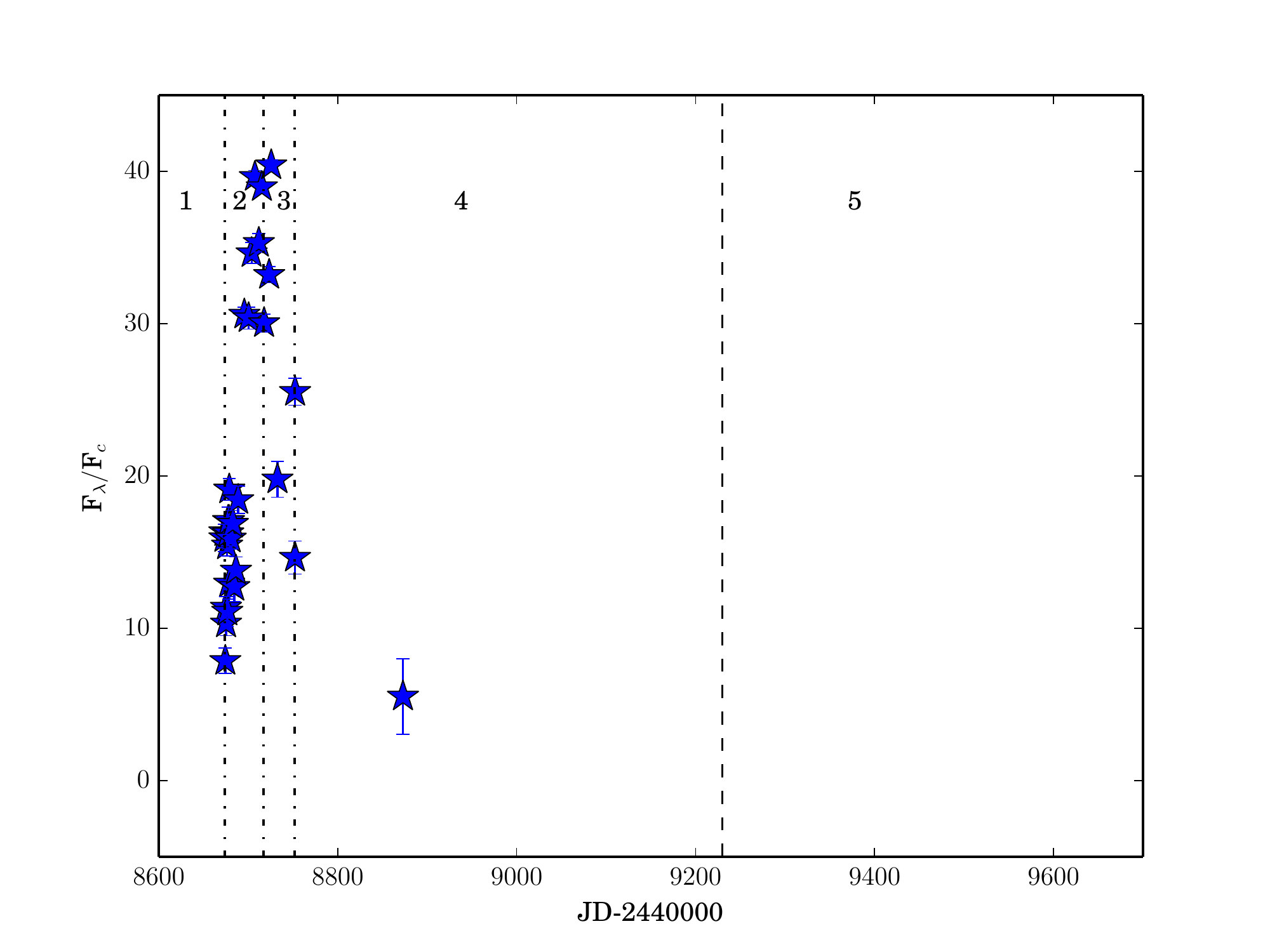}
\caption[V1974 Cyg Al III line spectral evolution]{V1974 Cyg Al III line spectral evolution. The numbers 1,2,3,4,and 5 correspond to the fireball, Fe optically thick, transition, nebular and quiescent phases, respectively.}
\label{fig:aliii}
\end{figure}

\begin{figure}
\includegraphics[height=14cm,width=13cm]{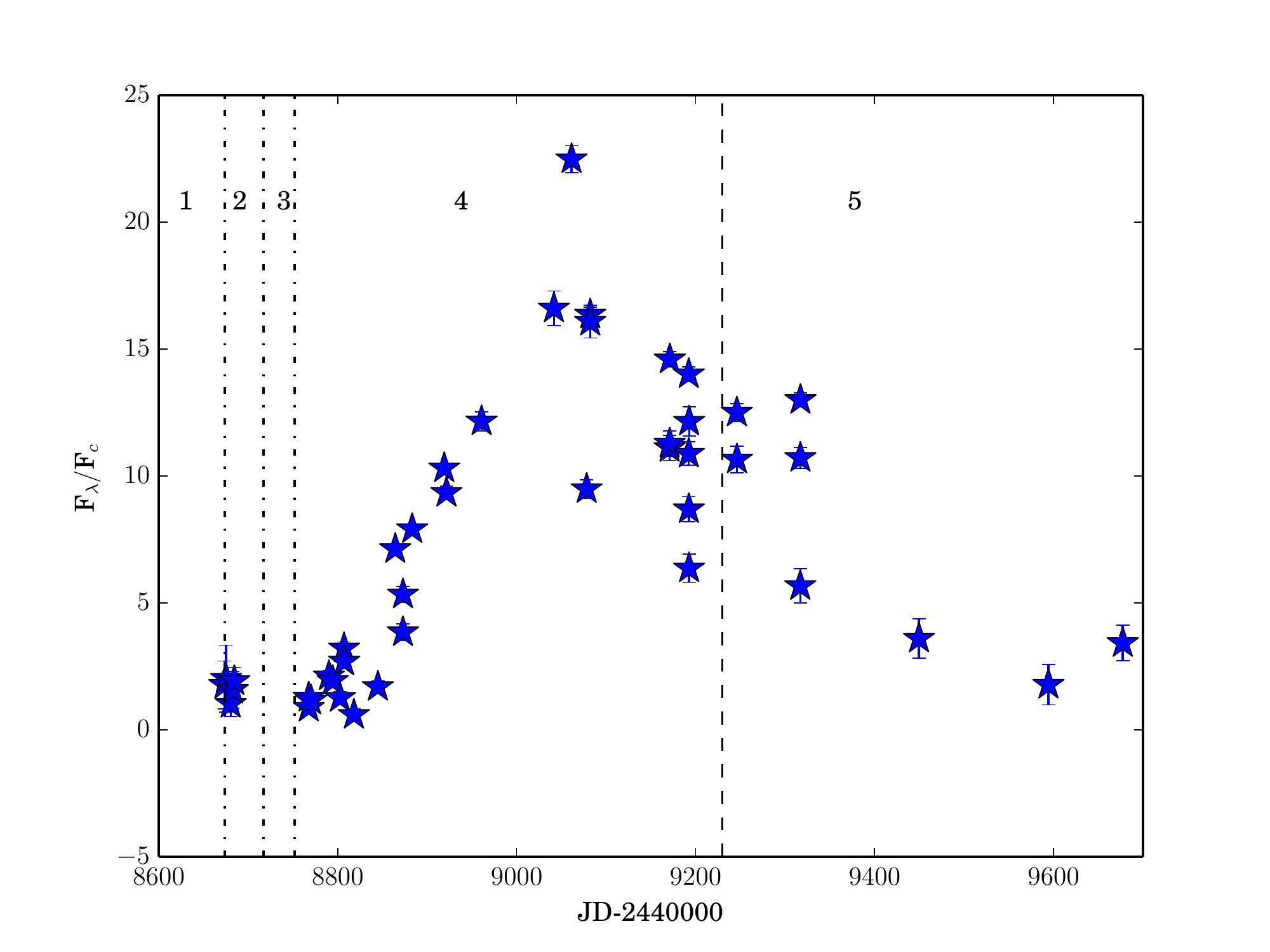}
\caption[V1974 Cyg N III{]} line spectral evolution.]{V1974 Cyg N III] line spectral evolution. The numbers 1,2,3,4,and 5 correspond to the fireball, Fe optically thick, transition, nebular and quiescent phases, respectively.}
\label{fig:niii_1974cyg}
\end{figure}

\begin{figure}
\includegraphics[height=14cm,width=13cm]{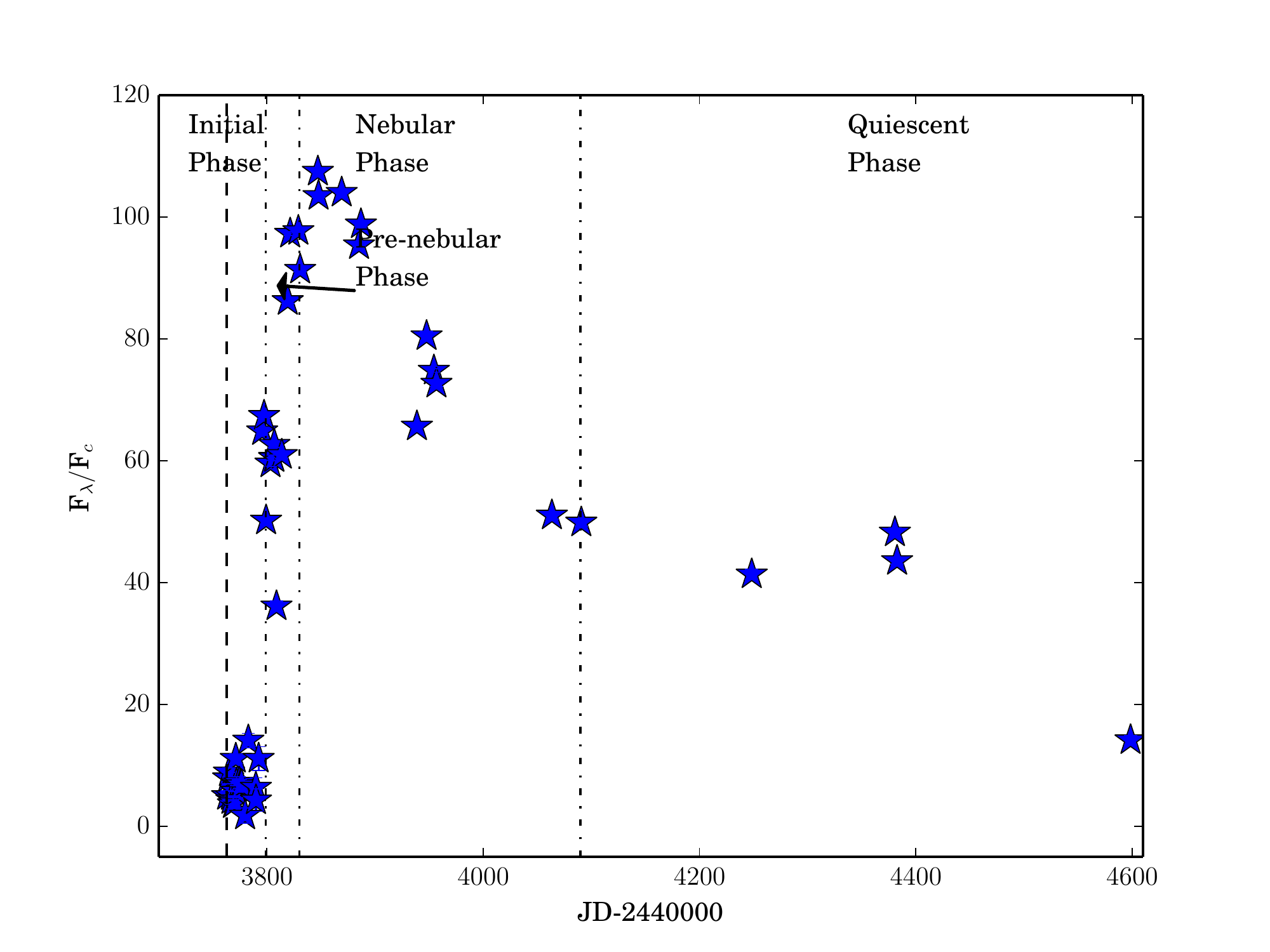}
\caption[V1974 Cyg N IV{]} line spectral evolution]{V1974 Cyg N IV] line spectral evolution. The numbers 1,2,3,4,and 5 correspond to the fireball, Fe optically thick, transition, nebular and quiescent phases, respectively.}
\label{fig:niv_1974cyg}
\end{figure}

\begin{figure}
\includegraphics[height=14cm,width=13cm]{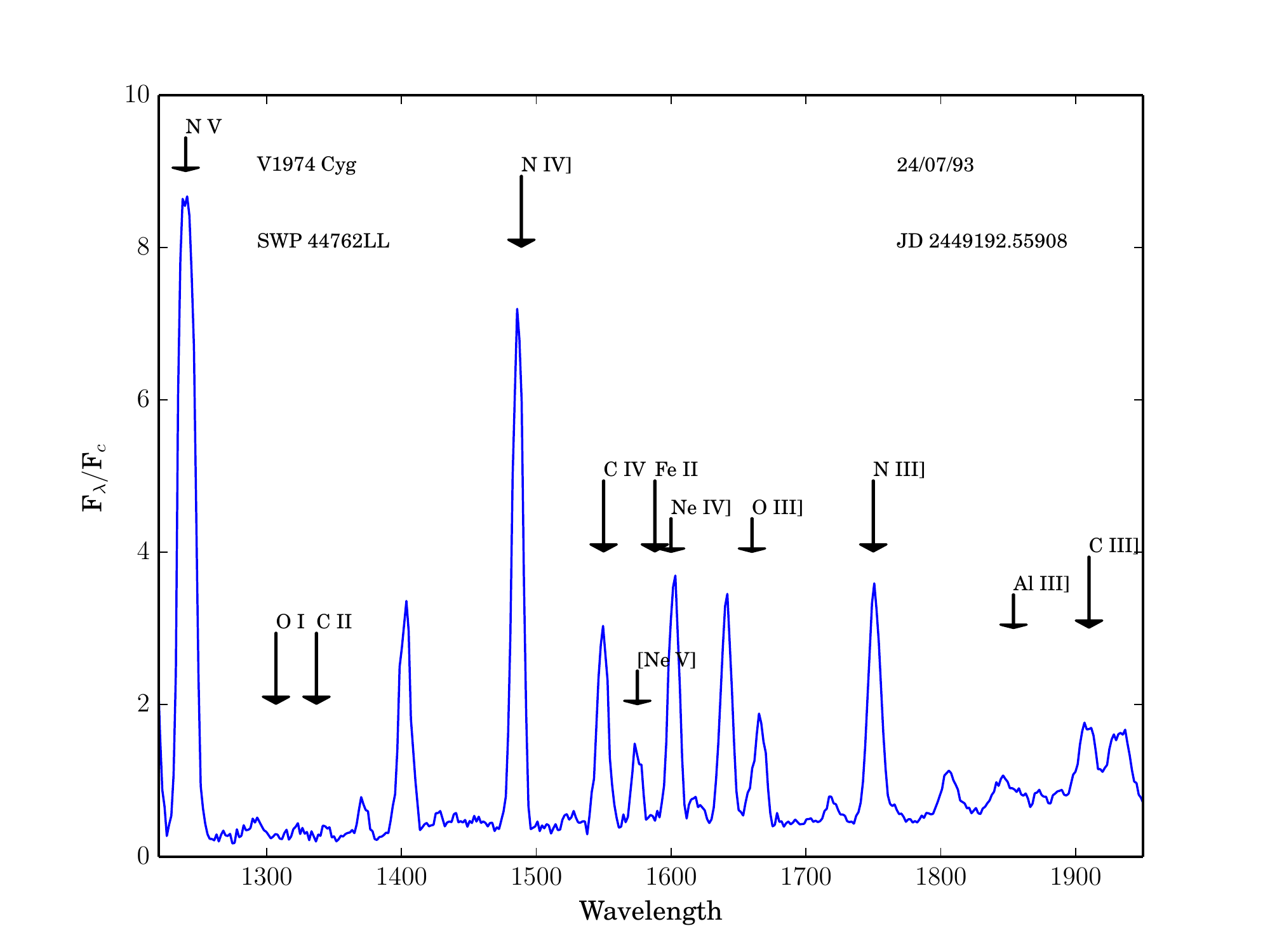}
\caption[V1974 Cyg N V line spectral evolution]{V1974 Cyg N V line spectral evolution. The numbers 1,2,3,4,and 5 correspond to the fireball, Fe optically thick, transition, nebular and quiescent phases, respectively.}
\label{fig:nv_1974}
\end{figure}

\begin{figure}
\includegraphics[height=14cm,width=13cm]{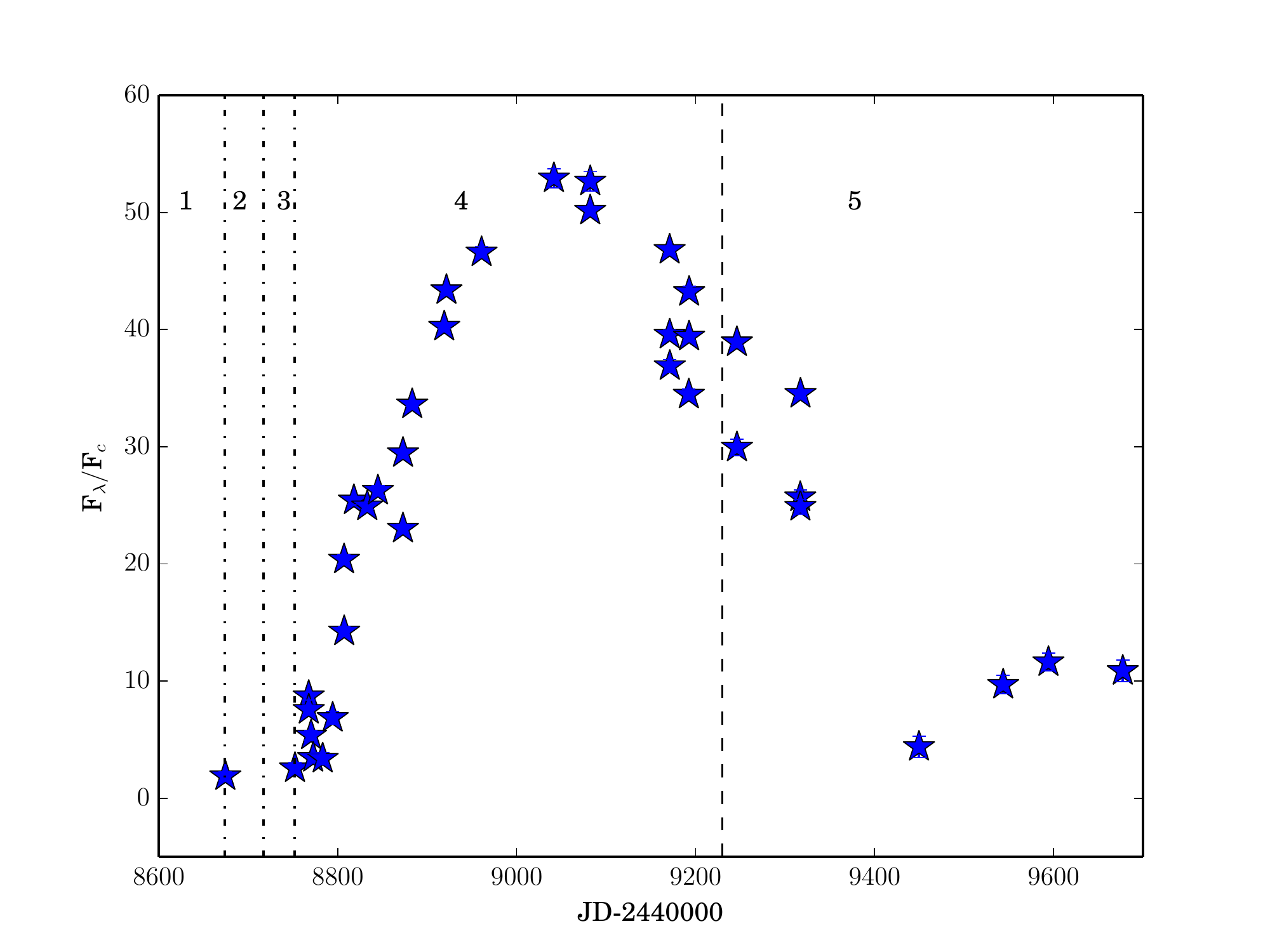}
\caption[V1974 Cyg {[}Ne V{]} line spectral evolution]{V1974 Cyg [Ne V] line spectral evolution. The numbers 1,2,3,4,and 5 correspond to the fireball, Fe optically thick, transition, nebular and quiescent phases, respectively.}
\label{fig:nev}
\end{figure}

\begin{figure}
\includegraphics[height=14cm,width=13cm]{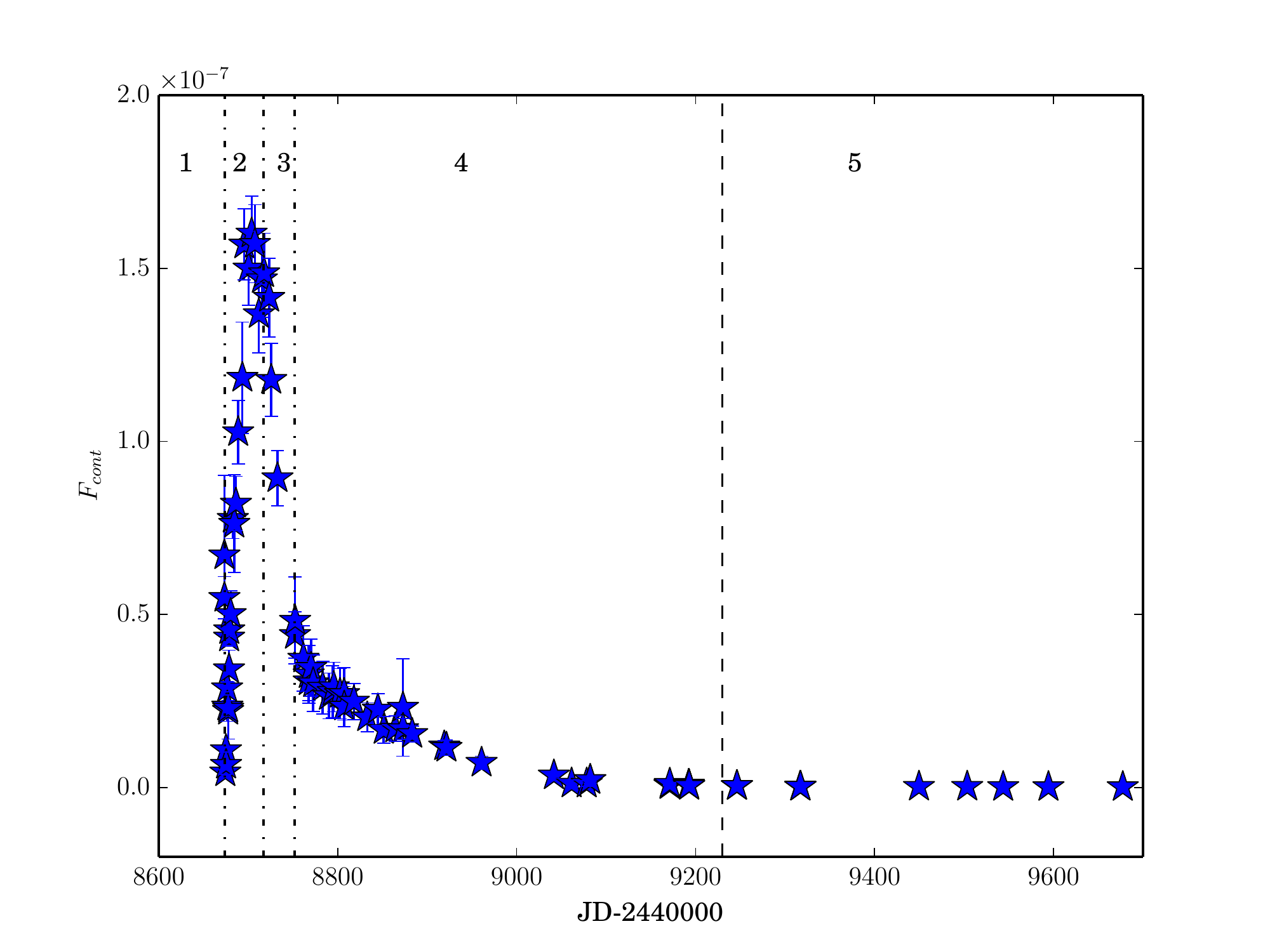}
\caption[V1974 Cyg UV shortwavelength continuum evolution]{V1974 Cyg UV shortwavelength continuum evolution. The numbers 1,2,3,4,and 5 correspond to the fireball, Fe optically thick, transition, nebular and quiescent phases, respectively.}
\label{fig:contflux}
\end{figure}

\begin{table}
\caption[V1974 Cyg Fe II and C II emission line parameters]{V1974 Cyg Fe II and C II emission line parameters. $(F_{\lambda}/F_c)$ is the normalized flux, $F_{\lambda}$ is the absolute flux and L is the Luminosity. Intermediate values are the mean of all the values calculated for a specific parameter.}
\label{tab:calc6}
\begin{center}
\resizebox{\textwidth}{!}{%
\begin{tabular}{|lcc|}\hline\hline                                                 
Value &                                          Fe II&                        C II                     \\
\hline
$(F_{\lambda}/F_c)_{\mathrm{max}}                                                         $&$ 64.7 \pm 0.4         $&$ 38.4 \pm 0.3   $\\ 
$F_{\lambda(\mathrm{max})}(  10^{-09} \mathrm{erg\, cm^{-2}\,s^{-1}\,\text{\AA}^{-1}})    $&$ 3.2     \pm 0.1      $&$ 4.5  \pm 0.1   $\\ 
$L_{\mathrm{max}} (  10^{36} \mathrm{erg\, s^{-1}})                                       $&$ 1.0     \pm 0.2      $&$ 1.4 \pm 0.2    $\\ 
$(F_{\lambda}/F_c)_{\mathrm{mid}}                                                         $&$ 13.6  \pm 0.8        $&$ 10.5 \pm 0.8   $\\ 
$F_{\lambda(\mathrm{mid})}(  10^{-10} \mathrm{erg\, cm^{-2}\,s^{-1}\,\text{\AA}^{-1}})    $&$ 6      \pm    2      $&$ 6.8 \pm 0.3    $\\ 
$L_{\mathrm{mid}} (  10^{35} \mathrm{erg\, s^{-1}})                                       $&$ 1.9      \pm 0.7     $&$ 2.1 \pm 0.4    $\\ 
$(F_{\lambda}/F_c)_{\mathrm{min}}                                                         $&$ 0.6 \pm 0.4          $&$ 2.1 \pm 0.6    $\\ 
$F_{\lambda(\mathrm{min})}(  10^{-13} \mathrm{erg\, cm^{-2}\,s^{-1}\,\text{\AA}^{-1}})    $&$ 8     \pm 3          $&$ 49 \pm 9       $\\ 
$L_{\mathrm{min}}  (  10^{32} \mathrm{erg\, s^{-1}})                                      $&$ 2     \pm 1          $&$ 15 \pm 4  $\\ 
\hline
\end{tabular}}
\end{center}
\end{table}

\begin{table}
\caption[V1974 Cyg O I and Al III emission line parameters]{V1974 Cyg O I and Al III emission line parameters. $(F_{\lambda}/F_c)$ is the normalized flux, $F_{\lambda}$ is the absolute flux and L is the Luminosity. Intermediate values are the mean of all the values calculated for a specific parameter.}
\label{tab:calc7}
\begin{center}
\resizebox{\textwidth}{!}{%
\begin{tabular}{|lcc|}\hline\hline                                                 
Value                                                                                      & O I             &              Al III\\
\hline
$(F_{\lambda}/F_c)_{\mathrm{max}}                                                         $&$ 143  \pm 1    $&$ 40.4 \pm 0.4     $\\ 
$F_{\lambda(\mathrm{max})}(  10^{-09} \mathrm{erg\, cm^{-2}\,s^{-1}\,\text{\AA}^{-1}})    $&$ 31.4 \pm 0.2  $&$ 7.5  \pm 0.2     $\\ 
$L_{\mathrm{max}} (  10^{36} \mathrm{erg\, s^{-1}})                                       $&$ 10  \pm 2     $&$ 2.3  \pm 0.4    $\\ 
$(F_{\lambda}/F_c)_{\mathrm{mid}}                                                         $&$ 27   \pm 1    $&$ 16.9 \pm 0.9     $\\ 
$F_{\lambda(\mathrm{mid})}(  10^{-10} \mathrm{erg\, cm^{-2}\,s^{-1}\,\text{\AA}^{-1}})    $&$ 9.7  \pm 0.2  $&$ 18   \pm 1       $\\ 
$L_{\mathrm{mid}} (  10^{35} \mathrm{erg\, s^{-1}})                                       $&$ 3.0  \pm 0.5  $&$ 6  \pm 1     $\\ 
$(F_{\lambda}/F_c)_{\mathrm{min}}                                                         $&$ 0.4  \pm 0.2  $&$ 6    \pm 3       $\\ 
$F_{\lambda(\mathrm{min})}(  10^{-13} \mathrm{erg\, cm^{-2}\,s^{-1}\,\text{\AA}^{-1}})    $&$ 7    \pm 5    $&$ 0.0011 \pm 0.0001$\\ 
$L_{\mathrm{min}}  (  10^{32} \mathrm{erg\, s^{-1}})                                      $&$ 2    \pm 1    $&$ 330  \pm 70      $\\ 
\hline
\end{tabular}}
\end{center}
\end{table}

\begin{table}
\caption[V1974 Cyg N III{]} and N IV{]} emission line parameters]{V1974 Cyg N III] and N IV] emission line parameters. $(F_{\lambda}/F_c)$ is the normalized flux, $F_{\lambda}$ is the absolute flux and L is the Luminosity. Intermediate values are the mean of all the values calculated for a specific parameter.}
\label{tab:calc8}
\begin{center}
\resizebox{\textwidth}{!}{%
\begin{tabular}{|lcc|}\hline\hline
Value 											 &     N III]  & N IV]                                  \\ \hline 
$(F_{\lambda}/F_c)_{\mathrm{max}}                                                       $&$ 104.1 \pm 0.3                $&$ 103.6 \pm 0.3     $\\
$F_{\lambda(\mathrm{max})} (  10^{-09} \mathrm{erg\, cm^{-2}\,s^{-1}\,\text{\AA}^{-1}}) $&$ 7.0 \pm 0.1                  $&$ 2.86 \pm 0.04     $\\
$L_{\mathrm{max}}(  10^{35}\mathrm{erg\, s^{-1}})                                       $&$ 21 \pm 4                     $&$ 9 \pm 1      $\\
$(F_{\lambda}/F_c)_{\mathrm{mid}}                                                       $&$ 33.9 \pm 0.6                 $&$ 68.7 \pm 0.5      $\\
$F_{\lambda(\mathrm{mid})} (  10^{-11} \mathrm{erg\, cm^{-2}\,s^{-1}\,\text{\AA}^{-1}}) $&$ 3 \pm 1                      $&$ 47.7 \pm 0.6      $\\
$L_{\mathrm{mid}}   (  10^{34} \mathrm{erg\, s^{-1}})                                   $&$ 10 \pm 4                     $&$ 1.5   \pm 0.3        $\\
$(F_{\lambda}/F_c)_{\mathrm{min}}                                                       $&$ 1.7 \pm 0.9                  $&$ 4.7 \pm 1.1       $\\
$F_{\lambda(\mathrm{min})} (  10^{-12} \mathrm{erg\, cm^{-2}\,s^{-1}\,\text{\AA}^{-1}}) $&$ 30 \pm 20                    $&$ 3.2 \pm 0.2       $\\
$L_{\mathrm{min}}  (  10^{33} \mathrm{erg\, s^{-1}})                                    $&$ 0.09 \pm 0.06                     $&$ 1.0 \pm 0.1       $\\
\hline                                                
\end{tabular}}
\end{center}
\end{table}

\begin{table}
\caption[V1974 Cyg N V and {[}Ne V{]} emission line parameters]{V1974 Cyg N V and [Ne V] emission line parameters. $(F_{\lambda}/F_c)$ is the normalized flux, $F_{\lambda}$ is the absolute flux and L is the Luminosity. Intermediate values are the mean of all the values calculated for a specific parameter.}
\label{tab:calc9}
\begin{center}
\resizebox{\textwidth}{!}{%
\begin{tabular}{|lcc|}\hline\hline                                                 
Value                                                                                      &                  N V&                       [Ne V]                         \\
\hline
$(F_{\lambda}/F_c)_{\mathrm{max}}                                                         $&$ 166.4 \pm 0.4    $& $22.5 \pm 0.5      $\\
$F_{\lambda(\mathrm{max})}(  10^{-09} \mathrm{erg\, cm^{-2}\,s^{-1}\,\text{\AA}^{-1}})    $&$ 2.9 \pm 0.1      $& $0.21 \pm 0.03  $\\
$L_{\mathrm{max}} (  10^{36} \mathrm{erg\, s^{-1}})                                       $&$ 0.9 \pm 0.1      $& $0.6 \pm 0.1 $\\
$(F_{\lambda}/F_c)_{\mathrm{mid}}                                                         $&$ 62.5 \pm 0.5     $& $6.4  \pm 0.6      $\\
$F_{\lambda(\mathrm{mid})}(  10^{-10} \mathrm{erg\, cm^{-2}\,s^{-1}\,\text{\AA}^{-1}})    $&$ 10.5 \pm 0.2     $& $41   \pm3    $\\
$L_{\mathrm{mid}} (  10^{35} \mathrm{erg\, s^{-1}})                                       $&$ 3.2 \pm 0.6      $& $0.13 \pm 0.02    $\\
$(F_{\lambda}/F_c)_{\mathrm{min}}                                                         $&$ 1.8 \pm 0.2      $& $0.6  \pm 0.2        $\\
$F_{\lambda(\mathrm{min})}(  10^{-13} \mathrm{erg\, cm^{-2}\,s^{-1}\,\text{\AA}^{-1}})    $&$ 23 \pm 3         $& $4       \pm 2$\\
$L_{\mathrm{min}}  (  10^{32} \mathrm{erg\, s^{-1}})                                      $&$ 7 \pm 1          $& $1.1     \pm 0.6    $\\
\hline
\end{tabular}}
\end{center}
\end{table}

\begin{table}
\caption{Times of maxima of different emission lines and their ionization potentials.}
\label{tab:ions}
\begin{center}
\resizebox{\textwidth}{!}{%
\begin{tabular}{|ccc|}\hline\hline 
Line  &  $\mathrm{T_{max}}$ (JD -2440000) & Ionization Potential (eV)\\\hline
Fe II &  8678             &   7.90                \\
C II  &  8682             &   11.26               \\
O I   &  8717             &   13.60               \\
Al III&  8725             &   18.83               \\
N III] & 8767		  &   29.60               \\
N IV]  &  8883             &   47.43               \\
N V   &  8919             &   77.47               \\\relax
[Ne V] &  9061             &   97.12               \\
\hline
\end{tabular}}
\end{center}
\end{table}  

\begin{figure}
\includegraphics[height=14cm,width=13cm]{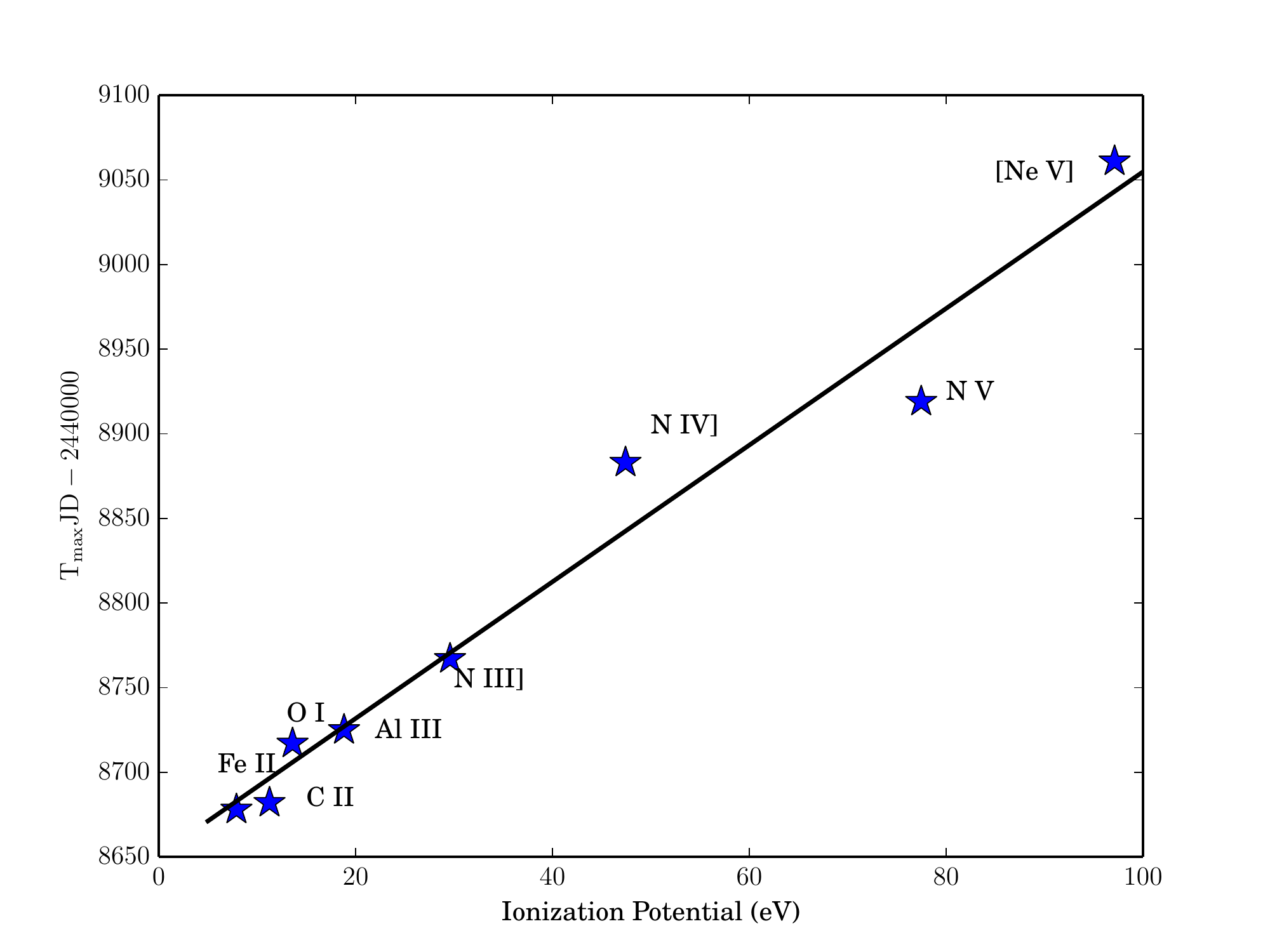}
\caption{Linear relation between the times of maxima of different emission lines versus their ionization potentials for V1974 Cyg.}
\label{fig:ionsfig}
\end{figure}

\section{Ultraviolet Spectroscopy Discussion}\label{sec:dis}
\subsection{PW Vul}\label{subsec:pwvul_dis}

In a previous study of the same IUE data, \citet{2005A&A...439..205C} obtained a similar trend for the evolution of the line fluxes of the O I and C III] lines (see Figs ~\ref{fig:oi} and ~\ref{fig:ciii} from this work and figure 2 from \citealp{2005A&A...439..205C}). Moreover, most of the measured line fluxes here determined are consistent with the fluxes measured by \citet{1991A&A...244..111A} for the nebular phase spectra.

\citet{2005A&A...439..205C}  studied three phases of the evolution of the UV spectrum of PW Vul. First, "lifting the iron curtain" (which they called "initial phase"). They found that this phase ended about 100 days after the visual maximum, when the O I line reached its maximum flux. Second, the pre-nebular phase which they found that it ended about 250 days after the visual maximum with the maximum flux of the C III] line. Our results are consistent with these two findings. The third and final phase they studied was the nebular phase.

We present the evolution of the fitted continuum flux in the whole short wavelength range in fig  ~\ref{fig:pwvul_continuum} where it can be seen that the evolution curve reached the maximum at JD $2445987 \pm 23$ which is consistent with the evolution curve of the continuum at 1455 \AA \, from \citet{2002A&A...384.1023C}.

We also calculate the expansion velocity of the nova by measuring the widths of the spectral lines. We find an average value of $\sim 1300 \, \mathrm{kms^{-1}}$ which is slightly higher than the average ejection speed of the CO2 model ($1200 \, \mathrm{kms^{-1}}$) of \citet{1998ApJ...494..680J}. This value is different from the value of $750 \, \mathrm{kms^{-1}}$ reported by \citet{2005A&A...439..205C} and \citet{1988ASSL..138..115C} where the authors report three different expansion velocities for the absorption lines (0, 750 and 1550 $\mathrm{kms^{-1}}$). The expansion velocity we calculated lies within the range of the these values.

During the early phases of the outburst, the mass accretion process onto the white dwarf is greatly reduced and only starts to become significant in the quiescent stage \citep{2006ApJS..167...59H}. We therefore measured the mass accretion rate ($\dot{M}_\mathrm{{acc}}$) from the last two spectra corresponding to the quiescent phase. The mass accretion rate determined from the C IV line on JD 2446367, about 611 days after the outburst, results $\dot{M}_\mathrm{{acc}} = 5 \pm 2 \times 10^{-9} \mathrm{M_{\odot} yr^{-1}}$ while the average value obtained using all the studied emission lines is $\dot{M}_\mathrm{{acc}} = 2 \pm 1  \times 10^{-9} \mathrm{M_{\odot} yr^{-1}}$. This latter value suggests a recurrence time of $3.01 \times 10^{4} \mathrm{yr}$ smaller than the accretion time for the CO2 model calculated by \citet{1998ApJ...494..680J} for novae of similar WD mass and expansion speed, where The recurrence time is the $\Delta M_{env}$\nomenclature{$\Delta M_{env}$}{The mass of the envelope at the end of the accretion phase} (the mass of the envelope at the end of the accretion phase) of the Model in \citet{1998ApJ...494..680J} model divided by the calculated mass accretion rate. The values we obtained for these properties (expansion velocity, accretion rate and accretion time) suggest that the model that best describes the nova is the CO2 model. \citet{1998ApJ...494..680J} model calculated the the abundances of different elements in the ejecta few days after the nova outburst. They identify PW Vul as a CO4 nova based on the agreement between the results of the model and the abundances calculated by \citet{1991A&A...244..111A} using the nebular phase spectra  and those calculated by \citet{1994A&A...291..869A} using observations taken about 180 days after the outburst.

Comparing our results to the model of \citet{2005ApJ...623..398Y} yielded that the average mass accretion rate and the recurrence time for PW Vul were close to that of a model with $M_\mathrm{{W\-D}} =1 \mathrm{M_{\odot}}$ and $\mathrm{T_{\odot}}$ of $10^7$ K although the average ejection speed was closer to the model with accretion rate of  $10^{-10}$ K.

\subsection{V1668 Cyg}\label{subsec:v1668cyg_dis}

The evolution curve of the normalized fluxes here mentioned for the O I, N III] and N V lines (Figures ~\ref{fig:oi_v1668cyg} -~\ref{fig:nv_v1668cyg}) shows the same trend reported by \citeauthor{2005A&A...439..205C}(2005, their figure 1). Moreover, our results are in good agreement with the values determined by \citet{1981MNRAS.197..107S} during the nebular phase.

The evolutionary curve of the fitted continuum is presented in fig ~\ref{fig:v1668cyg_continuum} and it can be seen that it reaches the maximum on JD $2443790 \pm 9 $ which is consistent with evolution of the continuum at 1455 \AA \, presented by \citet{2002A&A...384.1023C}.

By measuring the widths of the spectral lines, we calculated the expansion velocities of the nova, finding an average value of $\sim 1800 \, \mathrm{kms^{-1}}$ which is close to the average ejection velocity of the CO3 model of \citet{1998ApJ...494..680J}. \citet{1979A&A....74L..18C} reported a value of $1160 \, \mathrm{kms^{-1}}$ for the expansion velocity. They calculated this value from the blue-shift of the absorption component of the P Cygni profile of Mg II line.

\citet{2005A&A...439..205C} classified the evolution of V1668 Cyg into the same three stages in the analysis of PW Vul, where they found that the initial phase ended on JD  2443799 and that the pre-nebular phase ended at JD 2443830. It can be seen from the IUE observations that the last observation taken in the nebular phase was at JD 2444090 then the system enters the quiescent (post-nova) phase. 

\citet{2006ApJS..167...59H} assumed the wind stops $\sim$ 280 days after the outburst (marked by the dotted line in figs ~\ref{fig:oi_v1668cyg}-~\ref{fig:nv_v1668cyg}), i.e. about 50 days before the last observation available in the nebular phase. We considered that the system has entered the quiescent (post-nova) phase for the final four IUE observations, starting on JD 2444248, when the flux of the continuum and emission lines has decreased to 0.5\% of the maximum values.

We calculated  mass accretion rate for V1668 Cyg in the quiescent phase using equation~\ref{eq:mass_acc2}, since we have not found any observations reported in the literature detecting the presence of a magnetic field in V1668 Cyg, and adopting a white dwarf of mass $0.95 \mathrm{M_{\odot}}$ \citep{2006ApJS..167...59H} and a radius of $0.0084 \mathrm{R_{\odot}}$ (determined from equation ~\ref{eq:radius}). The maximum value for the mass accretion rate results $9 \pm 2 \times 10^{-10} \mathrm{M_{\odot} yr^{-1}}$, as determined from the flux of the C IV lines on JD 2444248, about 490 days after the outburst. The average value for the mass accretion rate in the quiescent stage determined from all studied lines, in the last four spectra, is $4 \pm 1 \times 10^{-10}  \mathrm{M_{\odot} yr^{-1}}$. This yields a recurrence time of $1.7 \times 10^{5} \mathrm{yr}$ which is consistent with the theoretical accretion rate calculated by \citet{1998ApJ...494..680J} for novae of similar WD mass and expansion speed. \citet{1998ApJ...494..680J} identify V1668 Cyg as a CO1  nova based on the agreement between the results of the model and the abundances calculated by \citet{1981MNRAS.197..107S} in the nebular phase or CO4 based on the abundances calculated by \citet{1994A&A...291..869A} using observations taken about 338 days after the outburst. We summarized some of the results discussed here for PW Vul and V1668 Cyg in \citet{2018AN....339..173H}.

For the model of \citet{2005ApJ...623..398Y} the average mass accretion rate, the recurrence time and the averge ejection speed of V1668 Cyg were close to the model with  $M_\mathrm{{W\-D}} =1.25 \mathrm{M_{\odot}}$ and $\mathrm{T_{\odot}}$ of $10^7$ K although in our calculations we assumed it had a much smaller mass.


\subsection{V1974 Cyg}\label{subsec:v1974cyg_dis}

The evolution of different line fluxes relative to the continuum $F_{\lambda}/F_c$, show different phases of the outburst. The early spectrum of V1974 belongs to the permitted Fe II "$P_{Fe}$" class of \citet{1992AJ....104..725W}, where the spectrum originates from optically thick wind ejected by the white dwarf \citep{1993AJ....106.2408S,1996AJ....111..869A,2003cvs..book.....W}. The Fe II 1588 \AA \, line was the first line to reach maximum on JD 2448678 during the iron curtain phase of the outburst, where the cooling of the ejecta leads to the recombination of the iron peak elements. During this phase most remaining emission lines have low fluxes since they are blanketed by the iron curtain. The second line to peak was the C II 1336 \AA \, line on JD 2448682. By the time this recombination line reaches maximum, the iron curtain is being lifted and the ionization is enhanced due to the retreat of the pseudo-photosphere. The iron curtain was completely lifted when the O I 1306 \AA \, line reached maximum on JD 2448717 which is consistent with the evolution of this line in \citet{2002A&A...384.1023C}, the maximum of this line marks the end of the iron optically thick phase (both the iron curtain and lifting the iron curtain phases). The maximum of this line coincides with the time when the Fe II line drops to very low fluxes (see Figs ~\ref{fig:feii} and ~\ref{fig:oi}). The outburst then enters the transition or pre-nebular phase and this is also clear from the optical observations of \citet{1995A&A...294..488R}. The Al III 1854 \AA \, line reached maximum on JD 2448725 and we can see from Fig ~\ref{fig:aliii} that this line can no longer be seen after JD 2448873 which means that all the aluminium in the ejecta has lost at least three electrons. The N III] 1750 \AA \, line reached maximum on JD 2448767 shortly after the end of the pre-nebular phase. The N IV] 1487 \AA \, line reached maximum on JD 2448883. The N V 1240 \AA \, line reached maximum on JD 2448919 in the nebular phase. The last line to peak in our sample was the [Ne V] 1575 \AA \, forbidden line on JD 2449061 in the nebular phase. At the times of maximum of the last two lines, the line opacity in the ejecta is now low and the ions are subject to harder radiation fields from the central white dwarf \citep{1996AJ....111..869A,2012BASI...40..185S}.  Tables ~\ref{tab:calc6} -~\ref{tab:calc9} contain maximum, intermediate and minimum values for the studied lines. It can be seen that the studied lines reach maxima in order of increasing ionization potential showing that the outburst enters higher ionization conditions as time passes \citep{2005A&A...439..205C}. Note the disappearance of lower ionization lines in the later phases of the outburst in Figs ~\ref{fig:feii}-~\ref{fig:aliii} and see Table ~\ref{tab:ions} and Fig. ~\ref{fig:ionsfig}. \citet{2004A&A...420..571C} studied the evolution of the integrated flux of four of the lines we studied (C II, Al III, O I and N V) during the first four stages of the outburst and our evolutionary curves show a similar trend to theirs. Our calculated integrated fluxes agree with those calculated by \citet{1996AJ....111..869A} in the nebular phase and those obtained by \citet{1996ApJ...463L..21S} in the late nebular and quiescent phases. We have also calculated the evolution of the fitted continuum flux of the whole short wavelength range (Fig ~\ref{fig:contflux}). The continuum flux reached a maximum value of $\sim 1.6 \pm 0.1 \times10^{-7}\, \mathrm{erg\, cm^{-2}\, s^{-1}}$ at JD $2448703 \pm 8$ which is consistent with the time of maximum of the continuum at 1455 \AA of \citet{2002A&A...384.1023C}. This happens while the iron curtain is lifting where the pseudo-photosphere is receding and the emission from the hotter central regions is now harder. The peak lags about 28 days after the visual maximum which shows the shift of the maximum emission towards shorter wavelengths \citep{1990LNP...369..115C,2002A&A...384.1023C}. The average luminosity of the studied lines is $\sim4.8 \pm 0.9\times 10^{35} \mathrm{erg\,s^{-1}}$ and the average continuum luminosity is $\sim 1.2\pm 0.3 \times 10^{37}\mathrm{erg\, s^{-1}}$.


We can see from Fig ~\ref{fig:ionsfig} that there is a linear relation between the time of maximum of the different lines and their ionization potentials. This releation was best fit by the following equation 

\begin{equation}\label{eq:ffit}
T_{max} = 4.038 \chi + 2448651
\end{equation}
where $\chi$ is the ionization potential. This linear relation shows that the emission in the ejecta is shifting towards higher energies.


The average expansion speed of nova outburst calculated from the velocities of the emission lines was $\sim 2000\, \mathrm{kms^{-1}}$. This, along with the adopted White dwarf mass makes V1974 Cyg lie between models ONe1 and ONe2 of \citet{1998ApJ...494..680J}.

V1974 Cyg was identified as a ONeMg nova by \citet{1993AJ....106.2408S} since its early spectra showed P Cygni profiles, a feature not present in CO novae. It also showed strong Ne emission lines in the nebular phase, namely the UV lines Ne IV] 1600 \AA  and [Ne V] 1575 \AA.
    
The mass accretion rate is very low compared to the mass loss rate and the hydrogen burning rate in the early stages of the outburst  and it only becomes significant in the quiescent phase of the nova \citep{2006ApJS..167...59H}. Therefore, we calculate the accretion rate in quiescence using equation ~\ref{eq:mass_acc} and we found a maximum value of $2.4 \pm 0.4 \times 10^{-9} \mathrm{M_{\odot} yr^{-1}}$ from the N V line 574 days after the discovery of the outburst. The average accretion rate calculated from all the lines was $2.1\pm 0.4 \times 10^{-10} \mathrm{M_{\odot} yr^{-1}}$.

\citet{1996AJ....111..869A} assumed the ejected mass during the outburst $\sim 5 \times 10^{-5} M_{\odot}$. This value along with our calculated average accretion rate suggests a recurrence time of $\sim 2.4  \times 10^5 yr$ which is slightly higher than to the accretion time for the ONe2 and ONe3 models calculated by \citet{1998ApJ...494..680J} for novae of $1.15 M_{\odot}$ white dwarfs. The values we obtained for th expansion velocity, accretion rate and accretion time suggest that the model that best describes the nova is the ONe2 model.

The best model that describes V1974 Cyg in the models calculated by \citet{2005ApJ...623..398Y} was the model with $M_\mathrm{{W\-D}} =1.25 \mathrm{M_{\odot}}$ and $\mathrm{T_{\odot}}$ of $10^7$ K since it yielded an average mass accretion rate, a recurrence time and an averge ejection speed close to the values we calculated despite that the mass we used in our calculations was smaller.



The average temperature calculated using Stefan Boltzmann law from the average continuum luminosity is $\sim 5.01 \times 10^5 K$   which is close to the value calculated by \citet{1996AJ....111..869A}. It is slightly higher than the value of $\sim 3 \times 10^5 K$ estimated by both \citet{1994ApJ...421..344S} from the bolometric luminosity and \citet{1996ApJ...456..788K} by fitting the ROSAT X-ray observations during the nebular phase.

Some of the results presnted here for V1974 Cyg were presented in \citet{2018Ap.....61...91H}. However, In that paper we used equation ~\ref{eq:mass_acc2} to calculate the mass accretion rate assuming that V1974 is a non-magnetic nova hence getting a larger value for $\dot{M}_\mathrm{{acc}}$ and a shorter recurrence time. However, in this thesis we calculate the mass accretion rate using equation ~\ref{eq:mass_acc} since \citet{1997A&A...318..908C} detected the presence of a strong magnetic field in the white dwarf of V1974 Cyg using HST images. The latter equation was used to calculate the mass accretion rate on white dwarfs of magnetic cataclysmic variables in a number of papers (see e.g. \citealt{2015ApJ...812...97S} and \citealt{2017NewA...52..122Z}).

\clearpage

\subsection{Comparison Between the Spectral Evolution of The Three Novae}\label{subsec:comp}

It can be concluded from the previous results that:

1- V1974 Cyg showed the strongest outburst among the three novae the highest $L_{\lambda} $ and average expansion speed. This is expected since it has an ONe white dwarf and this is consistent with the correlation between the speed of the nova and the mass of the white dwarf and \citep{warner2008Properties}. See table ~\ref{tab:comp} for a comparison between some of the parameters of the three novae.

2- The average value of $\dot{M}_\mathrm{{acc}}$ for PW Vul were the highest among the three novae in the quiescent phase, this is because $\dot{M}_\mathrm{{acc}}$ is inversely proportional to the mass of the primary (see equations ~\ref{eq:mass_acc2} and ~\ref{eq:mass_acc}).

3- PW Vul average ejection speed and accretion rate in quiescence puts it close to the CO2 model calculated by \citet{1998ApJ...494..680J} while the average ejection speed and accretion rate of V1668 Cyg place it closer to the CO3 model. The average ejection speed and accretion rate of V1974 Cyg Cyg suggests that the ONe2 model is the best model describing its outburst. The adopted masses of the primaries of the three systems is consistent with this. 

4- There is a difference between the models we identify for both novae (PW Vul and V1668 Cyg), from some of the dynamical properties ($M_{\mathrm{WD}}$, average mass accretion rate and average ejection velocity), and the models identified by \citet{1998ApJ...494..680J}. This difference probably resulted from that the model predicts the chemical composition few days after the outburst while the measured abundances were based on observations taken in later stages of the outburst.



\begin{table}[ht!]                                         
\caption{Parameters of The Three Novae.}
\label{tab:comp}
\begin{center}
\begin{tabular}{|lrrrr|}\hline\hline                
Parameter      &   PW Vul                                    & V1668 Cyg                    & V1974 Cyg        &            Reference \\  \hline                
$M_{\mathrm{WD}} (\mathrm{M_{\odot}})$      		     &   $0.83 $                    &$0.95 $           &$1.05 $           &             (1)\\
Speed Class         	    		     &   Slow                                      &fast        &fast       &                    (2)\\
$L_{\mathrm{max}} (10^{35} \mathrm{erg\, s^{-1}})$  		     &   $ 8.1 \pm 0.3$              &$ 23.2 \pm 0.3 $&$120  \pm 10 $ & (3)\\
$L_{avg} (10^{35} \mathrm{erg\, s^{-1}})    $       		     &   $ 1.9 \pm 0.7 $             &$ 5 \pm 1 $  &$ 4.8 \pm 0.9 $ & (3)\\
Average $v_{exp} (\mathrm{kms^{-1}})$        &   $\sim 1300$         &$\sim 1800$                    &$\sim 2000$      &              (3)\\
Ejected Mass $M_\mathrm{{ejec}} (\mathrm{M_{\odot}})$ &   $1.6\times 10^{-4}$&$5.5\times 10^{-5}$   &$5\times 10^{-5}$ &   (4)\\
Average $\dot{M}_\mathrm{{acc}} (10^{-10} \mathrm{M_{\odot} yr^{-1}})$&   $20 \pm 10  $            &$4 \pm 1  $ &$2.1\pm 0.4  $ & (3)\\
Recurrence Time ($10^5 yr$)     & 0.3    & 1.7 & 2.4&(3)\\
\hline                                                
\end{tabular}
\vspace{+0.5cm}
\caption*{Notes: $M_{\mathrm{WD}}$ is the mass of the white Dwarf, $L_{\mathrm{max}}$ is the maximum ultraviolet luminosity, $L_{avg}$ is the average ultraviolet luminosity, $v_{exp}$ is the expansion velocity and $\dot{M}_\mathrm{{acc}}$ is the mass accretion rate.
References: (1) \citet{2015ApJ...798...76H} for PW Vul, \citet{2006ApJS..167...59H} for V1668 Cyg and \citep{2005ApJ...631.1094H} for V1974 Cyg. (2) \citet{1988ApJ...329..894G} for PW Vul, \citet{1980A&A....81..157D} for V1668 Cyg and \citep{1993A&A...277..103C} for V1974 Cyg. (3) This work. (4) \citet{1997MNRAS.290...75S} for PW Vul, \citet{1981MNRAS.197..107S} for V1668 Cyg and \citet{1996AJ....111..869A}for V1974 Cyg.}                                     
\end{center}                                          
\end{table} 

\clearpage

\section{Optical Photometry}

\subsection{V1668 Cyg Photometric Light Curve Analysis}

 The light curves were fitted using the MultiTermFit algorithm of the astroML python based package \citep{astroML}. The light curves were best fitted by a sixth degree Fourier fit of the form

\begin{equation}\label{eq:fourierfit}
mag = a_0+ \sum_{i=1}^{6}a_n cos(nw)+b_n sin(nw)
\end{equation}

The fit coefficients for V, $\mathrm{R_c}$ and $\mathrm{I_c}$ filters are presented in table ~\ref{tab:v1668cygfit}.
\vspace{+0.5cm}
\begin{table}[h]
\caption{V1668 Cyg Fourier Fit  Coefficients in V, $\mathrm{R_c}$ and $\mathrm{I_c}$ filters.}
\begin{center}
\label{tab:v1668cygfit}
\begin{tabular}{|l|c|c|c|}\hline\hline
Coefficient &V&$R_c$&$\mathrm{I_c}$\\
\hline
$a_0$&  21.000 &     20.220 &      20.470\\
$a_1$&  0.121 &    0.117 &     0.139\\
$b_1$&-0.0536 &   0.044 &   -0.079\\
$a_2$& 0.0573 &      0.210 &     0.227\\
$b_2$&-0.0461 &  -0.031 &   -0.022\\
$a_3$& 0.0155 &    0.128 &     0.141\\
$b_3$&-0.024 &   0.0429 &    0.019\\
$a_4$&  0.174 &    0.092 &      0.247\\
$b_4$& 0.0154 &  -0.057 &   -0.050\\
$a_5$&  0.146 &    0.139 &    0.061\\
$b_5$& 0.066 &   0.065 &   -0.030\\
$a_6$& 0.021 &    0.029 &    0.099\\
$b_6$&-0.038 &    0.040 &    0.080\\
w  &  6.283  &    6.283  &     6.283 \\
\hline
\end{tabular}
\end{center}
\end{table}

The maximum of of the V light curve was about 20.6 mag and the minimum was about 24.5 mag making the amplitude of variation $\sim$ 3.9 mag. For the $\mathrm{R_c}$ light curve, the maximum was about 19.5 mag and the minimum was about 22.0 mag with an amplitude of variaton of $\sim$ 2.5 mag. The $\mathrm{I_c}$ light curve had maximum of about 19.9 mag and the a minimum about 22.2 mag with an amplitude of variation of $\sim$ 2.3 mag. The phase magnitude diagrams plotted using the new ephemeris of equation ~\ref{eq:v1668cyg} with the best fit are shown in Figs ~\ref{fig:v1668cygr}-~\ref{fig:v1668cygv}

\begin{figure}[h]
\centering
\includegraphics[height=14cm,width=13cm]{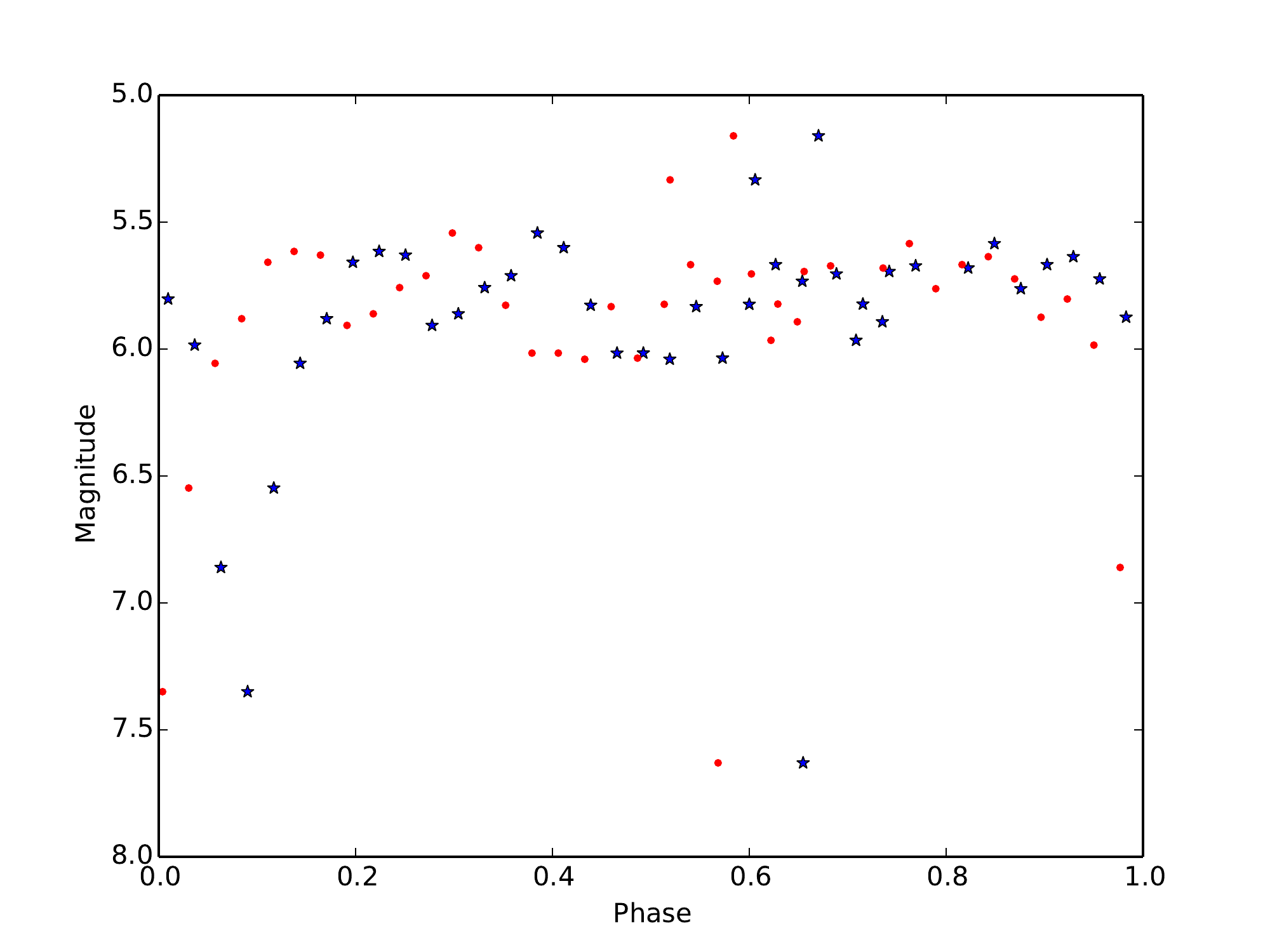}
\caption[V1668 Cyg $\mathrm{R_c}$ phase magnitude diagram with the best fit]{V1668 Cyg $\mathrm{R_c}$ phase magnitude diagram plotted using the new ephemeris of equation ~\ref{eq:v1668cyg} with the best fit. The solid line represents the best fit using equation ~\ref{eq:fourierfit}.}
\label{fig:v1668cygr}
\end{figure}

\begin{figure}[h]
\centering
\includegraphics[height=14cm,width=13cm]{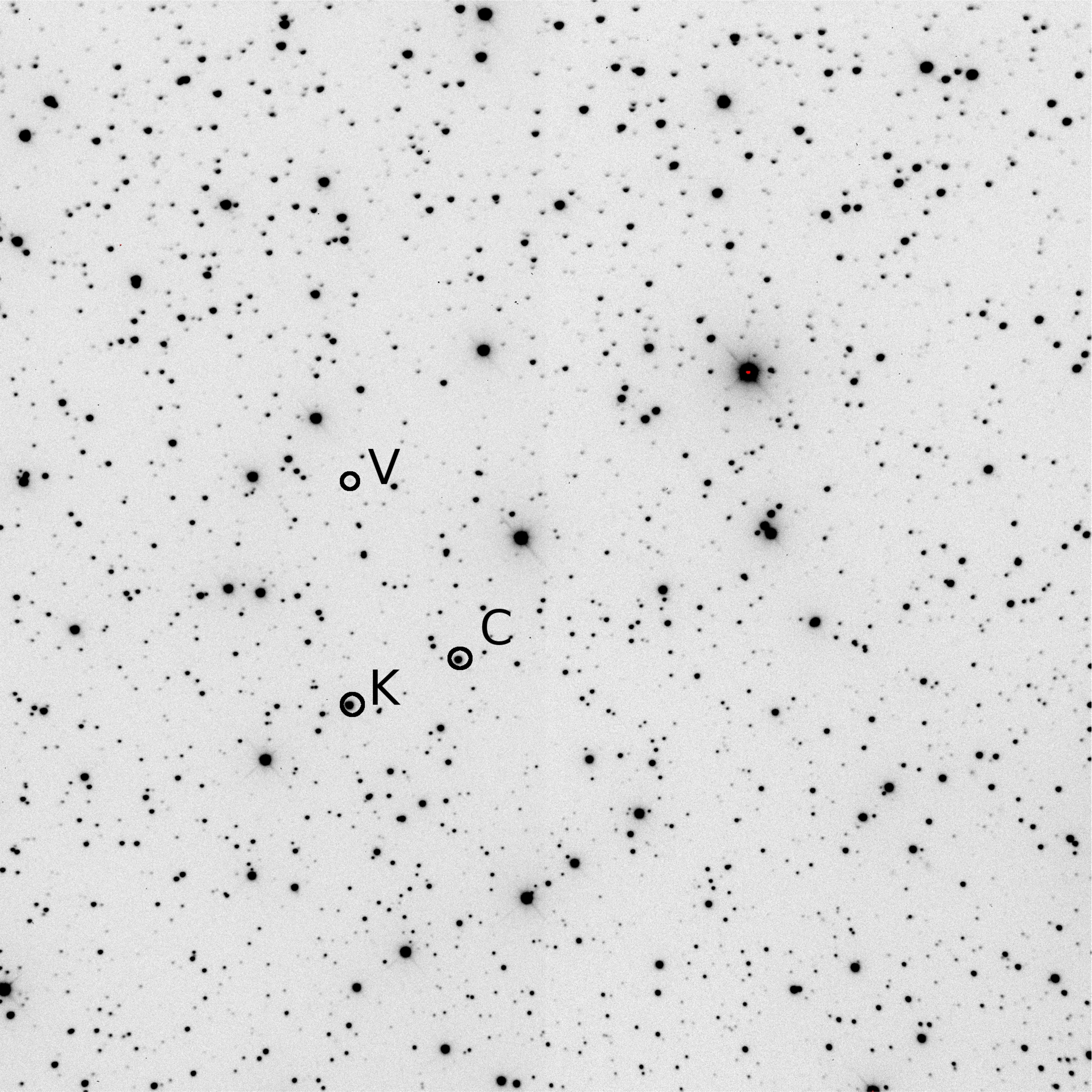}
\caption[V1668 Cyg $\mathrm{I_c}$ phase magnitude diagram with the best fit]{V1668 Cyg $\mathrm{I_c}$ phase magnitude diagram plotted using the new ephemeris of equation ~\ref{eq:v1668cyg} with the best fit. The solid line represents the best fit using equation ~\ref{eq:fourierfit}.}
\label{fig:v1668cygi}
\end{figure}

\begin{figure}[h]
\centering
\includegraphics[height=14cm,width=13cm]{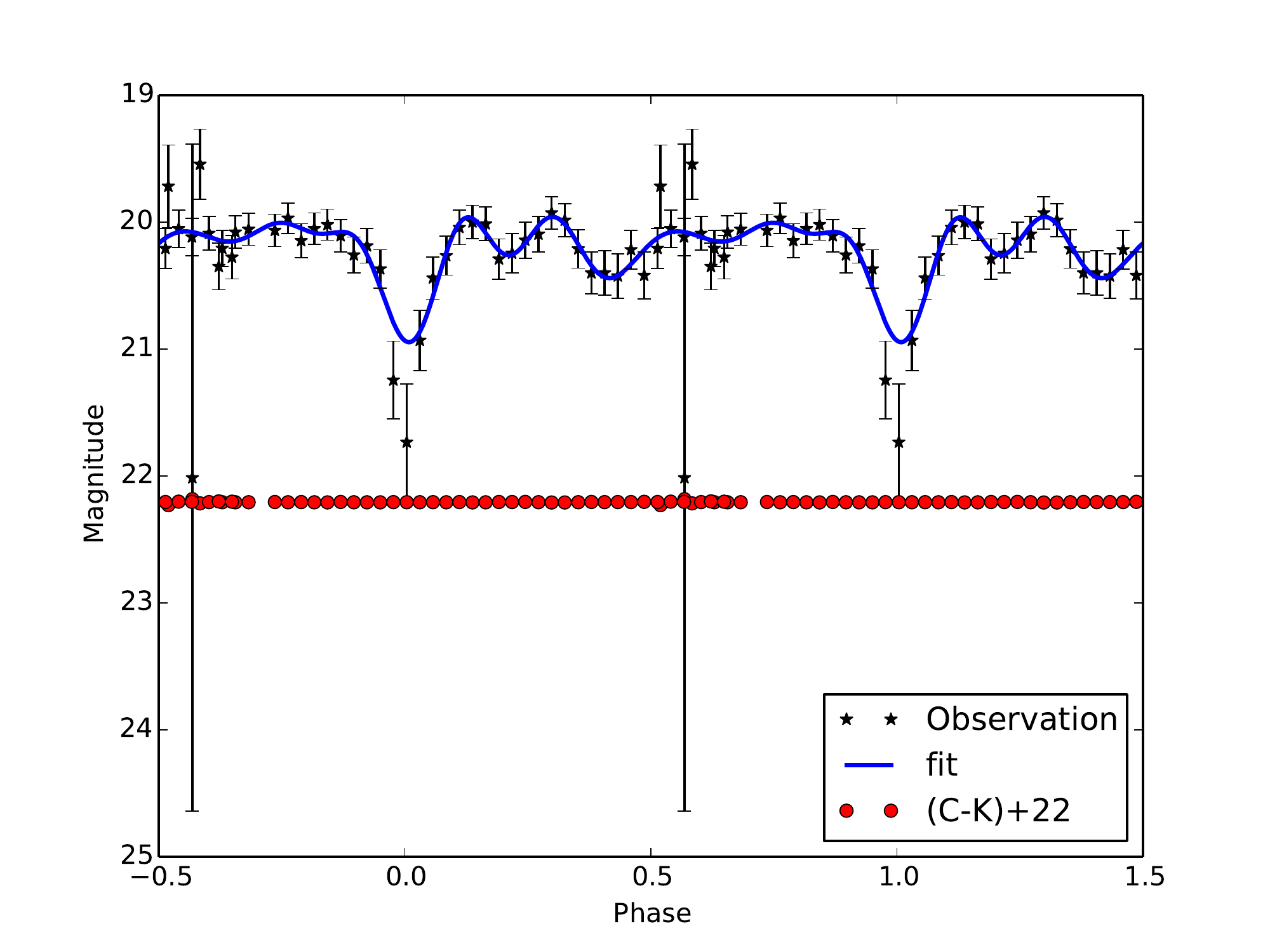}
\caption[V1668 Cyg V phase magnitude diagram with the best fit]{V1668 Cyg V phase magnitude diagram plotted using the new ephemeris of equation ~\ref{eq:v1668cyg} with the best fit. The solid line represents the best fit using equation ~\ref{eq:fourierfit}.}
\label{fig:v1668cygv}
\end{figure}

\clearpage
\subsection{PW Vul Photometric Light Curve Analysis}

The light curves were fitted by a sixth degree Fourier fit of the form of equation \ref{eq:fourierfit} using the same algorithm used for V1668 Cyg, and the coefficients are listed in table \ref{tab:pwvulfit}.

\begin{table}[h]
\caption{PW Vul Fourier Fit Coefficients in V, $\mathrm{R_c}$ and $\mathrm{I_c}$ filters.}
\begin{center}
\label{tab:pwvulfit}
\begin{tabular}{|l|c|c|c|c|}\hline\hline
coefficient &V&R&I&$\mathrm{I_c}$ 2015\\
\hline
$a_0$&      16.850 &       17.370&      15.750&      16.270 \\
$a_1$&     0.103 &     0.071&   -0.026&    0.064 \\
$b_1$&-   0.070 &     0.012&    0.010&   -0.057 \\
$a_2$&     0.064 &    -0.039&  0.0001&    0.055 \\
$b_2$&-  -0.017 &    -0.014&    0.068&   -0.050 \\
$a_3$&    0.029 &    -0.023&      0.028&   -0.044 \\
$b_3$&-  0.004 &    0.003&   0.003&   -0.020 \\
$a_4$&   -0.019 &  -0.001&    0.016&   -0.029 \\
$b_4$&   -0.013 &    -0.015&   -0.015&    0.021 \\
$a_5$&   -0.016 &   -0.005&   -0.023&   0.008 \\
$b_5$&   0.004 &    -0.013&     -0.016&    0.044 \\
$a_6$&    0.013 &     0.010&  -0.008&    0.026 \\
$b_6$&  -0.003 &   -0.002&    0.027&   0.003 \\
w  &     6.283  &      6.283 &     6.283 &     6.283  \\
\hline
\end{tabular}
\end{center}
\end{table}

The maximum of the V light curve was about 16.7 mag and the minimum was about 17.6 mag making the amplitude of variation $\sim$ 0.9 mag. The $\mathrm{R_c}$ light curve had its maximum about 16.9 mag and the minimum was about 17.8 mag with an amplitude of variation of $\sim$ 0.9 mag. For the $\mathrm{I_c}$ observations taken on 13-10-2015, the maximum magnitude was about 16.03 mag and the minimum was about 16.40 mag and the amplitude was $\sim$ 0.36 mag. The maximum magnitude of the $\mathrm{I_c}$ light curve (observations taken on 5-10-2016 ) was about 15.4 mag and the minimum was about 15.9 mag making the amplitude of variation  $\sim$ 0.5 mag. This can be seen in Figs ~\ref{fig:pwvuli2015}-~\ref{fig:pwvulv} showing the phase magnitude diagrams plotted using the new epoch at JD 2457309.229200 with the best fit.

\begin{figure}[h]
\centering
\includegraphics[height=14cm,width=13cm]{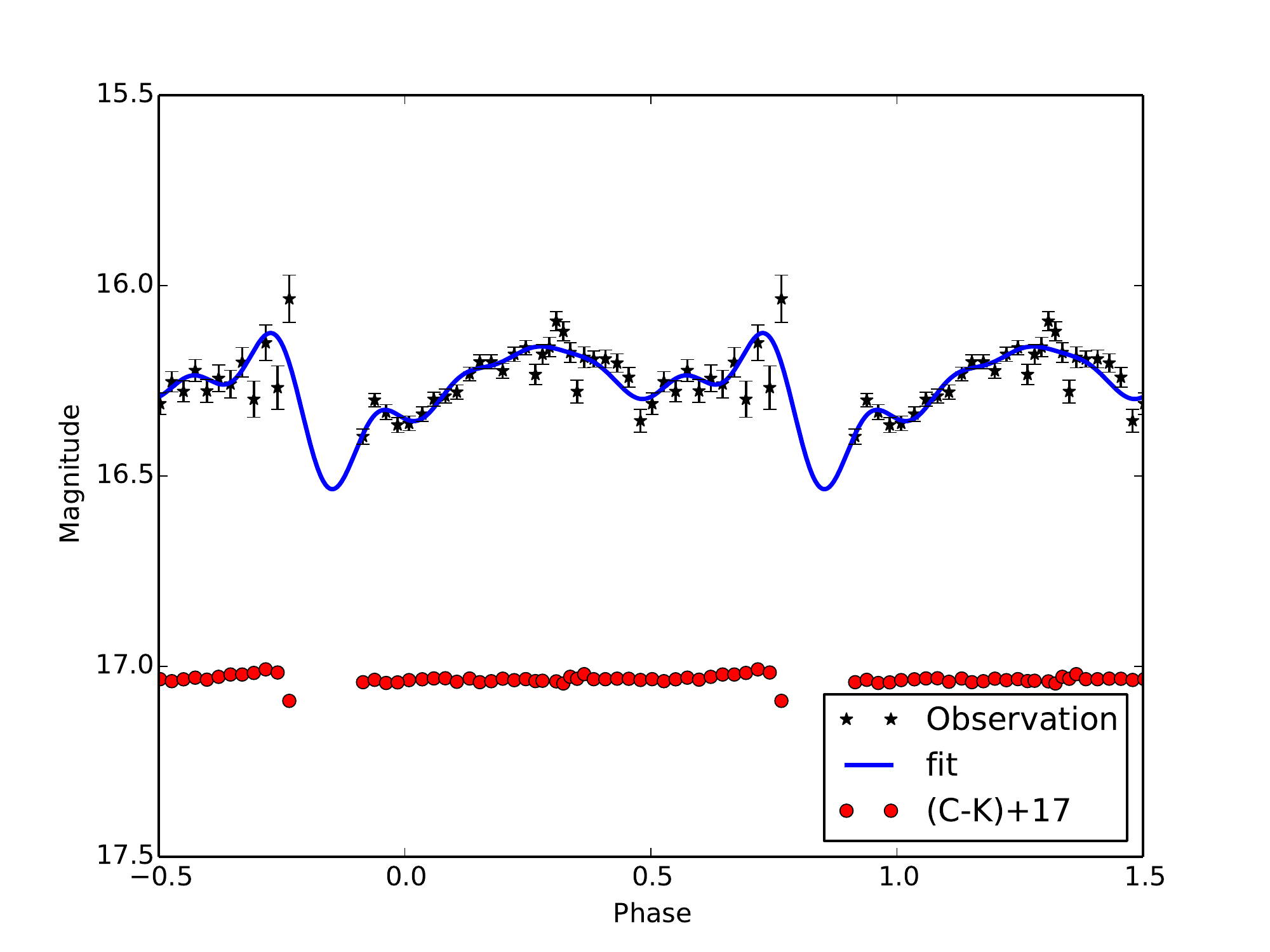}
\caption[PW Vul $\mathrm{I_c}$ (2015 observations) phase magnitude diagrams plotted using the new epoch with best fit]{PW Vul $\mathrm{I_c}$ (2015 observations) phase magnitude diagrams plotted using the new ephemeris of equation ~\ref{eq:pwvul} with the best fit. The solid line represents the best fit using equation ~\ref{eq:fourierfit}.}
\label{fig:pwvuli2015}
\end{figure}

\begin{figure}[h]
\centering
\includegraphics[height=14cm,width=13cm]{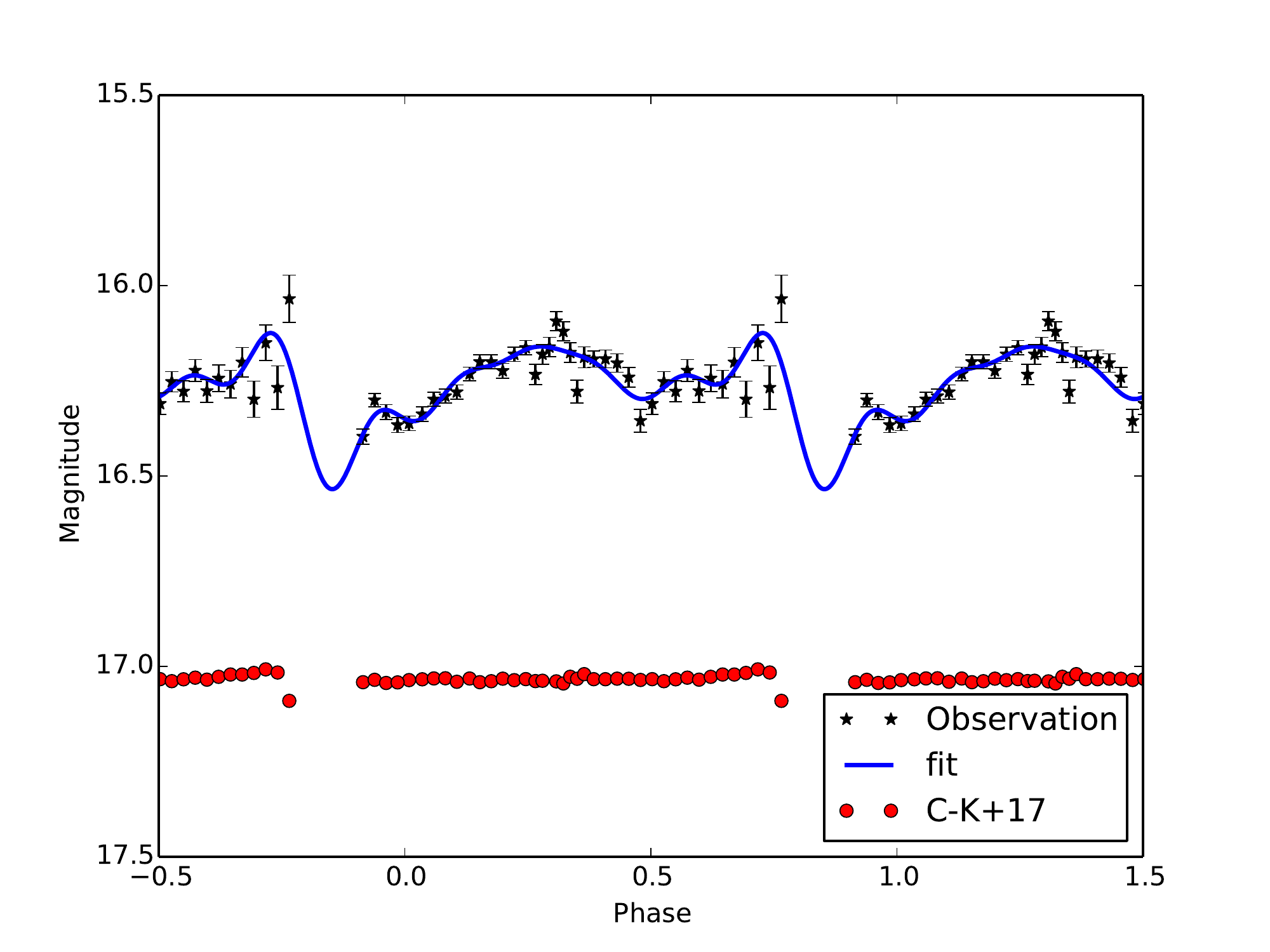}
\caption[PW Vul $\mathrm{I_c}$ (2016 observations) light Curve phase magnitude diagram with best fit]{PW Vul $\mathrm{I_c}$ (2016 observations) light Curve phase magnitude diagram plotted using the new ephemeris of equation ~\ref{eq:pwvul} with the best fit. The solid line represents the best fit using equation ~\ref{eq:fourierfit}.}
\label{fig:pwvuli}
\end{figure}

\begin{figure}[h]
\centering
\includegraphics[height=14cm,width=13cm]{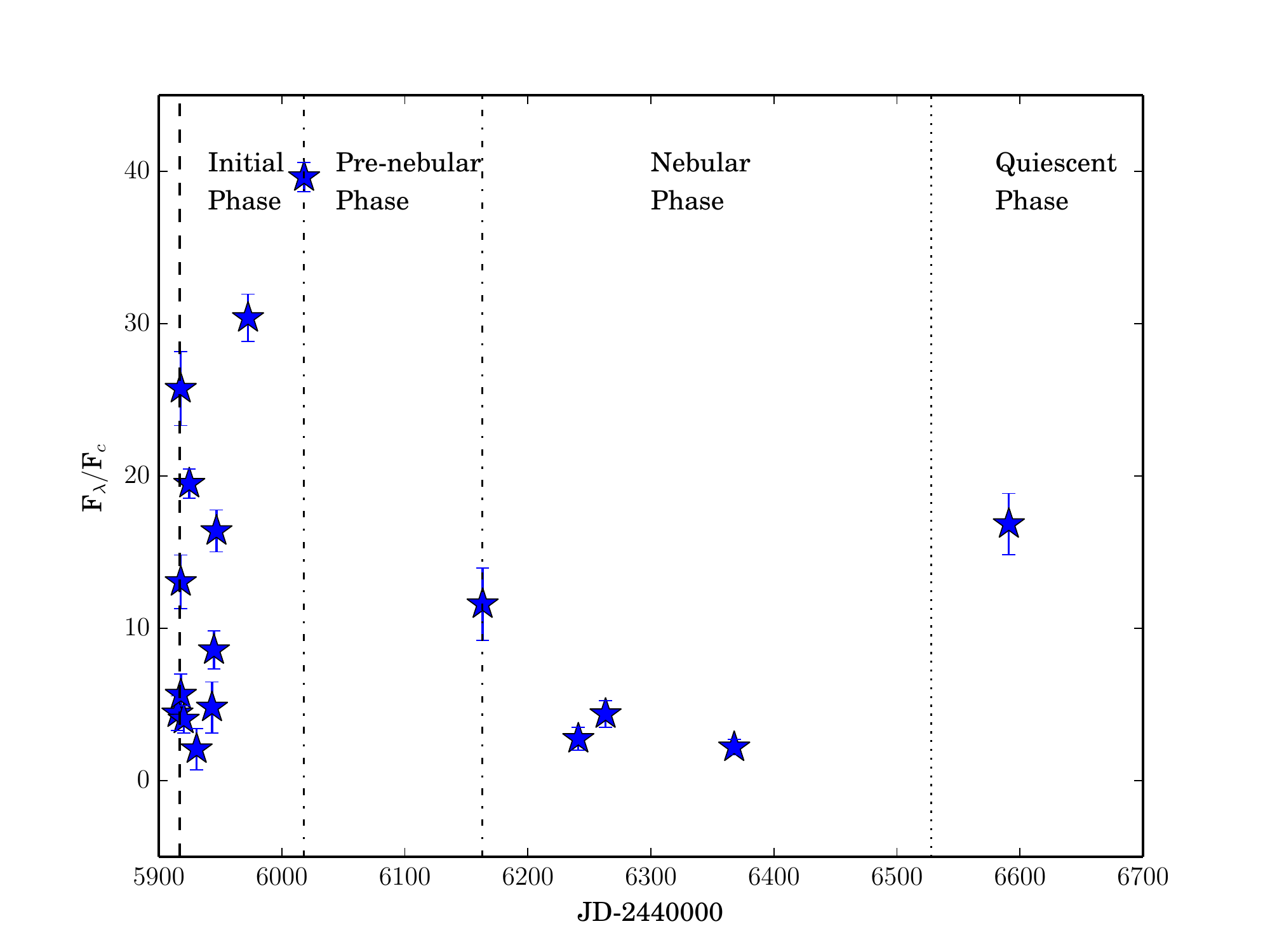}
\caption[PW Vul $\mathrm{R_c}$ phase magnitude diagram with best fit]{PW Vul $\mathrm{R_c}$ phase magnitude diagram plotted using the new ephemeris of equation ~\ref{eq:pwvul} with best fit. The solid line represents the best fit using equation ~\ref{eq:fourierfit}.}
\label{fig:pwvulr}
\end{figure}

\begin{figure}[h]
\centering
\includegraphics[height=14cm,width=13cm]{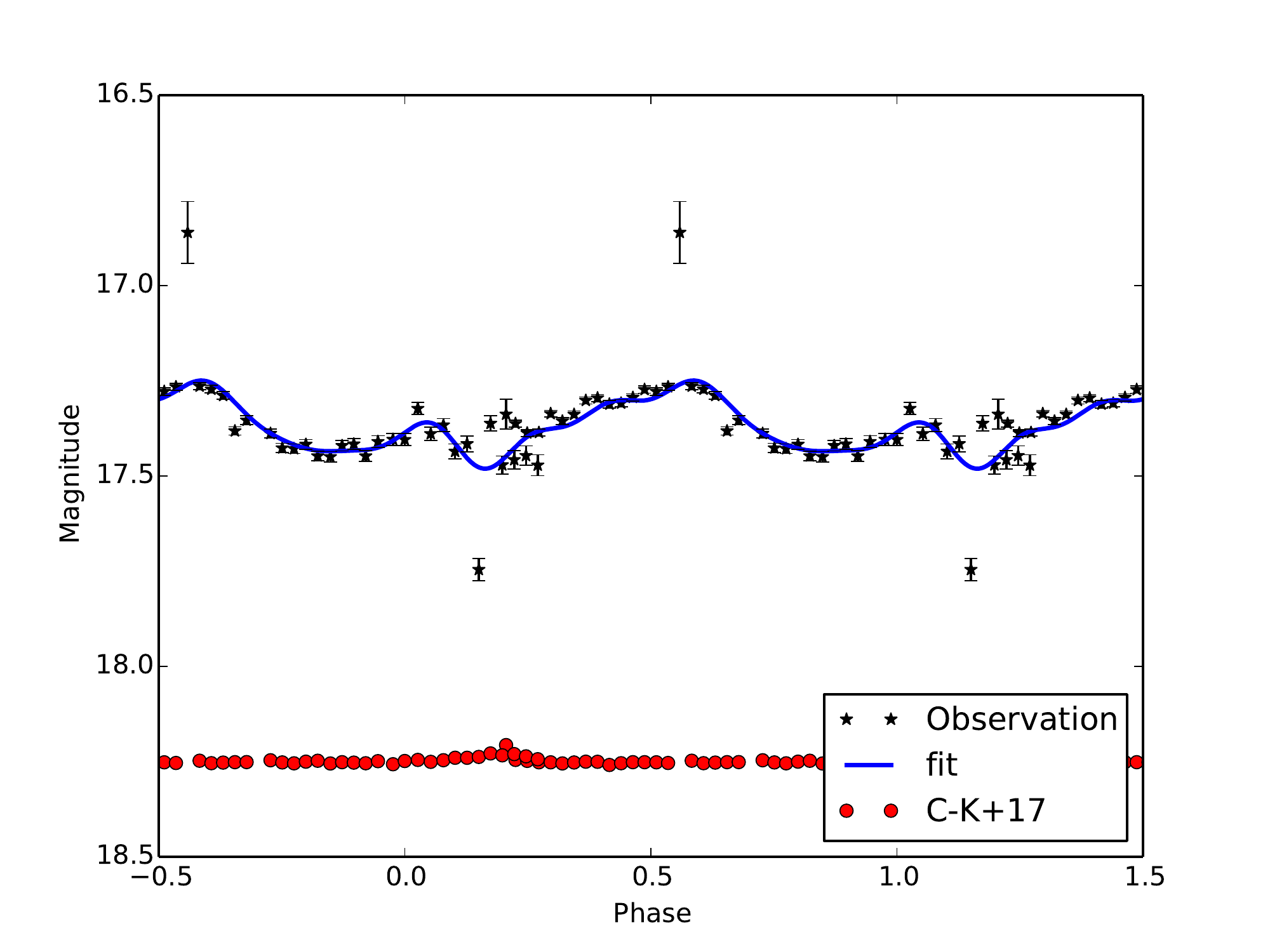}
\caption[PW Vul V phase magnitude diagram with best fit]{PW Vul V phase magnitude diagram plotted using the new ephemeris of equation ~\ref{eq:pwvul} with best fit. The solid line represents the best fit using equation ~\ref{eq:fourierfit}.}
\label{fig:pwvulv}
\end{figure}

\clearpage

\subsection{V1974 Cyg Photometric Light Curve Analysis}

The light curves were also fitted by a sixth degree Fourier fit of the form of equation \ref{eq:fourierfit} using the same algorithm used for V1668 Cyg and PW Vul, and the coefficients are listed in table \ref{tab:v1974cygfit}.
\vspace{+0.5cm}
\begin{table}[ht]
\caption{V1974 Cyg Fourier Fit Coefficients in V, $\mathrm{R_c}$ and $\mathrm{I_c}$ filters.}
\begin{center}
\label{tab:v1974cygfit}
\begin{tabular}{|l|c|c|c|c|}\hline\hline
coefficient &V&R&$\mathrm{I_c}$ 2015&$\mathrm{I_c}$ \\
\hline
$a_0$&       17.170 &      16.820 &    16.020 &     16.140 \\
$a_1$&     0.036 &    -0.318 &  0.046 &   0.012 \\
$b_1$&    -0.010 &     0.225 &  0.014 &  0.008 \\
$a_2$&   -0.006 &     0.186 &  0.032 &   0.033 \\
$b_2$&   -0.007 &    -0.394 &  0.011 &  -0.048 \\
$a_3$&    -0.010 &   -0.023 &  0.024 &   0.011 \\
$b_3$&     0.021 &     0.375 &   0.032 &   0.011 \\
$a_4$&    0.009 &   -0.068 & 0.007 &   0.017 \\
$b_4$&   -0.008 &    -0.260 &  0.040 &   0.014 \\
$a_5$&    -0.010 &    0.073 &-0.008 &  -0.011 \\
$b_5$&    0.003 &     0.147 &   0.041 &   0.015 \\
$a_6$&     0.016 &   -0.046 & -0.010 &  -0.014 \\
$b_6$&     0.014 &   -0.058 &  0.051 &   0.020 \\
w  &      6.283  &     6.283  &   6.283    &    6.283  \\
\hline
\end{tabular}
\end{center}
\end{table}

The maximum of of the V light curve was about 17.05 mag and the minimum was about 17.24 mag and the amplitude of variation was $\sim$ 0.19 mag. The maximum of of the $\mathrm{R_c}$ light curve was about 15.4 mag and the minimum was about 17.5 mag making the amplitude of variation $\sim$ 2.1 mag. For the $\mathrm{I_c}$ light curve (observations taken on 20-10-2015) the maximum was about 15.9 mag and the minimum was about 16.6 mag with an amplitude of $\sim$ 0.7 mag. The maximum of of the $\mathrm{I_c}$ light curve (observations taken on 30-7-2016) was about 15.9 mag and the minimum was about 16.2 mag making the amplitude $\sim$ 0.3 mag. The observed and fitted light curves are shown in Figs ~\ref{fig:v1974cygi2015}-~\ref{fig:v1974cygv}.

\begin{figure}[h]
\centering
\includegraphics[height=14cm,width=13cm]{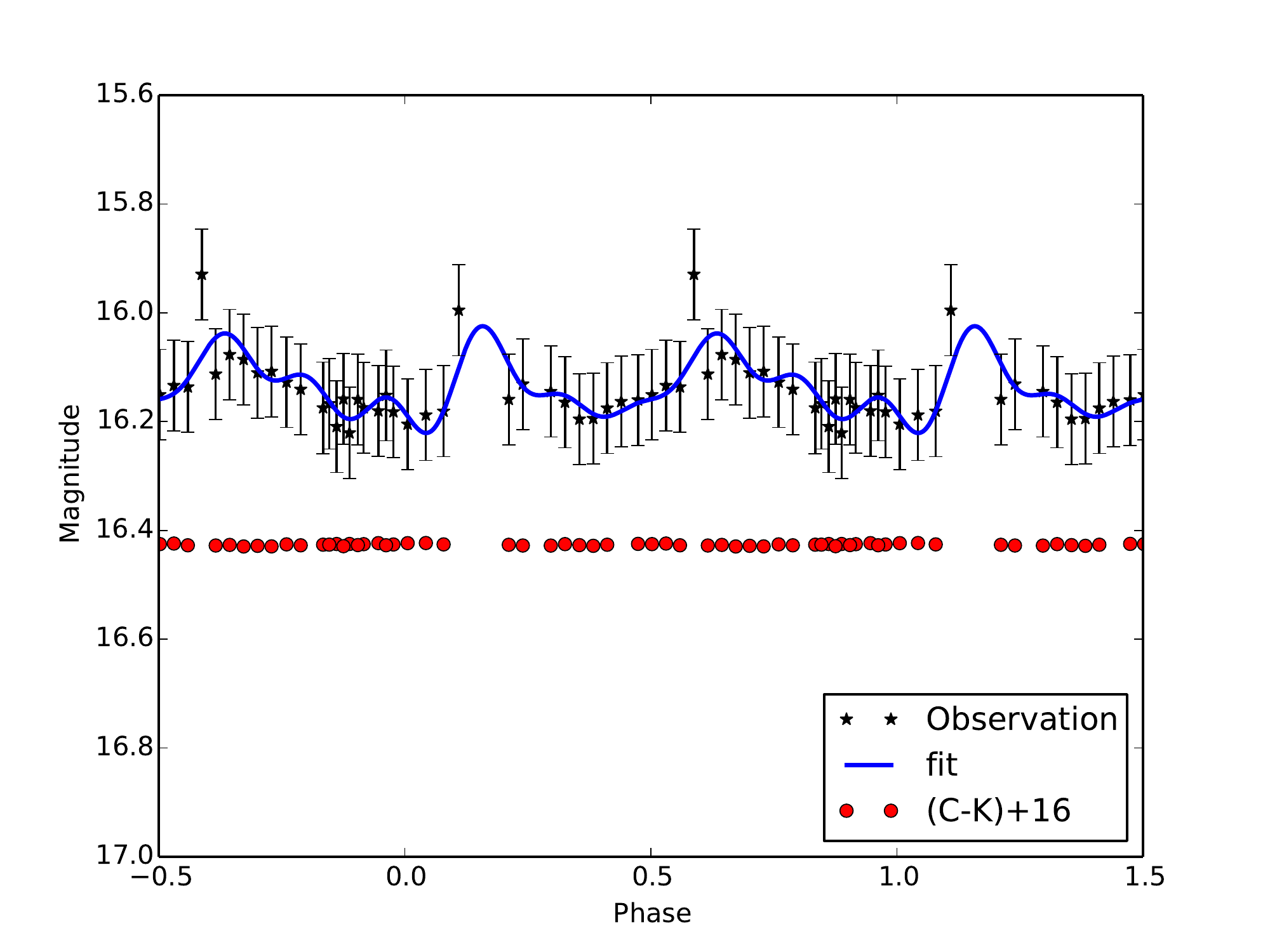}
\caption[V1974 Cyg $\mathrm{I_c}$ (2015 observations) phase magnitude diagram with the best fit]{V1974 Cyg $\mathrm{I_c}$ (2015 observations) phase magnitude diagram plotted using the new ephemeris of equation ~\ref{eq:v1974cyg} with the best fit. The solid line represents the best fit using equation ~\ref{eq:fourierfit}.}
\label{fig:v1974cygi2015}
\end{figure}

\begin{figure}[h]
\centering
\includegraphics[height=14cm,width=13cm]{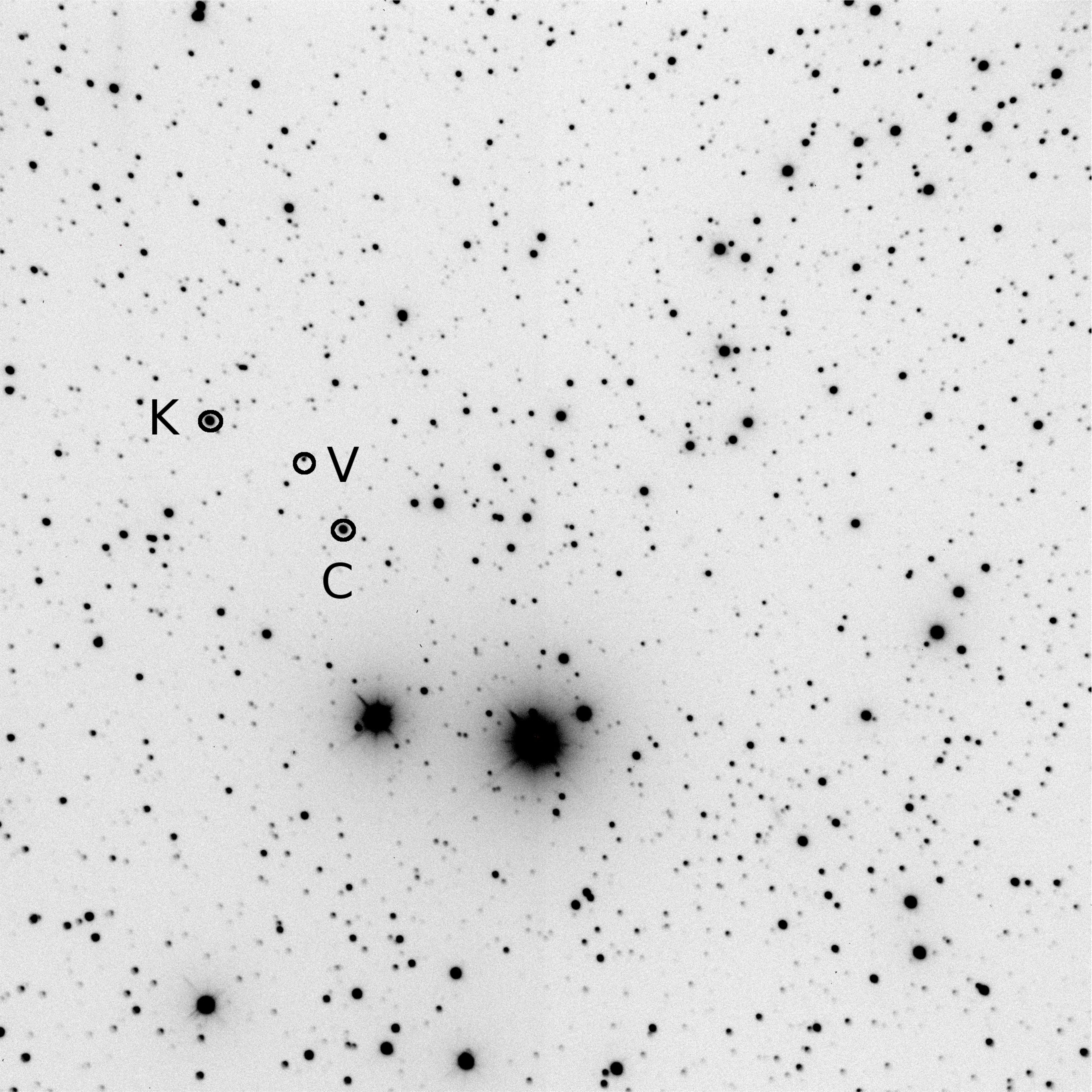}
\caption[V1974 Cyg $\mathrm{I_c}$ phase magnitude diagram with the best fit]{V1974 Cyg $\mathrm{I_c}$ phase magnitude diagram plotted using the new ephemeris of equation ~\ref{eq:v1974cyg} with the best fit. The solid line represents the best fit using equation ~\ref{eq:fourierfit}.}
\label{fig:v1974cygi2}
\end{figure}

\begin{figure}[h]
\centering
\includegraphics[height=14cm,width=13cm]{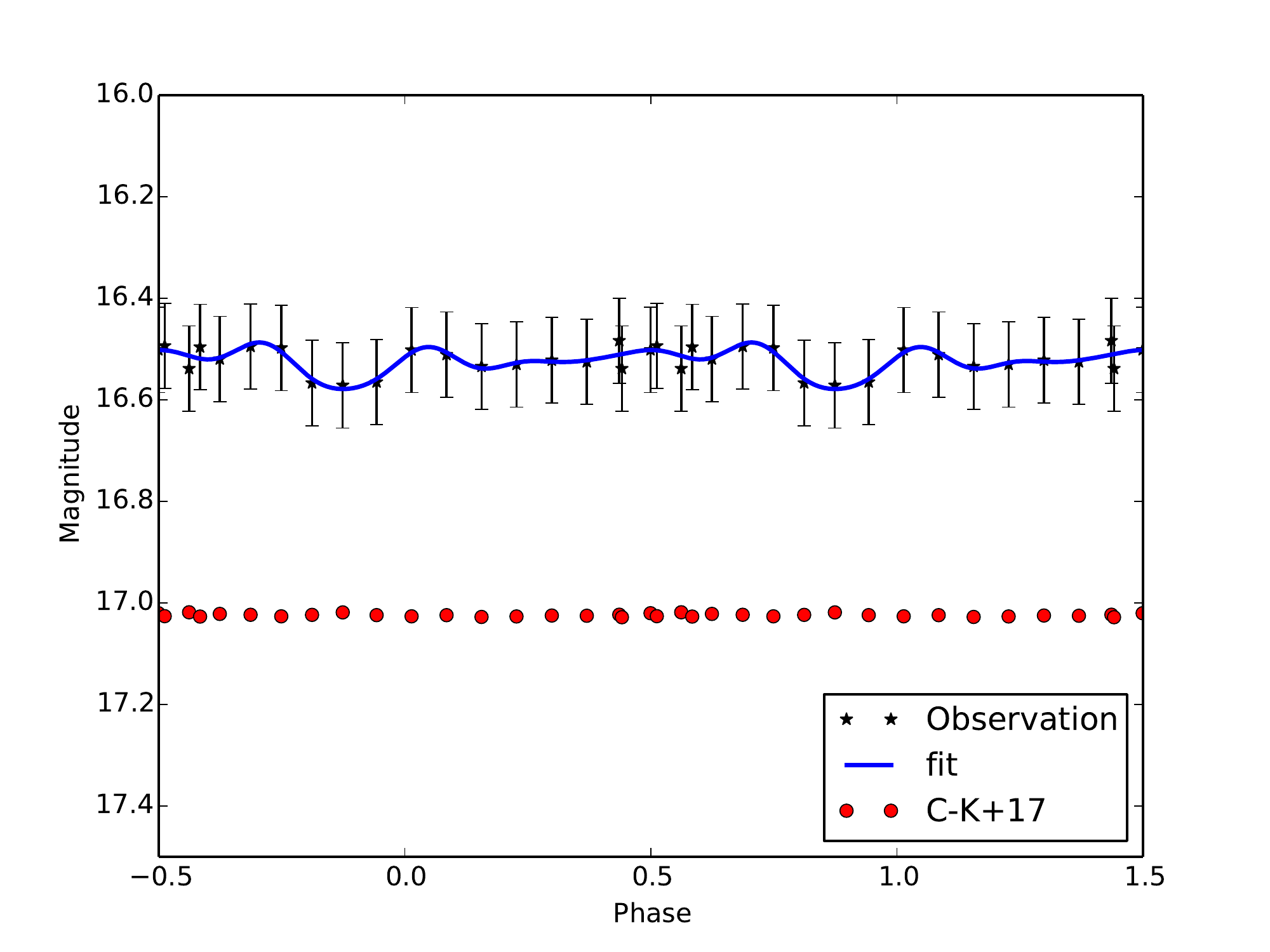}
\caption[V1974 Cyg $\mathrm{R_c}$ phase magnitude diagram with the best fit]{V1974 Cyg $\mathrm{R_c}$ phase magnitude diagram plotted using the new ephemeris of equation ~\ref{eq:v1974cyg} with the best fit. The solid line represents the best fit using equation ~\ref{eq:fourierfit}.}
\label{fig:v1974cygr}
\end{figure}

\begin{figure}[h]
\centering
\includegraphics[height=14cm,width=13cm]{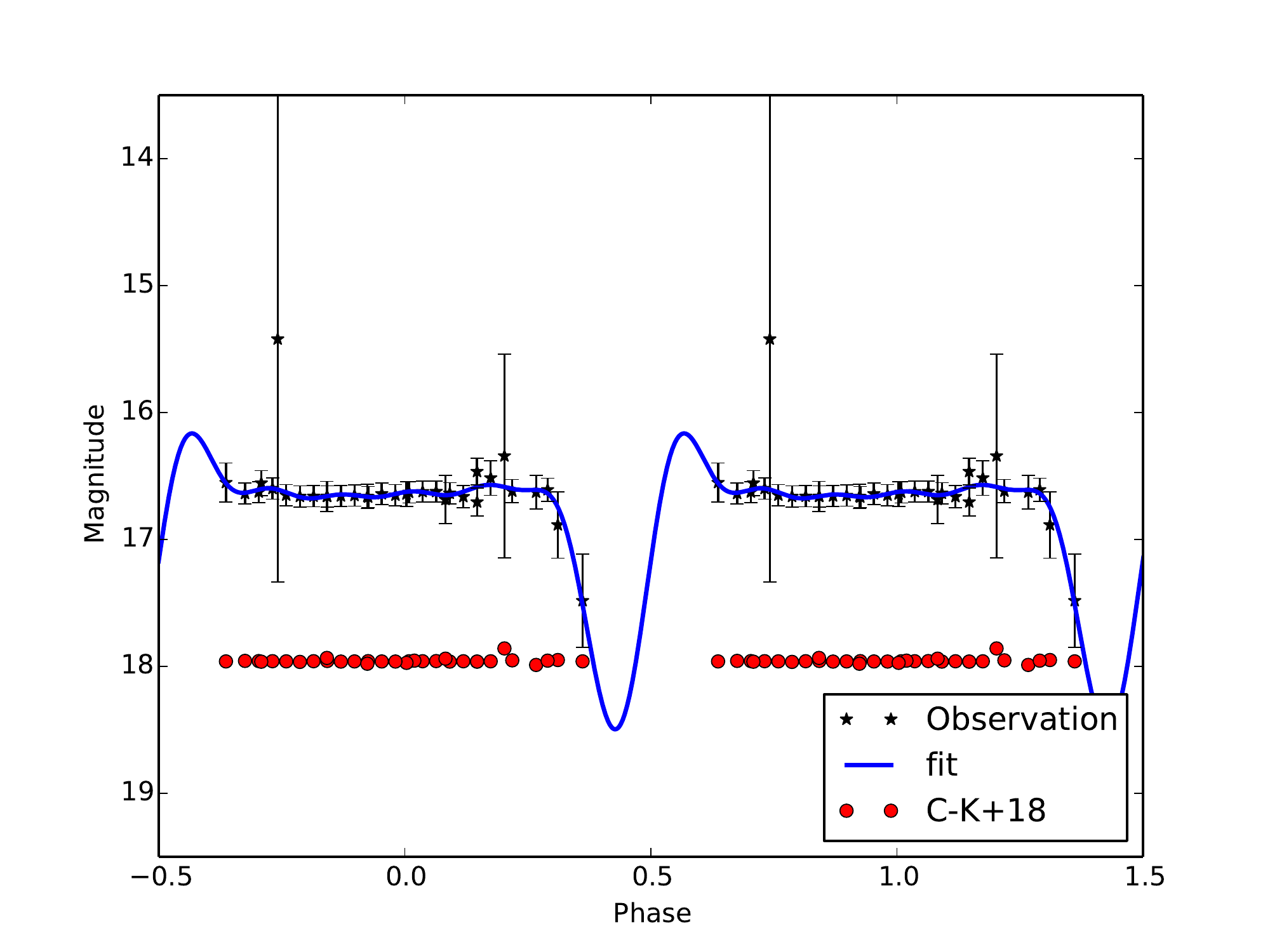}
\caption[V1974 Cyg V phase magnitude diagram with the best fit]{V1974 Cyg V phase magnitude diagram plotted using the new ephemeris of equation ~\ref{eq:v1974cyg} with the best fit. The solid line represents the best fit using equation ~\ref{eq:fourierfit}.}
\label{fig:v1974cygv}
\end{figure}

\clearpage

\begin{table}[h!]
\caption[V1974 O-C calculated using the times of minima]{V1974 Cyg O-C calculated using the times of minima using the ephemeris : $HJD_{min}=2457316.385920 + 0.081263 \times E$.}
\begin{center}
\label{tab:v1974cygomc}
\begin{tabular}{|l|c|c|c|c|}\hline\hline
$HJD_{min}$   &  E       & O-C (days)        & O-C (cycles)     & Reference\\
\hline
2449185.492   &  -100057.0 &  0.0383710004389    &  0.472182917674  & (1)  \\
2449267.560   &  -99047.0  &  0.0299410000443    &  0.368445664624  & (1)  \\
2449268.540   &  -99035.0  &  0.0348849999718    &  0.429285160182  & (1)  \\
2449268.623   &  -99034.0  &  0.0367220002227    &  0.451890777139  & (1)  \\
2449271.544   &  -98998.0  &  0.0326540004462    &  0.401831097132  & (1)  \\
2449271.631   &  -98997.0  &  0.0378910000436    &  0.466276165581  & (1)  \\
2449273.574   &  -98973.0  &  0.031278999988     &  0.384910721829  & (1)  \\
2449273.655   &  -98972.0  &  0.0310160005465    &  0.38167432345   & (1)  \\
2449276.503   &  -98937.0  &  0.0348110003397    &  0.428374541177  & (1)  \\
2449276.58    &  -98936.0  &  0.0302480002865    &  0.372223524686  & (1)  \\
2449276.663   &  -98935.0  &  0.0319850002415    &  0.393598565664  & (1)  \\
2449284.546   &  -98838.0  &  0.0322740003467    &  0.397154921018  & (1)  \\
2449284.626   &  -98837.0  &  0.0314110000618    &  0.386535078225  & (1)  \\
2449285.594   &  -98825.0  &  0.0239550005645    &  0.2947836108    & (1)  \\
2449286.501   &  -98814.0  &  0.0370620000176    &  0.456074720569  & (1)  \\
2449289.498   &  -98777.0  &  0.027331000194     &  0.336327728412  & (1)  \\
2449289.587   &  -98776.0  &  0.0352680003271    &  0.433998256612  & (1)  \\
2449293.483   &  -98728.0  &  0.0309439999983    &  0.380788304619  & (1)  \\
2449293.565   &  -98727.0  &  0.0315809999593    &  0.38862704994   & (1)  \\
2449299.498   &  -98654.0  &  0.0317820003256    &  0.39110050485   & (1)  \\
2449300.552   &  -98641.0  &  0.0296630002558    &  0.365024676123  & (1)  \\
2449302.510   &  -98617.0  &  0.0374510004185    &  0.460861651902  & (1)  \\
2449304.534   &  -98592.0  &  0.0295760002919    &  0.363954078632  & (1)  \\
2449306.483   &  -98568.0  &  0.0288639999926    &  0.355192399894  & (1)  \\
2449313.477   &  -98482.0  &  0.0333460001275    &  0.410346653797  & (1)  \\
2449314.528   &  -98469.0  &  0.0285270004533    &  0.351045376781  & (1)  \\
2449330.457   &  -98273.0  &  0.0298790000379    &  0.367682709695  & (1)  \\
2449331.520   &  -98260.0  &  0.0364600000903    &  0.448666675982  & (1)  \\
2450634.366   &  -82227.0  &  -0.00721899978817  &  -0.0888350145597& (2)  \\
2450639.431   &  -82165.0  &  0.0194749999791    &  0.239653962802  & (2)  \\
2450640.456   &  -82152.0  &  -0.0119440001436   &  -0.14697956196  & (2)  \\
2450641.389   &  -82141.0  &  0.0271630003117    &  0.334260368331  & (2)  \\
2450641.477   &  -82140.0  &  0.033900000155     &  0.417164024894  & (2)  \\
2450642.408   &  -82128.0  &  -0.0102559998631   &  -0.126207497424 & (2)  \\
2450642.495   &  -82127.0  &  -0.00451899971813  &  -0.0556095605396& (2)  \\
2450645.446   &  -82091.0  &  0.0210130000487    &  0.258580166234  & (2)  \\
2457316.386  &   0.0      &  0.0                &  0.0             & (3)  \\
2457600.298  &   3494.0   &  -0.0212989998981   &  -0.262099601272 & (3)  \\
2457601.298  &   3506.0   &  0.00364700006321   &  0.0448789739883 & (3)  \\
\hline
\end{tabular}
\end{center}
\caption*{References: (1) \citet{1995AcA....45..365S}, (2) \citet{2002AcA....52..273O}, (3) Present work.}
\end{table}

\citet{2002AcA....52..273O} calculated the times of minimum of V1974 in various observations and calculated the O-C based on a period of 0.084632 d. Here we use these times of minimum along with the times of minimum of our observations to calculate the O-C using the period of 0.081263 d and the epoch we found at JD 2457316.385920. The results are presented in table \ref{tab:v1974cygomc} and the O-C curve is plotted in Fig ~\ref{fig:v1974omcmin}.

\begin{figure}
\centering
\includegraphics[height=14cm,width=13cm]{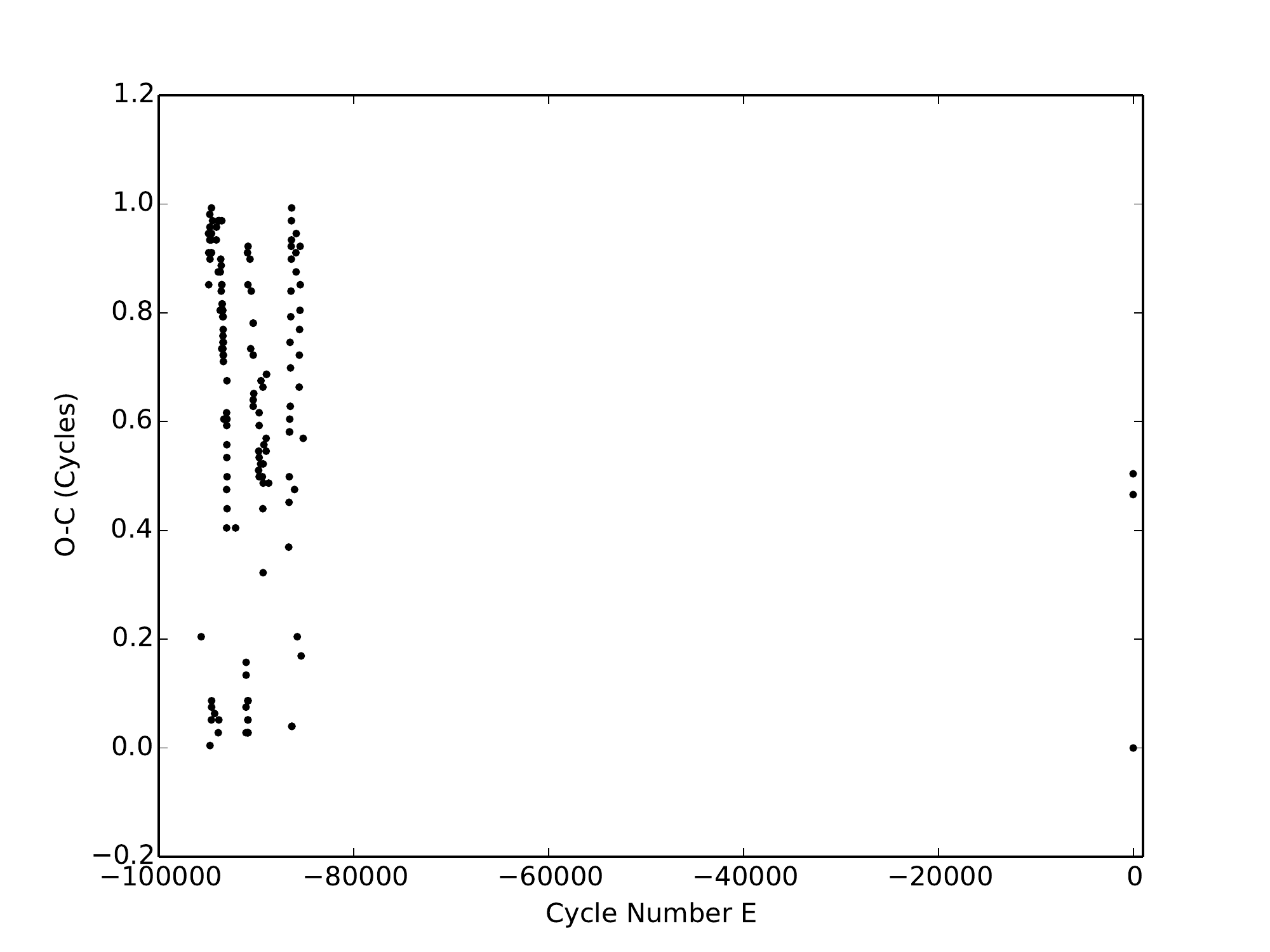}
\caption[V1974 O-C calculated using the times of minima]{V1974 Cyg O-C curve for the times of minimum according to table ~\ref{tab:v1974cygomc}.}
\label{fig:v1974omcmin}
\end{figure}

\citet{2002AcA....52..273O} reported the times of maximum of V1974 in various observations and calculated the O-C based on two periods of 0.0813 and 0.085. If we use these times of maximum and calculate the O-C from them using both periods taking the epoch at JD 2457601.3476135  and adding the times of maximum of our observations we get tables ~\ref{tab:v1974cygomctab2} and ~\ref{tab:v1974cygomctab3} and the O-C curve is plotted in Figures ~\ref{fig:v1974omcmaxtable2} and ~\ref{fig:v1974omcmaxtable3}, respectively.

\vspace{+0.5cm}
\begingroup
\small \renewcommand{\arraystretch}{0.8} \setlength{\tabcolsep}{3pt}
\begin{longtable}{|l|c|c|c|c|}
\caption[V1974 Cyg O-C for the times of maximum (shorter period)]{V1974 O-C for the times of maximum using the ephemeris : $HJD_{min}=2457601.3476135 + 0.081263 \times E$.}
\label{tab:v1974cygomctab2}
\renewcommand{\arraystretch}{0.7}
\cr \hline 
$HJD_{min}$   &  E       & O-C (days)        & O-C (cycles)     & Reference\\
\hline
  \endfirsthead
  \caption*{Continued. V1974 Cyg O-C for the times of maximum using the ephemeris : $HJD_{min}=2457601.3476135 + 0.081263 \times E$.}\\
  \hline\hline   
$HJD_{min}$   &  E       & O-C (days)        & O-C (cycles)     & Reference\\
  \hline
  \endhead
  \hline
  \endfoot
\endlastfoot
2449267.514 &  -102559.0 &-0.0195&  -0.2403 &(1)   \\
2449268.576 &  -102546.0 &-0.0142&  -0.1746 &(1)   \\
2449271.495 &  -102510.0 &-0.0202&  -0.2486 &(1)   \\
2449271.576 &  -102509.0 &-0.0209&  -0.2567 &(1)   \\
2449273.529 &  -102485.0 &-0.0180&  -0.2211 &(1)   \\
2449273.612 &  -102484.0 &-0.0160&  -0.1972 &(1)   \\
2449283.595 &  -102361.0 &-0.0280&  -0.3440 &(1)   \\
2449284.583 &  -102349.0 &-0.0146&  -0.1791 &(1)   \\
2449285.548 &  -102337.0 &-0.0247&  -0.3035 &(1)   \\
2449286.529 &  -102325.0 &-0.0189&  -0.2322 &(1)   \\
2449289.536 &  -102288.0 &-0.0181&  -0.2232 &(1)   \\
2449293.524 &  -102239.0 &-0.0119&  -0.1467 &(1)   \\
2449293.598 &  -102238.0 &-0.0195&  -0.2397 &(1)   \\
2449300.509 &  -102153.0 &-0.0158&  -0.1940 &(1)   \\
2449302.537 &  -102128.0 &-0.0187&  -0.2306 &(1)   \\
2449304.562 &  -102103.0 &-0.0250&  -0.3077 &(1)   \\
2449314.482 &  -101981.0 &-0.0190&  -0.2334 &(1)   \\
2449321.554 &  -101894.0 &-0.0166&  -0.2040 &(1)   \\
2449535.438 &  -99262.0  &-0.0056&  -0.0684 &(1)   \\
2449537.390 &  -99238.0  &-0.0038&  -0.0463 &(1)   \\
2449537.470 &  -99237.0  &-0.0050&  -0.0618 &(1)   \\
2449546.406 &  -99127.0  &-0.0075&  -0.0921 &(1)   \\
2449546.487 &  -99126.0  &-0.0077&  -0.0953 &(1)   \\
2449548.448 &  -99102.0  &0.0030 &  0.0375 &(1)    \\
2449548.524 &  -99101.0  &-0.0022& --0.0272 &(1)   \\
2449559.407 &  -98967.0  &-0.0079& --0.0970 &(1)   \\
2449559.490 &  -98966.0  &-0.0061& --0.0756 &(1)   \\
2449561.367 &  -98943.0  &0.0019 &  0.0235 &(1)    \\
2449561.445 &  -98942.0  &-0.0013&  -0.0166 &(1)   \\
2449561.533 &  -98941.0  &0.0054 &  0.0663 &(1)    \\
2449588.658 &  -98607.0  &-0.0100&  -0.1234 &(1)   \\
2449603.607 &  -98423.0  &-0.0126&  -0.1554 &(1)   \\
2449627.495 &  -98129.0  &-0.0147&  -0.1809 &(1)   \\
2449637.341 &  -98008.0  &-0.0010&  -0.0124 &(1)   \\
2449637.432 &  -98007.0  &0.0087 &  0.1075 &(1)    \\
2449642.534 &  -97944.0  &-0.0086&  -0.1054 &(1)    \\
2449646.269 &  -97898.0  &-0.0115&  -0.1411 &(1)    \\
2449646.358 &  -97897.0  &-0.0037&  -0.0458 &(1)    \\
2449659.356 &  -97737.0  &-0.0071&  -0.0876 &(1)    \\
2449659.440 &  -97736.0  &-0.0044&  -0.0539 &(1)    \\
2449659.526 &  -97735.0  &0.0004 &  0.0045 &(1)     \\
2449661.220 &  -97714.0  &-0.0121&  -0.1485 &(1)    \\
2449661.306 &  -97713.0  &-0.0073&  -0.0902 &(1)    \\
2449661.392 &  -97712.0  &-0.0026&  -0.0318 &(1)    \\
2449663.263 &  -97689.0  &-0.0005&  -0.0066 &(1)    \\
2449663.344 &  -97688.0  &-0.0008&  -0.0098 &(1)    \\
2449663.435 &  -97687.0  &0.0089 &  0.1101 &(1)     \\
2449669.284 &  -97615.0  &0.0073 &  0.0900 &(1)     \\
2449693.253 &  -97320.0  &0.0050 &  0.0614 &(3)     \\
2449713.476 &  -97071.0  &-0.0054&  -0.0669 &(1)    \\
2449717.463 &  -97022.0  &-0.0001&  -0.0014 &(1)    \\
2449771.658 &  -96355.0  &-0.0047&  -0.0576 &(2)    \\
2449897.438 &  -94807.0  &-0.0132&  -0.1624 &(2)    \\
2449903.369 &  -94734.0  &-0.0141&  -0.1733 &(2)    \\
2449908.418 &  -94672.0  &-0.0031&  -0.0385 &(2)    \\
2449925.396 &  -94463.0  &-0.0082&  -0.1009 &(2)    \\
2449930.431 &  -94401.0  &-0.0112&  -0.1384 &(2)    \\
2449972.286 &  -93886.0  &-0.0045&  -0.0552 &(2)    \\
2449973.419 &  -93872.0  &-0.0091&  -0.1121 &(2)    \\
2449990.408 &  -93663.0  &-0.0032&  -0.0392 &(2)    \\
2450012.344 &  -93393.0  &-0.0070&  -0.0867 &(2)    \\
2450018.359 &  -93319.0  &-0.0052&  -0.0639 &(2)    \\
2450041.689 &  -93032.0  &0.0036 &  0.0438 &(3)     \\
2450060.289 &  -92803.0  &-0.0047&  -0.0578 &(2)    \\
2450062.561 &  -92775.0  &-0.0079&  -0.0977 &(3)    \\
2450234.838 &  -90655.0  &0.0006 &  0.0068 &(3)     \\
2450343.399 &  -89319.0  &-0.0001&  -0.0013 &(3)    \\
2450361.351 &  -89098.0  &-0.0063&  -0.0774 &(4)    \\
2450640.563 &  -85662.0  &0.0009 &  0.0113 &(4)     \\
2457600.285 &  -13.0     &-0.0058&  -0.0713 & (5)   \\
2457600.367 &  -12.0     &-0.0053&  -0.0653 & (5)   \\
2457601.348 &  0.0       &0.0    &  0.0     & (5)   \\
\hline
\caption*{References: (1) \citet{1995AcA....45..365S}, (2) \citet{1996AcA....46..311O}, (3) \citet{1997PASP..109..114S}, (4) \citet{2002AcA....52..273O}, (5) Present work.}
\end{longtable}

\begin{figure}
\centering
\includegraphics[height=14cm,width=13cm]{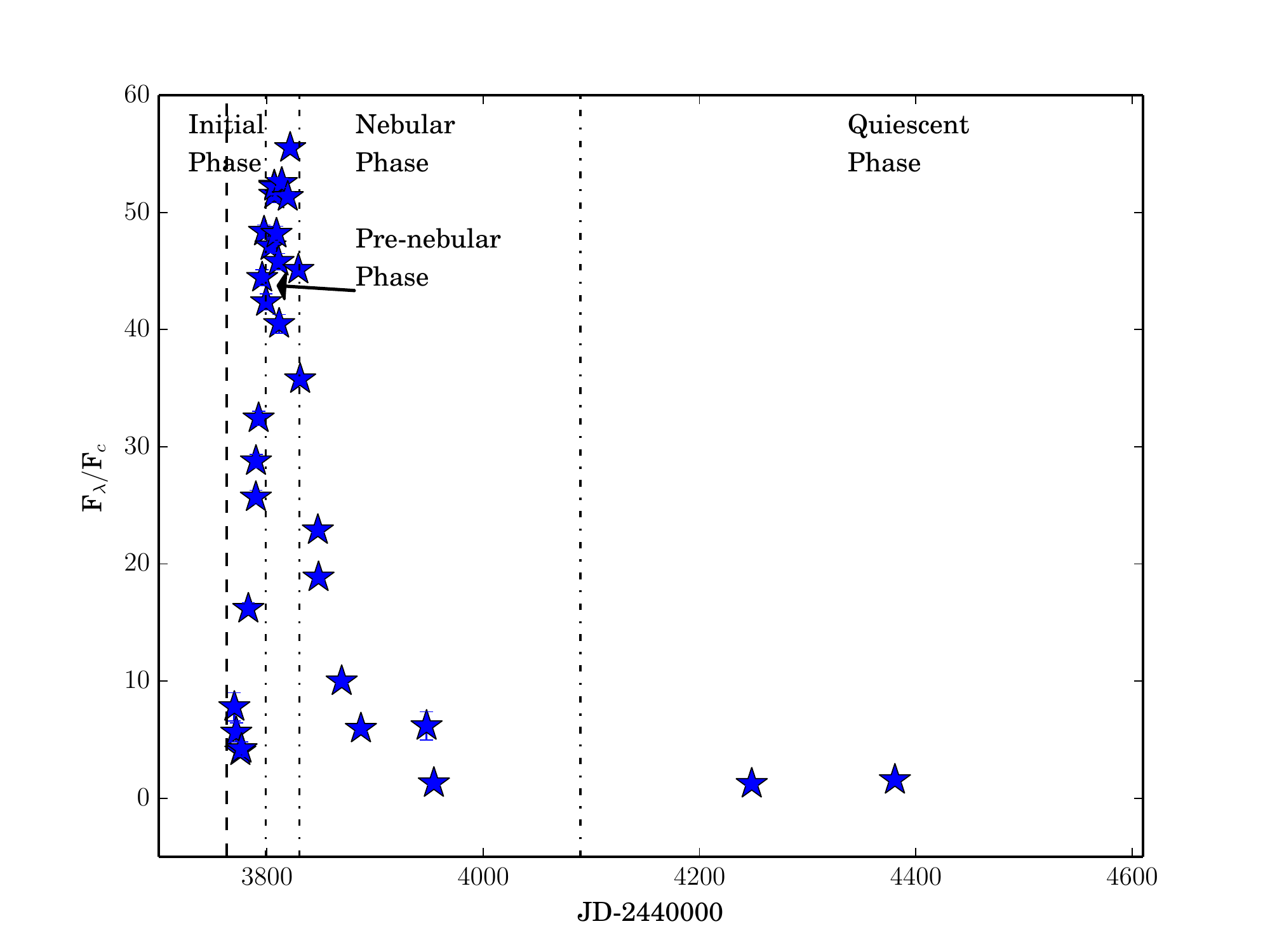}
\caption{V1974 Cyg O-C curve for the times of maximum according to table ~\ref{tab:v1974cygomctab2}. }
\label{fig:v1974omcmaxtable2}
\end{figure}

\clearpage
\vspace{+0.5cm}
\begingroup
\small \renewcommand{\arraystretch}{0.8} \setlength{\tabcolsep}{3pt}
\begin{longtable}{|l|c|c|c|c|}
\caption[V1974 Cyg O-C for the times of maximum (longer period)]{V1974 O-C for the times of maximum using the ephemeris : $HJD_{min}=2457601.3476135 + 0.085 \times E$.}
\label{tab:v1974cygomctab3}
\renewcommand{\arraystretch}{0.7}
\cr \hline 
$HJD_{min}$   &  E       & O-C (days)        & O-C (cycles)     & Reference\\
\hline
  \endfirsthead
  \caption*{Continued. V1974 Cyg O-C for the times of maximum using the ephemeris : $HJD_{min}=2457601.3476135 + 0.085 \times E$.}\\
  \hline\hline   
$HJD_{min}$   &  E       & O-C (days)        & O-C (cycles)     & Reference\\
  \hline
  \endhead
  \hline
  \endfoot
\endlastfoot
2449471.540&-95645.0 &0.0174&  0.2045 &  (1)              \\
2449535.438&-94894.0 &0.0804&  0.9457 &  (1)              \\
2449537.390&-94871.0 &0.0774&  0.9104 &  (1)              \\
2449537.470&-94870.0 &0.0724&  0.8516 &  (1)              \\
2449546.406&-94765.0 &0.0834&  0.9810 &  (1)              \\
2449546.487&-94764.0 &0.0794&  0.9340 &  (1)              \\
2449547.424&-94753.0 &0.0814&  0.9575 &  (1)              \\
2449548.448&-94740.0 &0.0004&  0.0045 &  (1)              \\
2449548.524&-94740.0 &0.0764&  0.8987 &  (1)              \\
2449559.407&-94612.0 &0.0794&  0.9340 &  (1)              \\
2449559.490&-94611.0 &0.0774&  0.9104 &  (1)              \\
2449560.437&-94599.0 &0.0044&  0.0516 &  (1)              \\
2449561.367&-94589.0 &0.0844&  0.9928 &  (1)              \\
2449561.445&-94588.0 &0.0774&  0.9104 &  (1)              \\
2449561.533&-94587.0 &0.0804&  0.9457 &  (1)              \\
2449562.395&-94576.0 &0.0074&  0.0869 &  (1)              \\
2449562.479&-94575.0 &0.0064&  0.0751 &  (1)              \\
2449570.800&-94478.0 &0.0824&  0.9693 &  (3)              \\
2449588.658&-94267.0 &0.0054&  0.0634 &  (1)              \\
2449603.607&-94092.0 &0.0794&  0.9340 &  (1)              \\
2449604.714&-94079.0 &0.0814&  0.9575 &  (3)              \\
2449620.270&-93896.0 &0.0824&  0.9693 &  (4)              \\
2449620.347&-93895.0 &0.0744&  0.8751 &  (4)              \\
2449620.445&-93893.0 &0.0024&  0.0281 &  (4)              \\
2449625.547&-93833.0 &0.0044&  0.0516 &  (1)              \\
2449627.495&-93811.0 &0.0824&  0.9693 &  (1)              \\
2449627.495&-93811.0 &0.0824&  0.9693 &  (3)              \\
2449637.262&-93696.0 &0.0744&  0.8751 &  (1)              \\
2449637.341&-93695.0 &0.0684&  0.8045 &  (1)              \\
2449637.432&-93694.0 &0.0744&  0.8751 &  (1)              \\
2449642.534&-93634.0 &0.0764&  0.8987 &  (1)              \\
2449646.269&-93590.0 &0.0714&  0.8398 &  (1)              \\
2449646.358&-93589.0 &0.0754&  0.8869 &  (1)              \\
2449650.595&-93539.0 &0.0624&  0.7340 &  (3)              \\
2449651.295&-93531.0 &0.0824&  0.9693 &  (4)              \\
2449651.370&-93530.0 &0.0724&  0.8516 &  (4)              \\
2449651.540&-93528.0 &0.0724&  0.8516 &  (3)              \\
2449653.662&-93503.0 &0.0694&  0.8163 &  (3)              \\
2449655.617&-93480.0 &0.0694&  0.8163 &  (3)              \\
2449659.356&-93436.0 &0.0684&  0.8045 &  (1)              \\
2449659.440&-93435.0 &0.0674&  0.7928 &  (1)              \\
2449659.526&-93434.0 &0.0684&  0.8045 &  (1)              \\
2449661.220&-93414.0 &0.0624&  0.7340 &  (1)              \\
2449661.306&-93413.0 &0.0634&  0.7457 &  (1)              \\
2449661.392&-93412.0 &0.0644&  0.7575 &  (1)              \\
2449663.263&-93390.0 &0.0654&  0.7693 &  (1)              \\
2449663.344&-93389.0 &0.0614&  0.7222 &  (1)              \\
2449663.435&-93388.0 &0.0674&  0.7928 &  (1)              \\
2449665.216&-93367.0 &0.0634&  0.7457 &  (4)              \\
2449665.298&-93366.0 &0.0604&  0.7104 &  (4)              \\
2449665.384&-93365.0 &0.0614&  0.7222 &  (4)              \\
2449669.284&-93319.0 &0.0514&  0.6045 &  (1)              \\
2449693.170&-93038.0 &0.0524&  0.6163 &  (4)              \\
2449693.243&-93037.0 &0.0404&  0.4751 &  (4)              \\
2449693.322&-93036.0 &0.0344&  0.4045 &  (4)              \\
2449694.273&-93025.0 &0.0504&  0.5928 &  (4)              \\
2449695.203&-93014.0 &0.0454&  0.5340 &  (1)              \\                                            
2449695.290&-93013.0 &0.0474&  0.5575 &  (4)              \\
2449696.235&-93002.0 &0.0574&  0.6751 &  (4)              \\
2449696.314&-93001.0 &0.0514&  0.6045 &  (4)              \\
2449697.240&-92990.0 &0.0424&  0.4987 &  (4)              \\
2449697.320&-92989.0 &0.0374&  0.4398 &  (4)              \\
2449771.607&-92115.0 &0.0344&  0.4045 &  (4)              \\
2449862.444&-91046.0 &0.0064&  0.0751 &  (2)              \\
2449862.525&-91045.0 &0.0024&  0.0281 &  (4)              \\
2449863.469&-91034.0 &0.0114&  0.1340 &  (4)              \\
2449863.556&-91033.0 &0.0134&  0.1575 &  (4)              \\
2449875.520&-90893.0 &0.0774&  0.9104 &  (4)              \\
2449877.400&-90870.0 &0.0024&  0.0281 &  (4)              \\
2449878.340&-90859.0 &0.0074&  0.0869 &  (4)              \\
2449878.420&-90858.0 &0.0024&  0.0281 &  (4)              \\
2449878.507&-90857.0 &0.0044&  0.0516 &  (4)              \\
2449879.340&-90848.0 &0.0724&  0.8516 &  (4)              \\
2449879.442&-90846.0 &0.0044&  0.0516 &  (4)              \\
2449879.530&-90845.0 &0.0074&  0.0869 &  (4)              \\
2449880.380&-90835.0 &0.0074&  0.0869 &  (4)              \\
2449880.451&-90835.0 &0.0784&  0.9222 &  (4)              \\
2449880.545&-90833.0 &0.0024&  0.0281 &  (4)              \\
2449897.449&-90635.0 &0.0764&  0.8987 &  (4)              \\
2449903.385&-90565.0 &0.0624&  0.7340 &  (2)              \\
2449908.409&-90506.0 &0.0714&  0.8398 &  (2)              \\
2449925.319&-90307.0 &0.0664&  0.7810 &  (2)              \\
2449925.391&-90306.0 &0.0534&  0.6281 &  (4)              \\
2449925.392&-90306.0 &0.0544&  0.6398 &  (2)              \\
2449925.484&-90305.0 &0.0614&  0.7222 &  (4)              \\
2449925.574&-90304.0 &0.0664&  0.7810 &  (4)              \\
2449930.408&-90247.0 &0.0554&  0.6516 &  (4)              \\
2449972.301&-89754.0 &0.0434&  0.5104 &  (2)              \\
2449973.409&-89741.0 &0.0464&  0.5457 &  (2)              \\
2449977.240&-89696.0 &0.0524&  0.6163 &  (2)              \\
2449977.318&-89695.0 &0.0454&  0.5340 &  (4)              \\
2449977.408&-89694.0 &0.0504&  0.5928 &  (4)              \\
2449977.485&-89693.0 &0.0424&  0.4987 &  (4)              \\
2449990.407&-89541.0 &0.0444&  0.5222 &  (4)              \\
2449993.225&-89508.0 &0.0574&  0.6751 &  (2)              \\
2450005.280&-89366.0 &0.0424&  0.4987 &  (4)              \\
2450009.270&-89319.0 &0.0374&  0.4398 &  (4)              \\
2450010.224&-89308.0 &0.0564&  0.6634 &  (4)              \\
2450011.470&-89293.0 &0.0274&  0.3222 &  (4)              \\
2450012.334&-89283.0 &0.0414&  0.4869 &  (4)              \\
2450012.422&-89282.0 &0.0444&  0.5222 &  (2)              \\
2450018.375&-89212.0 &0.0474&  0.5575 &  (4)              \\
2450038.264&-88978.0 &0.0464&  0.5457 &  (2)              \\
2450038.351&-88977.0 &0.0484&  0.5693 &  (4)              \\
2450041.251&-88943.0 &0.0584&  0.6869 &  (4)           \\
2450041.336& -88942.0&0.0584&  0.6869 &  (4)           \\
2450060.274& -88719.0&0.0414&  0.4869 &  (4)           \\
2450234.939& -86664.0&0.0314&  0.3693 &  (2)           \\
2450237.836& -86630.0&0.0384&  0.4516 &  (3)           \\
2450239.965& -86605.0&0.0424&  0.4987 &  (3)           \\
2450240.822& -86595.0&0.0494&  0.5810 &  (3)           \\
2450242.777& -86572.0&0.0494&  0.5810 &  (3)           \\
2450243.799& -86560.0&0.0514&  0.6045 &  (3)           \\
2450246.531& -86528.0&0.0634&  0.7457 &  (3)           \\
2450248.901& -86500.0&0.0534&  0.6281 &  (3)           \\ 
2450250.777& -86478.0&0.0594&  0.6987 &  (3)           \\ 
2450252.910& -86453.0&0.0674&  0.7928 &  (3)           \\ 
2450254.869& -86430.0&0.0714&  0.8398 &  (3)           \\ 
2450256.831& -86407.0&0.0784&  0.9222 &  (3)           \\ 
2450257.849& -86395.0&0.0764&  0.8987 &  (3)           \\ 
2450258.790& -86384.0&0.0824&  0.9693 &  (3)           \\ 
2450258.957& -86382.0&0.0794&  0.9340 &  (3)           \\ 
2450260.747& -86361.0&0.0844&  0.9928 &  (3)           \\ 
2450261.856& -86347.0&0.0034&  0.0398 &  (3)           \\ 
2450262.791& -86336.0&0.0034&  0.0398 &  (3)           \\ 
2450285.608& -86068.0&0.0404&  0.4751 &  (3)           \\ 
2450297.715& -85926.0&0.0774&  0.9104 &  (3)           \\ 
2450299.497& -85905.0&0.0744&  0.8751 &  (3)           \\                
2450300.608& -85892.0&0.0804&  0.9457 &  (3)           \\
2450309.640& -85785.0&0.0174&  0.2045 &  (3)           \\
2450326.679& -85585.0&0.0564&  0.6634 &  (3)           \\
2450327.704& -85573.0&0.0614&  0.7222 &  (3)           \\
2450329.663& -85550.0&0.0654&  0.7693 &  (3)           \\
2450333.406& -85506.0&0.0684&  0.8045 &  (3)           \\
2450334.691& -85491.0&0.0784&  0.9222 &  (3)           \\
2450335.790& -85478.0&0.0724&  0.8516 &  (3)           \\
2450343.382& -85388.0&0.0144&  0.1693 &  (3)           \\
2450361.351& -85177.0&0.0484&  0.5693 &  (5)           \\
2457600.285& -13.0   &0.0428&  0.5041 &  (6)           \\
2457600.367& -12.0   &0.0396&  0.4658 &  (6)           \\
2457601.348& 0.0     &0.0000&  0.0000 &  (6)           \\
\hline
\caption*{References: (1) \citet{1995AcA....45..365S}, (2) \citet{1996AcA....46..311O}, (3) \citet{1997PASP..109..114S}, (4) \citet{2002AcA....52..273O},(5) \citet{1997MNRAS.286..745R}, (6) Present work.}
\end{longtable}

\begin{figure}
\centering
\includegraphics[height=14cm,width=13cm]{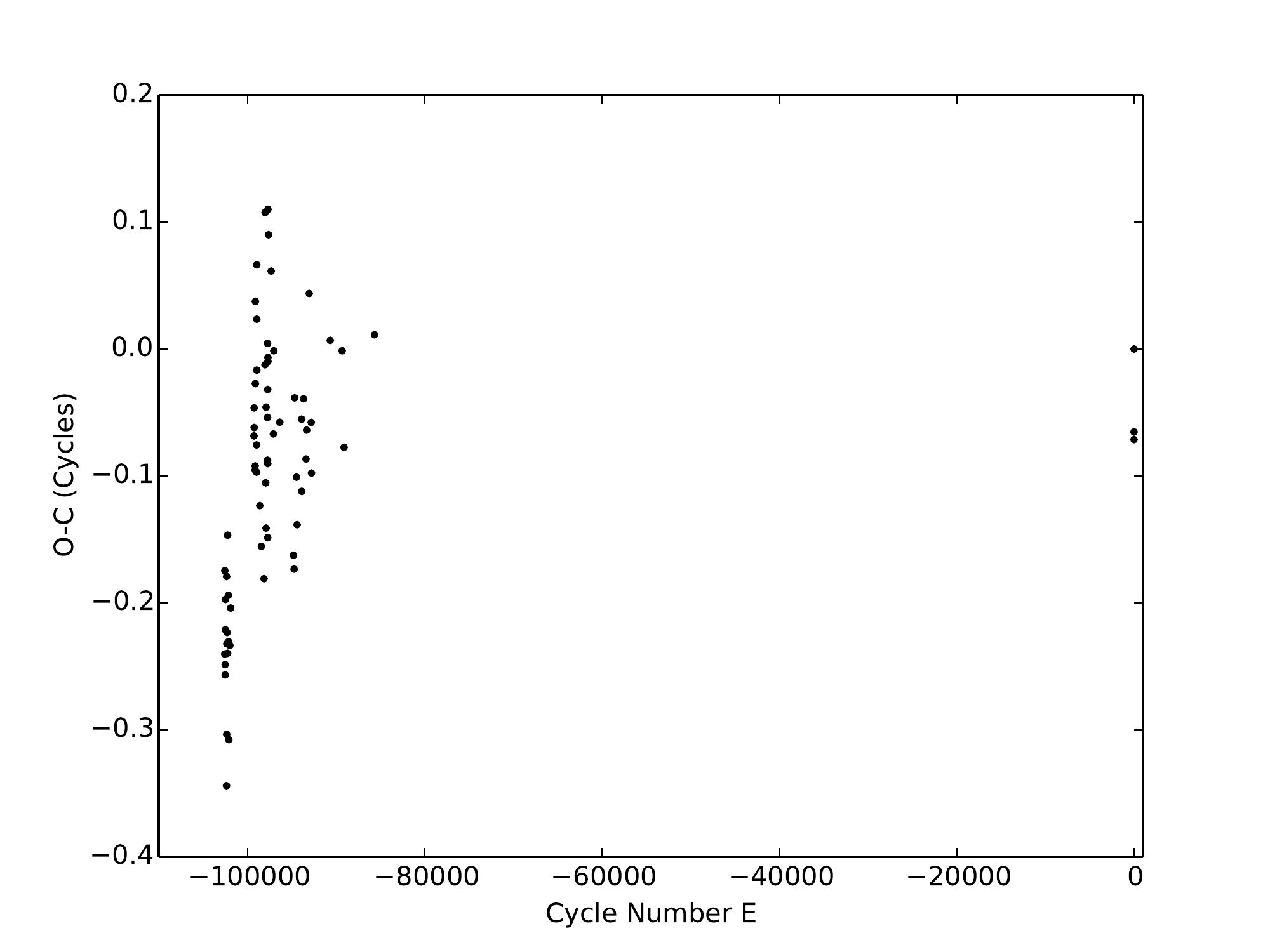}
\caption{V1974 Cyg O-C curve for the times of maximum according to table ~\ref{tab:v1974cygomctab3}.}
\label{fig:v1974omcmaxtable3}
\end{figure}

\clearpage
\section{Conclusion}\label{sec:conc}
\subsection{UV Spectroscopy}\label{subsec:conc1}

In this study we present the results of three classical novae and the main conclusions can be summarized as follows $\colon$

1 - The evolution of the continuum at 1455 is representative of the evolution of the continuum along the whole short wavelength range.

2 - The difference between the three systems in the WD mass ($0.83 \mathrm{M_{\odot}}$ for PW Vul, $0.95 \mathrm{M_{\odot}}$ for V1668 Cyg and $1.05 \mathrm{M_{\odot}}$ for V1974 Cyg) is the main cause behind the difference in the speed class and the spectral evolution. 

3 - The evolutionary curve of the normalized flux $(F_{\lambda}/F_c)$ for most of the lines is similar to the evolutionary curve of the absolute flux $(F_{\lambda})$.

4 - The change in the ionization conditions such as opacity, temperature and density of the envelope leads to the variation of $F_{\lambda}/F_c$ for different lines during  different phases of the outburst.


5 - Average expansion speeds, masses and average mass accretion rates in quiescence suggest that the models that best describe the three novae PW Vul, V1668 Cyg and V1974 Cyg are CO2, CO3 and ONe2 models of \citet{1998ApJ...494..680J}, respectively.

6 - The difference between the models we concluded for the two novae (PW Vul and V1668 Cyg) using the dynamical parameters and the models deduced from the chemical abundances needs to be resolved. This can probably be done if the models predict the chemical abundances in later days of the outburst when abundances can be determined from nebular line analysis. Also, the agreement between the chemical properties and the dynamical properties for each model has to be tested using accurate determinations of these properties from observations of nova outbursts.

7 - The temperature of the central white dwarf of V1974 Cyg estimated from the fitted continuum is consistent with temperatures estimated by other authors.

8 - The evolution of the spectral lines of V1974 Cyg follows the linear relation between the time of maximum and the ionization potential of \citet{2005A&A...439..205C} although this relation was derived for CO novae while V1974 is an ONe nova.

9 - The time of decline of continuum evolution curve of V1974 Cyg is consistent with the time when hydrogen burning ends according to the model of \citet{2005ApJ...631.1094H}.


10 - The evolutionary sequence described in \citet{2005A&A...439..205C} can be applied to V1974 Cyg adding the 'fireball' and 'Iron Curtain' phases before the initial phase.

\subsection{Optical Photometry}\label{subsec:conc2}

We observed the three post-novae (PW Vul, V1668 Cyg and V1974 Cyg) for several nights in 2015 and 2016 in the V, $\mathrm{R_c}$ and $\mathrm{I_c}$ filters using the 1.88 m telescope at the Kottamia Astronomical Observatory, and it can be concluded from the analysis of the optical photometric observations that $\colon$

1 - There is a phase shift in the light curves of the three systems. More photometric observations are required so that we can construct an O-C diagram and test the stability of the orbit.

2 - The amplitude of variation of the light curve of V1668 Cyg was $\sim$ 2.5 mag in the $\mathrm{R_c}$ filter, $\sim$ 2.3 mag in the $\mathrm{I_c}$ filter and $\sim$ 3.9 mag in the V filter.

3 - The amplitude of variation was $\sim$ 0.4 mag for the $\mathrm{I_c}$ light curve of PW Vul in the observations taken on 13-10-2015, $\sim$ 0.5 mag for the $\mathrm{I_c}$ filter observations taken on 5-10-2016, $\sim$ 0.9 mag for both the $\mathrm{R_c}$ and V filter observations.

4 - The amplitude of variation of V1974 Cyg light curve was $\sim$ 0.7 mag for the $\mathrm{I_c}$ observations taken on 20-10-2015, $\sim$ 0.3 mag for the $\mathrm{I_c}$ filter observations taken on 31-7-2016, $\sim$ 2.1 mag for the $\mathrm{R_c}$ filter observations taken on 30-7-2016 and $\sim$ 0.2 mag for the V filter observations.

5 - It can be seen that V1668 Cyg light curves showed the highest amplitude of variation among the three systems we studied. This is because V1668 Cyg is an eclipsing binary star.

6 - We detected three times of maxima and three times of minima in the light curves of V1974 cyg and we used them to calculate the O-C and added them to the values determined by other authors. Further observations are required to determine the stability of the orbit.

7 - All the photometric light curves of the three systems were best fitted by a Fourier function of the sixth degree (Equation ~\ref{eq:fourierfit}).

8 - More observations are needed to perform a full light curve analysis,including a period search, for the systems and determine the different orbital and physical parameters.

\bibliographystyle{an} 
\medskip
\renewcommand{\bibname}{References}
\bibliography{ghamed_thesis.bbl}
\includepdf[pages=-,pagecommand={},width=\textwidth]{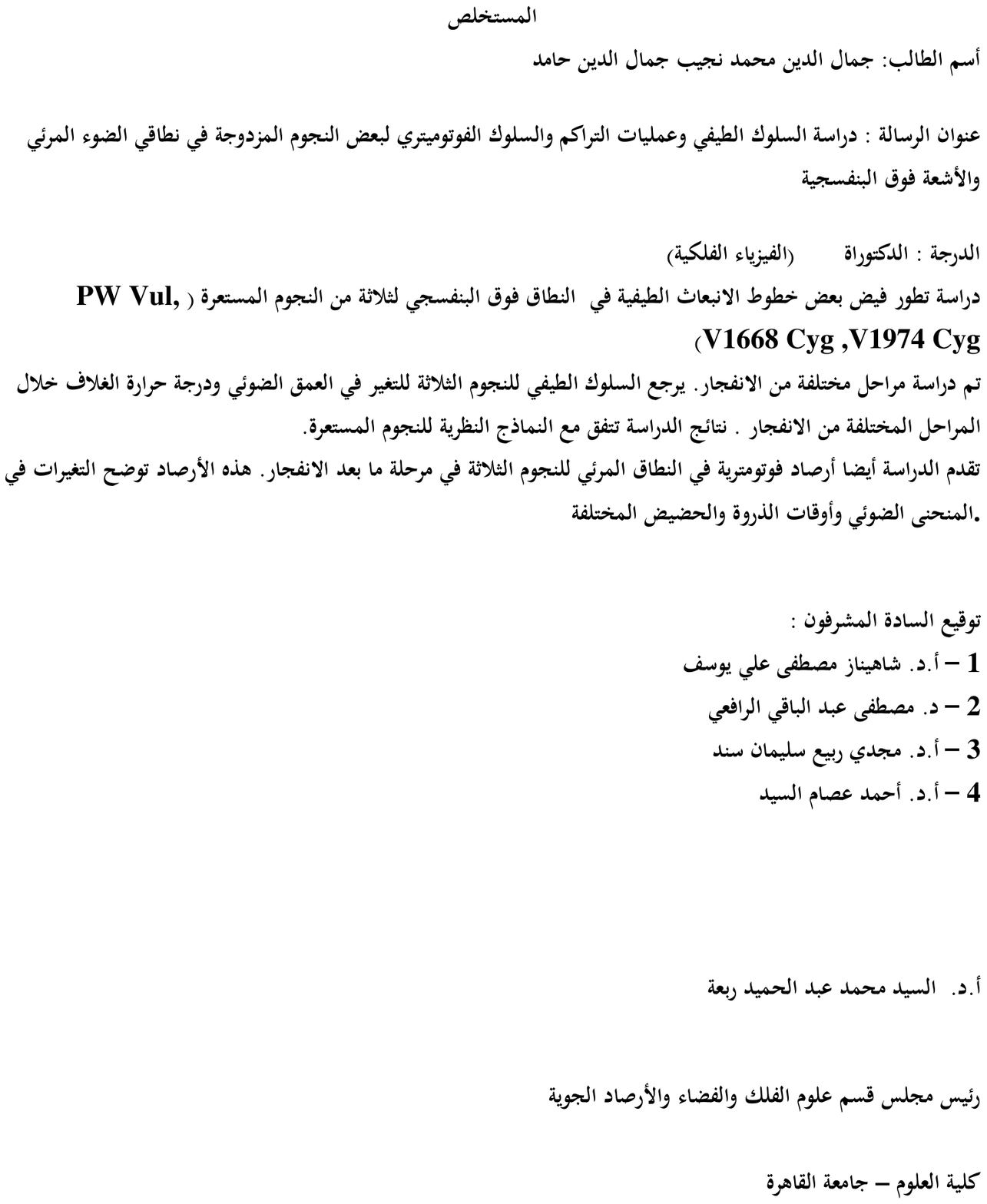}
\end{document}